\documentclass[
aps,
pra,
twocolumn,
superscriptaddress,
floatfix,
longbibliography
]{revtex4-1}
\usepackage[dvipdfmx]{graphicx}
\usepackage{dcolumn}
\usepackage{bm}
\usepackage{ulem} 
\usepackage{amsmath}
\usepackage{amssymb}
\usepackage{txfonts}
\usepackage{hyperref}
\usepackage{color} 
\usepackage{xcolor}
\usepackage{listings}
\newcommand{\bs}   {\boldsymbol}

\newcommand{\Tr}{{\rm Tr}}
\newcommand{\e}{{\rm e}}
\newcommand{\imag}{{\rm i}}
\newcommand{\dd}{{\rm d}}

\hypersetup{
  colorlinks=true,
  linkcolor=[rgb]{0.60,0.00,0.00},
  citecolor=[rgb]{0.00,0.00,0.60},
  urlcolor=[rgb]{0.00,0.00,0.60},
  setpagesize=false
}

\definecolor{codegreen}{rgb}{0,0.6,0}
\definecolor{codegray}{rgb}{0.5,0.5,0.5}
\definecolor{codepurple}{rgb}{0.58,0,0.82}
\definecolor{backcolour}{rgb}{0.95,0.95,0.92}

\lstdefinestyle{mystyle}{
    backgroundcolor=\color{backcolour},   
    commentstyle=\color{codegreen},
    keywordstyle=\color{magenta},
    numberstyle=\tiny\color{codegray},
    stringstyle=\color{codepurple},
    basicstyle=\ttfamily\footnotesize,
    breakatwhitespace=false,         
    breaklines=true,                 
    captionpos=b,                    
    keepspaces=true,                 
    numbers=left,                    
    numbersep=5pt,                  
    showspaces=false,                
    showstringspaces=false,
    showtabs=false,                  
    tabsize=2
}

\begin{document}

\title{
  Energy-filtered random-phase states as 
  microcanonical thermal pure quantum states 
}

\author{Kazuhiro~Seki}
\affiliation{Quantum Computational Science Research Team, RIKEN Center for Quantum Computing (RQC), Saitama 351-0198, Japan}

\author{Seiji~Yunoki}
\affiliation{Quantum Computational Science Research Team, RIKEN Center for Quantum Computing (RQC), Saitama 351-0198, Japan}
\affiliation{Computational Materials Science Research Team, RIKEN Center for Computational Science (R-CCS),  Hyogo 650-0047,  Japan}
\affiliation{Computational Quantum Matter Research Team, RIKEN Center for Emergent Matter Science (CEMS), Saitama 351-0198, Japan}
\affiliation{Computational Condensed Matter Physics Laboratory, RIKEN Cluster for Pioneering Research (CPR), Saitama 351-0198, Japan}

\begin{abstract}
  We propose a method to calculate finite-temperature properties of a quantum many-body 
  system for a microcanonical ensemble by introducing a pure quantum state named here an energy-filtered 
  random-phase state, which is also a potentially promising application of near-term quantum computers. 
  In our formalism, a microcanonical ensemble is specified by two parameters, i.e., 
  the energy of the system and its associated energy window. Accordingly, 
  the density of states is
  expressed as a sum of Gaussians centered at the target energy
  with its spread corresponding to the width of the energy window.  
  We then show that the thermodynamic quantities such as entropy and temperature are calculated 
  by evaluating the trace of the time-evolution operator and the trace of the time-evolution operator 
  multiplied by the Hamiltonian of the system. We also describe how these traces can be evaluated using
  random diagonal-unitary circuits appropriate to quantum computation.  
  The pure quantum state 
  representing 
  our microcanonical ensemble is related to a state of the form 
  introduced by Wall and Neuhauser for
  the filter diagonalization method
  \href{https://doi.org/10.1063/1.468999}{[M. R. Wall and D. Neuhauser, J. Chem. Phys. {\bf 102}, 8011 (1995)]},
  and therefore we refer to it as an energy-filtered random-phase state. 
  The energy-filtered random-phase state is essentially a Fourier transform 
  of a time-evolved state whose initial state is prepared as a random-phase state,
  and the cut-off time in the time-integral for the Fourier transform 
  sets the inverse of the width of the energy window.
  The proposed method is demonstrated numerically by calculating
  thermodynamic quantities for the one-dimensional spin-1/2 Heisenberg model
  on small clusters up to 28 qubits, showing that
  the method is most efficient for the target energy around which 
  the dense distribution of energy eigenstates is found.
\end{abstract}

\date{\today}

\maketitle

\section{Introduction}

In order to study thermal properties of quantum many-body systems
based on the thermodynamic ensembles in statistical mechanics such as 
the microcanonical,
canonical, or
grand-canonical ensemble, 
one might have to calculate 
the number of microstates,
the partition function, or
the grand partition function,
as their logarithms give 
the thermodynamic potentials which
characterize thermodynamic states.
Generally, the microcanonical ensemble formalism is suitable for treating an isolated system,
while the canonical or grand canonical ensemble formalism is useful for describing a system
coupled to an environment. 
For quantum many-body systems,
these thermodynamic ensembles are expressed
by the corresponding density matrices, 
which are statistical mixtures of pure quantum states~\cite{Fano1957}. 
On the other hand,
it is often computationally useful to formulate 
statistical mechanics based directly on pure quantum states,
instead of explicitly calculating the density matrices. 
The pure-state-based approaches make use
of the fact that the statistical expectation
values of observables are given by the trace of some operators,  
e.g., a product of the thermal density matrix and an observable. 
Indeed, the trace can be evaluated 
without knowing the eigenvalues of these operators.  
Depending on how the trace is evaluated, 
the pure-state-based approaches might be classified
into two classes. 

A well known class of the pure-state-based approach
is the purification, where the system size is doubled and
the thermal expectation values are calculated as
an inner product of pure states in a whole (double-sized) system~\cite{Fano1957}.
In particular, the thermofield-double states~\cite{Suzuki1985}
are such pure states that are in accordance with the canonical ensemble, 
implemented for example with the density-matrix-renormalization-group method~\cite{Feiguin2005}
and recently with the variational Monte Carlo method~\cite{Nomura2021}. 
It is also worth noting that, 
given a pure state of a whole system,  
similarity (or in particular cases equivalence) 
between the reduced density matrix of a subsystem 
obtained by bipartitioning the whole system
and the thermal density matrix of the subsystem
has been recently explored~\cite{Li2008,Poilblanc2010,Qi2012,Dalmonte2018,Giudici2018,MendesSantos2020,Seki2020thermal}.

Another class of the pure-state-based formalism
makes use of random states~\cite{Tolman}
to evaluate the trace of observables~\cite{Drabold1993,Iitaka2004,Weisse2006,Jin2021}. 
In this class, several different methods for 
the different thermodynamic ensembles have been developed. 
The methods for the canonical ensemble include
the finite-temperature Lanczos method~\cite{Jaklic1994,Prelovsek} and 
the canonical thermal-pure-quantum(TPQ)-state formalism~\cite{Sugiura2013}.  
The methods for the microcanonical ensemble include
the microcanonical Lanczos method~\cite{Long2003,Okamoto2018} and 
the TPQ-state formalism~\cite{Sugiura2012}.

Here, we briefly review state-preparation schemes of the latter class. 
Let $\hat{\cal H}$ be a Hamiltonian describing the system of interest. 
In the canonical formalism, for a given inverse temperature $\beta$, 
a canonical TPQ state is obtained by applying 
the positive-definite Hermitian operator $\e^{-\beta \hat{\cal H}/2}$
to a random-phase state,  
for which 
either the Lanczos method~\cite{Jaklic1994}, 
the block Lanczos method~\cite{Seki2020ftlm},
the polynomial-expansion method~\cite{Weisse2006}, or
the Suzuki-Trotter decomposition~\cite{Suzuki1990} is often employed. 
In the microcanonical formalism, for a given energy $E$, 
the lowest eigenstate of $(\hat{\cal H}-E)^2$, 
which can be obtained via a Lanczos method, 
is used as a TPQ state in Refs.~\cite{Long2003,Okamoto2018},
while the repeated application (i.e., the power iteration)
of the properly rescaled Hamiltonian to a random-phase state 
is used to obtain a TPQ state in Ref.~\cite{Sugiura2012}. 

As compared to the canonical-TPQ-state-preparation schemes, 
the microcanonical TPQ-state-preparation schemes seem
rather indirect in the sense that
these methods do not explicitly specify a width of an energy window around a given target energy $E$, 
which could however be important particularly for finite-size systems where  
the distribution of the energy eigenvalues is so sparse that
the number of microstates depends crucially 
not only on the target energy $E$
but also on the energy window.
Indeed, the width of the energy window in the microcanonical Lanczos method~\cite{Long2003}
can be estimated from the standard deviation 
of the energy expectation value with respect to the
so-obtained lowest eigenstate of $(\hat{\cal H}-E)^2$, 
thus not being an input parameter, and moreover it does depend on the
number of Lanczos iterations. 
In addition, the power iteration used in the TPQ-state formalism~\cite{Sugiura2012}
is by construction sequential, starting with a high-energy TPQ state to which the properly rescaled 
Hamiltonian power is applied as many times as required to reach a TPQ state with a desired energy. 
While the sequential construction of TPQ states may be  
advantageous when the thermal properties of the system 
in the whole energy range are intended to be examined, 
it is still highly desirable to 
prepare a TPQ state of an arbitrary target energy 
with an arbitrary energy window especially 
when the energy range of interest is limited.

In this paper, we propose a method to calculate
finite-temperature properties of many-body systems  
for a microcanonical ensemble, which may find 
a potential application of near-term quantum computers.
In our formalism, a microcanonical ensemble is specified with a target energy 
as well as a width of the energy window by expressing the density of states  
as a sum of Gaussians centered at the 
target energy with its spread associated with the width of the energy window. 
Using the Fourier representation of the Gaussian, we can then show that 
thermodynamic quantities such as entropy and temperature 
can be calculated by evaluating the trace of the time-evolution operator, which is thus unitary, 
and the trace of the time-evolution operator multiplied by the Hamiltonian of the system. 
We also describe how these traces can be evaluated 
using random diagonal-unitary circuits suitable for quantum computation.  
The pure state representing 
our microcanonical ensemble is closely related to a state 
introduced in the filter-diagonalization method in quantum chemistry~\cite{Wall1995}  
and thus is named here an energy-filtered random-phase state.  
The energy-filtered random-phase state is described essentially 
by the Fourier transform of a time-evolved random-phase state with 
its dynamics governed by the Hamiltonian of the system and 
the cut-off time in the time-integral for the Fourier transform is associated with the inverse of 
the width of the energy window specifying the microcanonical ensemble. 
We demonstrate the proposed method by numerically calculating
thermodynamic quantities of the one-dimensional spin-1/2 Heisenberg model on small clusters 
and show that the proposed method 
is most effective for the target energy around which a larger
number of energy eigenstates exist.  

The rest of this paper is organized as follows.
In Sec.~\ref{sec:def}, 
we define the density of states and, accordingly, the entropy and the inverse temperature 
that are evaluated from the density of states.  
In Sec.~\ref{sec:form}, we formulate the proposed method  
suitable for numerical simulations on classical computers as well as 
quantum computers. 
By further examining our formalism as a pure-state approach to a microcanonical
ensemble, we show in Sec.~\ref{sec:form_purestate} that the energy-filtered state with an initial random state can be considered 
as a microcanonical TPQ state. 
In Sec.~\ref{sec:method},
we introduce a concrete Hamiltonian of the system for the demonstration of the proposed method, and describe the preparation 
of initial random states and the treatment of the time-evolution operator, both of which are 
appropriate to quantum computation. 
In Sec.~\ref{sec:results},
we demonstrate our proposed method by numerically calculating
thermodynamic quantities of the one-dimensional spin-1/2 Heisenberg model 
up to 28 qubits.
Finally, the paper is concluded with discussion 
in Sec.~\ref{sec:conclusions}.
Additional details on
the cumulative number of states, 
the energy-filtered random-phase state,
translation to the canonical ensemble, and 
the random-phase-product states as well as
additional numerical results 
are provided in
Appendixes. 
Throughout the paper, we set $\hbar=1$ and the Boltzmann constant $k_B=1$.

\section{Preliminaries} \label{sec:def}
In this section,
we define the density of states as well as the entropy and
the inverse temperature. 
We also give a few remarks on the choice of the energy window 
because the number of states within the energy window is a central quantity 
in the microcanonical ensemble. 

\subsection{
  Density of states, entropy, and inverse temperature} \label{sec:dos}

Let $\hat{\cal H}$ be the Hamiltonian describing the system of interest
on a Hilbert space of dimension $D$, 
and let $\{E_n,|E_n\rangle\}_{n=0}^{D-1}$ be the eigenpairs of
$\hat{\cal H}$ such that $\hat{\cal H}|E_n\rangle = E_n |E_n\rangle$.
Without loss of generality, we assume that
$E_0 \leqslant E_1 \leqslant \cdots \leqslant E_{D-1}$.
For given input parameters $E$ and $\tau$, 
which have the dimensions of energy and time, respectively,
we define the density of states $g_\tau(E)$ at energy $E$ as 
\begin{equation}
  g_\tau(E) \equiv 
  \frac{\tau}{\sqrt{\pi}}
  \sum_{n=0}^{D-1}
  \e^{-(E_n-E)^2\tau^{2}}.
  \label{eq:dos}
\end{equation}
It is plausible to call $g_\tau(E)$ the density of states because
$g_\tau(E)$ is reduced to a sum of delta functions
in the limit of $\tau \to \infty$ as 
$
\lim_{\tau\to \infty}g_\tau(E)=
\sum_{n=0}^{D-1}
\delta(E_n-E),
$
which is the conventional form of the density of states. 
Here, the Gaussian representation of the delta function
$\delta(x-\mu)=
\lim_{\sigma\to 0}\frac{1}{\sqrt{2\pi}\sigma} \e^{-\frac{(x-\mu)^2}{2\sigma^2}}$
is used.

In accordance with the microcanonical ensemble, 
we define the entropy $S_\tau(E)$ as the logarithm of the number of states within a given energy window, i.e., 
\begin{alignat}{1}
  S_\tau(E)
  &\equiv \ln{\left[g_\tau(E) \frac{\sqrt{\pi}}{\tau} \right]},  \label{eq:S}
\end{alignat}
and the inverse temperature $\beta_\tau(E)$ as the derivative of $S_\tau(E)$ with respect to $E$, i.e., 
\begin{alignat}{1}
  \beta_\tau(E)
  &\equiv \frac{\partial S_\tau (E)}{\partial E}
  =
  2\tau^2 \left[
    {\cal E}_{\tau}(E)-E
  \right],
  \label{eq:beta}
\end{alignat}
where
\begin{equation}
  {\cal E}_{\tau}(E)\equiv 
  \frac{\sum_{n=0}^{D-1} E_n \e^{-(E_n-E)^2 \tau^2}}
       {\sum_{n=0}^{D-1} \e^{-(E_n-E)^2 \tau^2}}.
       \label{eq:E}
\end{equation}
We introduced the factor $\frac{\sqrt{\pi}}{\tau}$ in Eq.~(\ref{eq:S})
as the width of the energy window, which 
makes the argument in the logarithm dimensionless.
Hereafter, we refer to $E$ as the target energy and $\tau$ as the filtering time 
for reasons discussed in Sec.~\ref{sec:EFS}.

\subsection{Trace formalism}
Let us introduce a dimensionless positive-definite Hermitian operator 
\begin{equation}
  \hat{G}_\tau(E) \equiv 
  \e^{-(\hat{\cal H}-E)^2\tau^2}. 
  \label{eq:Gop}
\end{equation}
We can then readily show that 
the density of states,
the entropy, and
the inverse temperature in Eqs.~(\ref{eq:dos})--(\ref{eq:beta}) 
are now expressed 
as 
\begin{alignat}{1}
  \label{eq:gTr}
  g_\tau(E) &= \frac{\tau}{\sqrt{\pi}} \Tr\left[\hat{G}_\tau(E)\right],\\
  \label{eq:STr}
  S_\tau(E) &= \ln{\left(\Tr\left[\hat{G}_\tau(E)\right] \right)}, 
\end{alignat}
and 
\begin{alignat}{1}
  \label{eq:betaTr}
  \beta_\tau(E) &=2\tau^2\left(
  \frac{\Tr\left[\hat{\cal H}\hat{G}_\tau(E)\right]}{\Tr\left[\hat{G}_\tau(E)\right]}
  -E
    \right), 
\end{alignat}
respectively. Namely, these quantities are all expressed in terms of
$\Tr\left[\hat{G}_\tau(E)\right]$ and $\Tr\left[\hat{\cal H}\hat{G}_\tau(E)\right]$. 
Therefore, owning to these trace operations, the microscopic energy eigenvalues $\{E_n\}_{n=0}^{D-1}$
have ostensibly disappeared in the expressions of the density of states and the thermodynamic quantities 
$S_\tau(E)$ and $\beta_\tau(E)$, as it should. 

Note that $\hat{\rho}_{\rm mic}\equiv \hat{G}_\tau(E)/\Tr[\hat{G}_\tau(E)]$
can be regarded as a density matrix, and a brief consideration on
such a Gaussian form of the density matrix, not limited to
the microcanonical ensemble, can be found in Ref.~\cite{Tolman}. 
Moreover, analytical properties of this ensemble have been
explored in detail in Refs.~\cite{Challa1988,Yoneta2019}.
Note also that the energy expectation value in Eq.~(\ref{eq:E}) is 
expressed as
\begin{equation}
  {\cal E}_{\tau}(E)
  = \frac{\Tr\left[\hat{\cal H}\hat{G}_\tau(E)\right]}{\Tr\left[\hat{G}_\tau(E)\right]}
  =\Tr\left[\hat{\rho}_{\rm mic}\hat{\cal H}\right].
  \label{eq:ETr}
\end{equation}
Equations~(\ref{eq:gTr})-(\ref{eq:ETr}) summarize
the density of states,
the entropy,
the inverse temperature, and
the energy expectation value, 
which are the central quantities 
considered in this study.

\subsection{Energy window and number of states in it}~\label{sec:window}
We now make comments on our choice of the width of the energy window, i.e., 
the factor $\frac{\sqrt{\pi}}{\tau}$ in Eq.~(\ref{eq:S}), 
and the number of states within the energy window. 
Approximately,  
$g_\tau(E) \frac{\sqrt{\pi}}{\tau}=
\Tr[\hat{G}_\tau(E)]=
\sum_{n=0}^{D-1}\e^{-(E_n-E)^2\tau^{2}}$
in the logarithm of Eqs.~(\ref{eq:S}) and (\ref{eq:STr})
counts the number of energy eigenstates $|E_n\rangle$ such that 
\begin{equation}
  E-\frac{\delta E}{2} \leqslant E_n < E+\frac{\delta E}{2},
  \label{eq:range}
\end{equation}
where $\delta E$ is the width of the energy window here chosen as 
\begin{equation}
  \delta E = \frac{\sqrt{\pi}}{\tau}. 
  \label{eq:width}
\end{equation}
Based on the dimensional analysis, $1/\tau$ dependence of $\delta E$ is essential, but 
the factor $\sqrt{\pi}$ in Eq.~(\ref{eq:width}) is not.
Here, the factor in $\delta E$ is selected as in Eq.~(\ref{eq:width})
so as for the number of states in the zero-filtering-time limit, i.e., $\tau\to0$, corresponding to
the infinitely large width of the energy window, to satisfy 
\begin{equation}
  \lim_{\tau \to 0} g_\tau(E) \delta E =
  \lim_{\tau \to 0} \Tr[\hat{G}_{\tau}(E)]=D, 
  \label{eq:nos_tau0}
\end{equation}
independently of $E$. 
In Appendix~\ref{app:cumulative}, we show that
the cumulative number of states is given by a sum of complementary error functions. 
Finally, we note that 
$\delta E = \sqrt{\pi}/\tau\, (\sim 1.772/\tau)$ has nothing to do with
the ``two sigma'' $2/\sqrt{2}\tau = \sqrt{2}/\tau\, (\sim 1.414/\tau)$
nor the full width at half maximum (FWHM, 
$2\sqrt{2\ln2}\frac{1}{\sqrt{2}\tau}\sim 1.665/{\tau}$)
of the Gaussian in Eq.~(\ref{eq:Gop}).

\section{Formalism}\label{sec:form}
In this section, we describe how to evaluate 
$\Tr[\hat{G}_{\tau}(E)]$ and
$\Tr[\hat{\cal H}\hat{G}_{\tau}(E)]$
without explicitly calculating the energy eigenvalues $\{E_n\}_{n=0}^{D-1}$
of the Hamiltonian.
In particular, we shall provide a formalism
to calculate these quantities by exploiting quantum computers. 
To this end,
we first express $\hat{G}_\tau(E)$
in terms of the time-evolution operator, which is thus 
compatible with quantum computation. 
We then discuss quantum circuits for the trace evaluation.

\subsection{Fourier representation of $\hat{G}_{\tau}(E)$}
Since $\hat{G}_\tau(E)$ is a Gaussian in the energy domain,
it can be expressed by a Fourier transform of a Gaussian in the time domain as 
\begin{equation}
  \hat{G}_\tau(E)=
  \frac{1}{2\sqrt{\pi}\tau}
  \int_{-\infty}^{\infty} \dd t\ 
  \e^{-\frac{t^2}{4\tau^2}}
  \e^{\imag E t}
  \hat{U}(t),
  \label{eq:GFT}
\end{equation}
where 
\begin{equation}
  \hat{U}(t)\equiv \e^{-\imag \hat{\cal H}t} 
  \label{eq:ut}
\end{equation}
is the time-evolution operator
generated by the time-independent Hamiltonian $\hat{\cal H}$. 
Therefore, 
$\Tr[\hat{G}_\tau(E)]$ and  $\Tr[\hat{\cal H}\hat{G}_\tau(E)]$
can be evaluated by integrating
$\Tr[\hat{U}(t)]$ and $\Tr[\hat{\cal H}\hat{U}(t)]$
over time $t$, respectively.
In the following, we discuss how these traces can be
evaluated using random-phase states, which are also suitable for quantum computation.

\subsection{Trace evaluation by random sampling}
\label{sec:trace}
Here, we briefly review the fact that
the trace of an operator can be evaluated
by taking random average 
of its diagonal matrix elements 
of the operator   
with respect to random-phase states~\cite{Iitaka2004}, 
focusing particularly on random-phase states prepared by unitary operators, 
which is thus compatible with quantum computation, especially, 
random-diagonal circuits and diagonal-unitary designs~\cite{Nakata2014}.
We also briefly discuss the corresponding variance~\cite{Jin2021}.

The state called a 
random-phase state~\cite{Iitaka2004} or a
phase-random state with equal amplitudes~\cite{Nakata2012} has the form of
\begin{equation}
  |\Phi_{r}\rangle= \frac{1}{\sqrt{D}}\sum_{b=0}^{D-1}
  \e^{\imag \theta_{b,r}}|b\rangle,
  \label{eq:rps}
\end{equation}
where $\{|b\rangle\}_{b=0}^{D-1}$ are the computational-basis states
satisfying $\langle b|b^\prime\rangle=\delta_{bb^\prime}$ and
$\sum_{b=0}^{D-1}|b\rangle\langle b|=\hat{I}$, 
$\{\theta_{b,r}\}_{b=0}^{D-1}$ are random variables 
drawn uniformly from $[0,2\pi)^D$,
and $r$ specifies the set of random variables.
To be specific, we consider an $N$-qubit system with $D=2^N$
and assume that $\{|b\rangle\}_{b=0}^{D-1}$
consists of the eigenstates of Pauli-$Z$ operators.
A basis state $|b\rangle$ possesses 
a bit string of length $N$ associated with
the binary representation of $b$.
Then, $|\Phi_r\rangle$ in Eq.~(\ref{eq:rps}) can be written as
\begin{equation}
  |\Phi_r\rangle =
  \hat{V}_r |+\rangle,
  \label{eq:rps2}
\end{equation}
where
\begin{equation}
  \hat{V}_r
  \equiv
  \sum_{b=0}^{D-1}
  \e^{\imag \theta_{b,r}}
  |b\rangle
  \langle b|
  \label{eq:Vr}
\end{equation}
is a diagonal-unitary operator and
\begin{equation}
  |+\rangle \equiv
  \hat{H}^{\otimes N}|0\rangle^{\otimes N}
  =
  \frac{1}{\sqrt{D}}\sum_{b=0}^{D-1}|b\rangle
\end{equation}
with $\hat{H}$ being the Hadamard gate such that $\hat{H}|0\rangle = \frac{1}{\sqrt{2}}
(|0\rangle + |1\rangle)$. Here, $|0\rangle$ ($|1\rangle$)
is the eigenstate of the Pauli-$Z$ operator with eigenvalue $+1$ ($-1$).

We can now relate the trace operation and the random average
(also see Ref.~\cite{Iitaka2004}).
Let us consider the random average of 
some quantity $O_r$,
which is a functional of 
the random diagonal-unitary operators $\hat{V}_r$ and $\hat{V}_r^\dag$, 
and denote it as 
\begin{equation}
  \langle \langle O_r \rangle \rangle_R = \frac{1}{R}\sum_{r=1}^{R} O_r. 
\end{equation}
where $R$ is the number of samples. 
Note that $O_r$ can be either a c-number or an operator. 
The ideal (i.e., $R\to\infty$) random average of $O_r$ is given as 
\begin{alignat}{1}
  \mathbb{E}_D[O_r]
  &\equiv \lim_{R\to\infty}
  \langle \langle O_r \rangle \rangle_R \notag \\
  &=\frac{1}{(2\pi)^D}
  \int_{0}^{2\pi}  \dd \theta_{0,r}
  \int_{0}^{2\pi}  \dd \theta_{1,r}
  \cdots
  \int_{0}^{2\pi}  \dd \theta_{D-1,r} 
    O_r, 
\end{alignat}
where the second line follows from
the assumption that 
$\{\theta_{b,r}\}_{b=0}^{D-1}$
are drawn uniformly from $[0,2\pi)^D$. 
The subscript $D$ in $\mathbb{E}_D$
is to indicate that the number of random variables is $D$. 
It then follows that the ideal random average of the diagonal matrix element
$A_{r}\equiv \langle+| V_r^\dag \hat{A} V_r |+\rangle$ of operator $\hat{A}$
can be given in terms of $\Tr[\hat{A}]$, i.e.,  
\begin{alignat}{1}
  \mathbb{E}_D\left[A_{r}\right]
  & =
  \frac{1}{D}
  \sum_{b^\prime=0}^{D-1}\sum_{b=0}^{D-1}
  \langle b^\prime|\hat{A} | b\rangle
  \mathbb{E}_D\left[
    \e^{-\imag (\theta_{b^\prime,r}-\theta_{b,r})}
    \right]\notag \\
  &= \frac{1}{D}\Tr \left(\hat{A} \right).
  \label{eq:Trace}
\end{alignat}
Here, the second equality follows from the identity
\begin{equation}
  \mathbb{E}_D\left[\e^{-\imag (\theta_{b^\prime,r}-\theta_{b,r})}\right]=\delta_{b b^\prime}.
  \label{eq:Krondelta}
\end{equation}
This proves the statement described at the beginning of this section. 

It is also important to understand how the variance is converged when the trace is evaluated by 
the random samplings. 
Since a product of four random-phase factors satisfies the following identity~\cite{Nakata2012,Jin2021}: 
\begin{equation}
  \mathbb{E}_D
  \left[\e^{-\imag \left(\theta_{i,r} - \theta_{j,r} - \theta_{k,r} + \theta_{l,r}\right)}\right]
  = \delta_{ij} \delta_{kl} + \delta_{ik} \delta_{jl} - \delta_{ij} \delta_{jk} \delta_{kl}, 
\end{equation}
we can show that
the covariance between
$A_{r}  =\langle \Phi_r |\hat{A}|\Phi_r\rangle$ and 
$B_{r}^*=\langle \Phi_r |\hat{B}^\dag|\Phi_r\rangle$ for operators $\hat{A}$ and $\hat{B}$ 
is given by
\begin{alignat}{1}
  &\mathbb{E}_D\left[A_{r} B_{r}^*\right]
  -
  \mathbb{E}_D\left[A_{r}\right]
  \mathbb{E}_D\left[B_{r}^*\right]
  \notag \\
  =&
  \frac{1}{D^2}
  \left\{
    \sum_{i,j,k,l}
    \mathbb{E}_D
    \left[\e^{-\imag \left(\theta_{i,r} - \theta_{j,r} - \theta_{k,r} + \theta_{l,r}\right)}\right]
    A_{ij} B_{kl}^*
  -\Tr\left(\hat{A}\right)
   \Tr\left(\hat{B}\right)^*
   \right\}
  \notag \\
  =&
  \frac{1}{D^2}\left[
    \Tr\left(\hat{A}\hat{B}^\dag\right)
    -\sum_{i=0}^{D-1}A_{ii}B_{ii}^*\right],
  \label{eq:cov}
\end{alignat}
where
$A_{ij}=\langle i|\hat{A}|j\rangle$ and
$B_{kl}^*=\langle l|\hat{B}^\dag|k\rangle                  
=\langle k|\hat{B}|l\rangle^*$
are the matrix elements of $\hat{A}$ and $\hat{B}^\dag$ in the computational basis
states $|i\rangle$, $|j\rangle$, $|k\rangle$, and $|l\rangle$.
Notice in Eq.~(\ref{eq:cov}) that the covariance depends on 
the computational basis used. 
If $\hat{A}$ and $\hat{B}$ are Hermitian, then the covariance is real.
Moreover, when $\hat{B}=\hat{A}$, 
Eq.~(\ref{eq:cov}) is reduced to the variance of $A_{r}$.
Most remarkably, the (co)variance
decreases exponentially in $N$ as $\sim 1/D=1/2^{N}$,
assuming that the numerator in Eq.~(\ref{eq:cov})
is $O(D)$. 

\subsection{Diagonal-unitary $t$-designs for trace evaluation}
In order to represent the diagonal unitary $\hat{V}_r$ in Eq.~(\ref{eq:Vr}) on a quantum circuit 
composed of one- and two-qubit gates,
an exponentially large number of gates is
required as it contains $D\,(=2^N)$ random variables.
This implies that the preparation of the
random-phase state $|\Phi_r\rangle$ in Eq.~(\ref{eq:rps2}) becomes exponentially difficult 
with increasing $N$. 
However, as we shall show in this section, 
we can apply diagonal-unitary $t$-designs~\cite{Nakata2014}
with $t\geqslant 1$ for the trace evaluation and  
with $t\geqslant 2$ for the same purpose with exponentially small (co)variance, 
which therefore avoid such difficulty.

In order to connect 
the trace evaluation with the diagonal-unitary $t$-design,
let us express the state in the form of a density matrix. 
Namely, by substituting the identity
$\langle + |\hat{V}_r^\dag \hat{A} \hat{V}_r | + \rangle
=\Tr\left[\hat{V}_r |+\rangle \langle +| \hat{V}_r^\dag \hat{A} \right]
$ into Eq.~(\ref{eq:Trace}), 
we obtain
\begin{equation}
  \Tr \left(\hat{A}\right) = D 
  \Tr\left(
    \mathbb{E}_D
    \left[
    \hat{V}_r
    |+\rangle \langle +|
    \hat{V}_r^\dag
    \right]
    \hat{A}
    \right).
    \label{eq:Trace2}
\end{equation}
It also follows directly from Eq.~(\ref{eq:Krondelta}) that 
the ideal random average maps the pure state
$\hat{V}_r |+\rangle \langle +|\hat{V}_r^\dag$
to the maximally mixed state, i.e., 
\begin{equation}
  \mathbb{E}_D
  \left[
    \hat{V}_r
    |+\rangle \langle +|
    \hat{V}_r^\dag
    \right]
  =\frac{\hat{I}}{D}, 
  \label{eq:mixed}
\end{equation}
which indeed satisfies Eq.~(\ref{eq:Trace2}). 

The property in Eq.~(\ref{eq:mixed})  
is not unique to the random-phase states of the form $\hat{V}_r|+\rangle$, 
and there are varieties of random states which have a property
that its random average gives the maximally mixed state~\cite{Tolman}. 
Now, we seek for states
that can be obtained by replacing the diagonal unitary $\hat{V}_r$ 
with some much simpler unitary $\hat{v}_{r}$ satisfying 
\begin{equation}
  \mathbb{E}_{N_v}
  \left[
    \hat{v}_r
    |+\rangle \langle +|
    \hat{v}_r^\dag
    \right]
  =\frac{\hat{I}}{D},
  \label{eq:mixed2}
\end{equation}
so that $\hat{v}_r$ can be implemented efficiently on quantum computers. 
Here, $N_v$ is the number of random variables $\{\theta_{k,r}\}_{k=0}^{N_v-1}$ in the unitary $\hat{v}_r$, 
and 
$\mathbb{E}_{N_v}$ denotes the corresponding ideal random average. 
As per the definitions
of the $\epsilon$-approximate unitary $t$-designs and 
of the $\epsilon$-approximate state $t$-designs in Ref.~\cite{Nakata2014}, 
it suffices for the trace evaluation that
$\hat{v}_r$ is a (diagonal-)unitary $t$-design with $t\geqslant 1$.

However, in practical simulations using either classical computers or quantum computers, 
the number $R$ of samples should be finite. 
This implies that the states after sampling
$\hat{V}_r|+\rangle$ and $\hat{v}_r|+\rangle$, i.e., 
$
\hat{\rho}_{V_r|+\rangle,R} \equiv \frac{1}{R}\sum_{r=1}^R
[
  \hat{V}_r
  |+\rangle \langle +|
  \hat{V}_r^\dag
  ]$
and 
$
\hat{\rho}_{v_r|+\rangle,R} \equiv \frac{1}{R}\sum_{r=1}^R
    [
  \hat{v}_r
  |+\rangle \langle +|
  \hat{v}_r^\dag
  ]$,
respectively, are in general different
from each other and also from their limits
$\lim_{R\to\infty}\hat{\rho}_{V_r|+\rangle,R}=
 \lim_{R\to\infty}\hat{\rho}_{v_r|+\rangle,R}=\hat{I}/D$
(also see Ref.~\cite{Iwaki2022} for a related discussion).
Therefore, the unitary $\hat{v}_r$ selected can affect
statistical errors of the random sampling for finite $R$,
even if it satisfies Eq.~(\ref{eq:mixed2}).
Indeed, as we shall discuss in Sec.~\ref{sec:results} 
along with numerical results,
(diagonal-)unitary $t$-designs with $t\geqslant 2$~\cite{Dankert2009}
are required to suppress statistical errors. 
Concrete examples of the unitary $\hat{v}_r$ 
related to the diagonal-unitary $t$-design with $t=1$ and $2$ 
will be given in Sec.~\ref{sec:initial}.

\subsection{Summary of formulation}~\label{sec:form_summary}
Here, we summarize the formalism that is suitable for quantum computation. 
As in Eqs.~(\ref{eq:gTr})--(\ref{eq:betaTr}), we have expressed 
the density of states $g_\tau(E)$, the entropy $S_\tau(E)$, and the inverse temperature $\beta_\tau(E)$ 
in terms of $\Tr[\hat{G}_\tau(E)]$ and $\Tr[\hat{\cal H}\hat{G}_\tau(E)]$.  
With random sampling, 
$\Tr[\hat{G}_\tau(E)]$ and 
$\Tr[\hat{\cal H}\hat{G}_\tau(E)]$ can now be evaluated as 
\begin{equation}
  \Tr[\hat{G}_\tau(E)]
  =
  \frac{D}{2\sqrt{\pi}\tau}
  \int_{-\infty}^{\infty} \dd t\ 
  \e^{-\frac{t^2}{4\tau^2}}
  \e^{\imag E t}
  \mathbb{E}_{N_v} \left[K_r(t)\right],
  \label{eq:TrG2}
\end{equation}
and
\begin{equation}
  \Tr[\hat{\cal H}\hat{G}_\tau(E)]
  =
  \frac{D}{2\sqrt{\pi}\tau}
  \int_{-\infty}^{\infty} \dd t\ 
  \e^{-\frac{t^2}{4\tau^2}}
  \e^{\imag E t}
  \mathbb{E}_{N_v} \left[L_r(t)\right],
  \label{eq:TrHG2}
\end{equation}
respectively. 
Here, 
$K_r(t)$ and
$L_r(t)$ are matrix elements of 
$\hat{U}(t)$ and $\hat{\cal H}\hat{U}(t)$ defined as
\begin{equation}
  K_r(t) \equiv \langle +|\hat{v}_r^\dag \hat{U}(t) \hat{v}_r |+\rangle
  \label{eq:Kr}
\end{equation}
and 
\begin{equation}
  L_r(t) \equiv 
  \langle +|\hat{v}_r^\dag \hat{\cal H}\hat{U}(t) \hat{v}_r |+\rangle, 
  \label{eq:Lr}
\end{equation}
respectively.
Unlike $\hat{V}_r$,
the unitary $\hat{v}_r$ can be implemented
efficiently on a quantum computer. 

Assuming that we know a form of  $\hat{\cal H}$ composed of  
a linear combination of unitary operators,  
an example being given later in Eq.~(\ref{eq:Ham_SWAP}), 
the matrix elements $K_r(t)$ and $L_r(t)$ can be estimated
efficiently using quantum computers. 
Since $K_r(-t)=K_r(t)^*$ and $L_r(-t)=L_r(t)^*$, it is sufficient to 
evaluate $K_r(t)$ and $L_r(t)$ only for $t \geqslant 0$
for estimating $\Tr[\hat{G}_\tau(E)]$ and $\Tr[\hat{\cal H}\hat{G}_\tau(E)]$ in 
Eqs.~(\ref{eq:TrG2}) and (\ref{eq:TrHG2}), respectively.
However, in practice,  the ideal random averages 
$\mathbb{E}_{N_v} \left[K_r(t)\right]$ and 
$\mathbb{E}_{N_v} \left[L_r(t)\right]$
in Eqs.~(\ref{eq:TrG2}) and (\ref{eq:TrHG2}), respectively,
should be replaced with the random averages 
$\langle \langle K_r(t)\rangle \rangle_R$ and 
$\langle \langle L_r(t)\rangle \rangle_R$
with a finite number $R$ of samples. 
Namely, by defining  
\begin{alignat}{1}
  {\cal N}_{\tau,r}(E)
  \equiv 
  \frac{1}{2\sqrt{\pi}\tau}
  \int_{-\infty}^{\infty} \dd t\ 
  \e^{-\frac{t^2}{4\tau^2}}
  \e^{\imag E t}
  K_r(t)
  \label{eq:Nr}
\end{alignat}
and
\begin{alignat}{1}
  {\cal H}_{\tau,r}(E)
  \equiv 
  \frac{1}{2\sqrt{\pi}\tau}
  \int_{-\infty}^{\infty} \dd t\ 
  \e^{-\frac{t^2}{4\tau^2}}
  \e^{\imag E t}
  L_r(t), 
  \label{eq:Hr}
\end{alignat}
we approximate $\Tr[\hat{G}_\tau(E)]$ and $\Tr[\hat{\cal H}\hat{G}_\tau(E)]$
respectively as  
\begin{alignat}{1}
  \Tr[\hat{G}_\tau(E)]&\approx
  D\langle \langle {\cal N}_{\tau,r}(E) \rangle \rangle_R
  \label{eq:TrG_R}
\end{alignat}
and
\begin{alignat}{1}
  \Tr[\hat{\cal H}\hat{G}_\tau(E)]&\approx
  D\langle \langle {\cal H}_{\tau,r}(E) \rangle \rangle_R. 
  \label{eq:TrHG_R}
\end{alignat}

Accordingly, 
the density of states $g_\tau(E)$,
the entropy $S_\tau(E)$, and
the inverse temperature $\beta_\tau(E)$ in Eqs.~(\ref{eq:gTr})--(\ref{eq:betaTr}) 
are approximated by
those truncated at a finite number $R$ of samples, denoted as 
$g_{\tau,R}(E)$,
$S_{\tau,R}(E)$, and
$\beta_{\tau,R}(E)$, respectively: 
\begin{alignat}{1}
  g_{\tau,R}(E) 
  & \equiv
    \frac{\tau}{\sqrt{\pi}}
  D \langle \langle {\cal N}_{\tau,r}(E) \rangle\rangle_R,
  \label{eq:g_filterTPQ}
  \\
  S_{\tau,R}(E)
  &\equiv
  \ln {\left[D \langle \langle {\cal N}_{\tau,r}(E) \rangle \rangle_R\right]}, 
  \label{eq:S_filterTPQ}
\end{alignat}
and
\begin{alignat}{1}
  \beta_{\tau,R}(E)
  &\equiv
  2\tau^2
  \left[{\cal E}_{\tau,R}(E) -E\right], 
  \label{eq:beta_filterTPQ}
\end{alignat}
where 
\begin{equation}
  {\cal E}_{\tau,R}(E) \equiv
  \frac{\langle \langle {\cal H}_{\tau,r}(E)\rangle \rangle_R}
       {\langle \langle {\cal N}_{\tau,r}(E)\rangle \rangle_R}.
       \label{eq:E_filterTPQ}
\end{equation}
These quantities necessarily acquire statistical errors. 
Obviously, in the limit of $R\to\infty$, 
$g_{\tau,R}(E)$, $S_{\tau,R}(E)$, $\beta_{\tau,R}(E)$, and ${\cal E}_{\tau,R}(E)$ are reproduced to 
$g_{\tau}(E)$, $S_{\tau}(E)$, $\beta_{\tau}(E)$, and ${\cal E}_{\tau}(E)$, respectively.
Equations~(\ref{eq:g_filterTPQ})--(\ref{eq:E_filterTPQ}) 
are the central equations of this study, corresponding to Eq.~(\ref{eq:gTr})--(\ref{eq:ETr}),  
and these are appropriate to simulations using 
quantum computers (as well as classical computers). 

\section{Energy-filtered random-phase state as a TPQ state}~\label{sec:form_purestate}
In this section, we shall discuss a connection between the formalism developed in Sec.~\ref{sec:form} and 
a state introduced in the filter-diagonalization method~\cite{Wall1995}.
While the equations summarized in Sec.~\ref{sec:form_summary} suffice
for numerical simulations, 
the reformulation based on an energy-filtered state allows us to recognize the proposed method as 
a microcanonical counterpart of the canonical TPQ-state formalism~\cite{Sugiura2013}.

\subsection{Energy-filtered random-phase state}~\label{sec:EFS}
Let $|\phi_r\rangle$ be a normalized random-phase state
suitable for the trace evaluation, e.g., 
$|\phi_r\rangle=\hat{v}_r|+\rangle$ or 
$|\phi_r\rangle=\hat{V}_r|+\rangle$.
It follows from Eq.~(\ref{eq:Gop})
that the positive square root of $\hat{G}_\tau(E)$ is given by
\begin{equation}
  \left[\hat{G}_{\tau}(E)\right]^{\frac{1}{2}}
  =\e^{-\frac{1}{2}(\hat{\cal H}-E)^2\tau^2}
  =\hat{G}_{\frac{\tau}{\sqrt{2}}}(E).
\end{equation}
Let us now introduce an unnormalized state $|\psi_{\tau,r}(E)\rangle$ defined as
\begin{alignat}{1}
  |\psi_{\tau,r}(E)\rangle
  &\equiv
  \left[\hat{G}_{\tau}(E)\right]^{\frac{1}{2}}|\phi_r\rangle. 
  \label{eq:TPQ}
\end{alignat}
Using the Fourier representation for
$\hat{G}_{\frac{\tau}{\sqrt{2}}}(E)$ [see Eq.~(\ref{eq:GFT})],  
we can readily show that the state $|\psi_{\tau,r}(E)\rangle$ introduced above can be written as 
\begin{alignat}{1}
    |\psi_{\tau,r}(E)\rangle=\frac{1}{\sqrt{2\pi} \tau}\int_{-\infty}^{\infty}\dd t \
  \e^{-\frac{t^2}{2\tau^2}}
  \e^{\imag E t}
  \hat{U}(t)|\phi_{r}\rangle. 
  \label{eq:filterTPQ}
\end{alignat}
It is now clear that the state  $|\psi_{\tau,r}(E)\rangle$ is essentially a
Fourier transform of the time-evolved state
$|\phi_r(t)\rangle \equiv  \hat{U}(t)|\phi_r\rangle$ from the initial random state $|\phi_r\rangle$,
but the interval of the time integral is
effectively cut-off around the filtering time $|t|\sim \tau$ 
by the Gaussian in the integrand.
Here we call $|\psi_{\tau,r}(E)\rangle$ an energy-filtered random-phase state. 
The state of the form in Eq.~(\ref{eq:filterTPQ})
is originally introduced in a different context to extract eigenvalues and eigenstates 
of $\hat{H}$ at any desired energy range around a given energy $E$, 
known as the filter-diagonalization method~\cite{Wall1995}. 

To examine the properties of the energy-filtered random-phase state $|\psi_{\tau,r}(E)\rangle$,  
let us expand the initial state $|\phi_r\rangle$
by the energy eigenstates as 
\begin{equation}
  |\phi_{r}\rangle = \sum_{n=0}^{D-1} c_{n,r} |E_n\rangle,
  \label{eq:initial}
\end{equation}
where $\{c_{n,r}\}_{n=0}^{D-1}=\{\langle E_n|\phi_r\rangle\}_{n=0}^{D-1}$ are complex coefficients 
satisfying the normalization condition
$\langle \phi_{r}|\phi_{r}\rangle = \sum_{n=0}^{D-1}|c_{n,r}|^2=1$.
By substituting Eq.~(\ref{eq:initial}) to Eq.~(\ref{eq:TPQ}), we obtain
\begin{alignat}{1}
  |\psi_{\tau,r}(E)\rangle
  &= \sum_{n=0}^{D-1}
  \e^{-\frac{1}{2}(E_n-E)^2\tau^2} c_{n,r} |E_n\rangle.
  \label{eq:psi_tauE}
\end{alignat}
At finite $\tau$, the energy-filtered state
is a linear combination of the energy eigenstates
whose eigenvalues are concentrated around $E$ with 
a spread of $\sim \frac{1}{\tau}$.
In the zero-filtering-time limit (i.e., $\tau \to 0$),
the energy-filtered state coincides with the initial state, 
\begin{equation}
  \lim_{\tau\to 0}|\psi_{\tau,r}(E)\rangle=|\phi_{r}\rangle.
  \label{eq:zero_filter}
\end{equation}
In the infinite-filtering-time limit (i.e., $\tau \to \infty$),
the energy-filtered state multiplied by $\frac{\tau}{\sqrt{2\pi}}$
can be formally written as a linear combination of
energy eigenstates weighted by delta functions, 
\begin{alignat}{1}
  \lim_{\tau \to \infty}
  \frac{\tau}{\sqrt{2\pi}}
  |\psi_{\tau,r}(E)\rangle&=
  \sum_{n=0}^{D-1} \delta(E-E_n) c_{n,r} |E_n\rangle \nonumber \\
  &=\delta(E-\hat{H})|\phi_r\rangle.
  \label{eq:infinite_filter}
\end{alignat}
These observations rationalize calling
$E$ the target energy and $\tau$ the filtering time.
It should be emphasized that,
unlike Eq.~(\ref{eq:zero_filter}),
Eq.~(\ref{eq:infinite_filter}) is
merely a formal expression and
$\lim_{\tau \to \infty}|\psi_{\tau,r}(E)\rangle$ is not in practical use.

\subsection{Squared norm and expectation value of the Hamiltonian}
In terms of the energy-filtered random-phase state $|\psi_{\tau,r}(E)\rangle$, 
$\Tr[\hat{G}_\tau(E)]$ and $\Tr[\hat{\cal H}\hat{G}_\tau(E)]$ can be
expressed simply as 
\begin{alignat}{1}
  \Tr[\hat{G}_\tau(E)]&=
  D\mathbb{E}_{N_v}\left[\langle \psi_{\tau,r}(E) | \psi_{\tau,r}(E) \rangle\right]
  \label{eq:TrGfilter}
\end{alignat}
and
\begin{alignat}{1}
  \Tr[\hat{\cal H}\hat{G}_\tau(E)]&=
  D\mathbb{E}_{N_v}\left[\langle \psi_{\tau,r}(E) |\hat{\cal H}| \psi_{\tau,r}(E) \rangle\right],
  \label{eq:TrHGfilter}
\end{alignat}
where 
the Hermiticity of $[\hat{G}_\tau(E)]^{\frac{1}{2}}$ and the identity 
$\hat{\cal H}\hat{G}_\tau(E)=
[\hat{G}_\tau(E)]^{\frac{1}{2}}
\hat{\cal H}
[\hat{G}_\tau(E)]^{\frac{1}{2}}$ are used. The latter 
follows from the fact that $\hat{G}_\tau(E)$ commutes with $\hat{\cal H}$.
Equations~(\ref{eq:TrGfilter}) and (\ref{eq:TrHGfilter})
indicates that the density of states as well as the thermodynamic quantities such as the entropy and 
the inverse temperature can be
evaluated simply by sampling
the squared norm of the energy-filtered random-phase state $|\psi_{\tau,r}(E)\rangle$ and
the expectation value of the Hamiltonian with respect to
the energy-filtered random-phase state $|\psi_{\tau,r}(E)\rangle$. 

It follows from
$\langle \psi_{\tau,r}(E)|\psi_{\tau,r}(E)\rangle = \langle \phi_r |\hat{G}_\tau(E)|\phi_r\rangle$ and
$\langle \psi_{\tau,r}(E)|\hat{\cal H}|\psi_{\tau,r}(E)\rangle = \langle \phi_r |\hat{\cal H}\hat{G}_\tau(E)|\phi_r\rangle$ that 
${\cal N}_{\tau,r}(E)$ and
${\cal H}_{\tau,r}(E)$
defined in Eqs.~(\ref{eq:Nr}) and (\ref{eq:Hr}) are
equivalent to the squared norm of $|\psi_{\tau,r}(E)\rangle$ and
the expectation value of the Hamiltonian, respectively, i.e., 
\begin{alignat}{1}
  {\cal N}_{\tau,r}(E)
  &=
  \langle \psi_{\tau,r}(E)|\psi_{\tau,r}(E)\rangle
  \label{eq:norm}
\end{alignat}
and
\begin{alignat}{1}
  {\cal H}_{\tau,r}(E)
  &=
  \langle \psi_{\tau,r}(E) |\hat{\cal H} |\psi_{\tau,r}(E) \rangle.
  \label{eq:H} 
\end{alignat}
Since $0<\e^{-(E_n-E)^2\tau^{2}}\leqslant 1$,
the squared norm
${\cal N}_{\tau,r}(E)=\sum_{n=0}^{D-1}|c_{n,r}|^2\e^{-(E_n-E)^2\tau^{2}}$
satisfies $0 < {\cal N}_{\tau,r}(E) \leqslant 1$,
and the equality is achieved when $\tau=0$. 
In Appendix~\ref{app:equiv}, we give another proof of 
the equivalence between Eqs.~(\ref{eq:Nr}) and~(\ref{eq:norm})
starting with the time-integral form of
the energy-filtered random-phase state $|\psi_{\tau,r}(E)\rangle$ in Eq.~(\ref{eq:filterTPQ}).

\subsection{Energy-filtered random-phase state as a microcanonical counterpart of the canonical TPQ state}\label{sec:microTPQ}
Our construction of the energy-filtered random-phase state $|\psi_{\tau,r}(E)\rangle$ in Eq.~(\ref{eq:TPQ}) 
reveals its formal similarity to the canonical TPQ state~\cite{Sugiura2013},
as we shall explain in this section. 
In the microcanonical ensemble, the number of states $\Tr\left[\hat{G}_\tau(E)\right]$ 
within a given energy window, specified here by $E$ and $\tau$, is the most fundamental quantity. 
It then follows from the statistical property of the random-phase state $|\phi_r\rangle$ in Eq.~(\ref{eq:mixed2}) that 
the corresponding energy-filtered random-phase state $|\psi_{\tau,r}(E)\rangle$ in Eq.~(\ref{eq:TPQ})
satisfies
\begin{alignat}{1}
  D \lim_{R\to\infty} \langle \langle |\psi_{\tau,r}(E) \rangle \langle \psi_{\tau,r}(E)|\rangle \rangle_R &
  = \hat{G}_\tau(E) 
\end{alignat}
and
\begin{alignat}{1}
  D
  \lim_{R\to\infty}
  \langle \langle\, \langle \psi_{\tau,r}(E)|\psi_{\tau,r}(E)\rangle\, \rangle \rangle_R &  
  = \Tr\left[\hat{G}_\tau(E)\right].
\end{alignat}
Accordingly, the density matrix can be given as
\begin{equation}
\label{eq:rho_E}
  \hat{\rho}_{\tau}(E) \equiv \frac{\hat{G}_{\tau}(E)}{\Tr\left[\hat{G}_{\tau}(E)\right]}
  =
  \frac
      {\lim_{R\to\infty} \langle \langle |\psi_{\tau,r}(E) \rangle \langle \psi_{\tau,r}(E)| \rangle \rangle_R}
      {\lim_{R\to\infty} \langle \langle\, \langle \psi_{\tau,r}(E)|\psi_{\tau,r}(E)\rangle\, \rangle \rangle_R}.
\end{equation}
Obviously, $\hat{\rho}_\tau(E)$ does not evolve in time because $\hat{G}_\tau(E)$ commutes with $\hat{\cal H}$,
and hence it is an equilibrium state. 
It should be emphasized that as opposed to the TPQ-state formalism~\cite{Sugiura2012}, 
there is no need to sequentially prepare, for example,
a state associated with lower $E$ from a state with higher $E$ because $E$ and $\tau$ are the input parameters in our case. 

In the canonical ensemble, the partition function $\Tr\left[\e^{-\beta \hat{\cal H}}\right]$
for a given inverse temperature $\beta$ plays a central role, and 
the corresponding canonical TPQ state is given by $|\beta_r \rangle\equiv \e^{-\beta \hat{\cal H}/2}|\phi_r\rangle$~\cite{Sugiura2013}. 
It is then easy to show that the canonical TPQ state $|\beta_r \rangle $ satisfies
\begin{alignat}{1}
  D\lim_{R\to\infty}\langle \langle |\beta_r \rangle \langle \beta_r| \rangle \rangle_R &= \e^{-\beta \hat{\cal H}} 
\end{alignat}
and 
\begin{alignat}{1}
  D\lim_{R\to\infty}\langle \langle\, \langle \beta_r|\beta_r\rangle\, \rangle \rangle_R &= \Tr\left[\e^{-\beta \hat{\cal H}}\right].
\end{alignat}
Accordingly, the density matrix is given as
\begin{equation}
  \hat{\rho}_{\rm can}(\beta) \equiv \frac{\e^{-\beta\hat{\cal H}}}{\Tr\left[\e^{-\beta\hat{\cal H}}\right]}
  =
  \frac
      {\lim_{R\to\infty}\langle \langle |\beta_r \rangle \langle \beta_r| \rangle \rangle_R}
      {\lim_{R\to\infty}\langle \langle\, \langle \beta_r|\beta_r\rangle\, \rangle \rangle_R}.
\end{equation}
These formal similarities suggest that the energy-filtered random-phase state $|\psi_{\tau,r}(E)\rangle$ 
in Eq.~(\ref{eq:TPQ}) 
can be considered as a microcanonical counterpart of the canonical TPQ state $|\beta_r\rangle$.
In Appendix~\ref{app:canonical}, we show how the
present microcanonical ensemble can be translated to the canonical ensemble.

\section{Model and method}\label{sec:method}

\subsection{Hamiltonian}
The pure-state formalism for the microcanonical ensemble based on the energy-filtered random-phase state 
proposed here is now validated numerically by examining 
the spin-1/2 antiferromagnetic Heisenberg model defined by the following Hamiltonian:  
\begin{eqnarray}
  \hat{\mathcal{H}} =
  J
  \sum_{\langle i,j \rangle}
  \hat{\mathcal{P}}_{ij}.
  \label{eq:Ham_SWAP}
\end{eqnarray}
Here, 
$J>0$ is the antiferromagnetic exchange interaction, 
$\langle i,j \rangle$
runs over all nearest-neighbor pairs of qubits $i$ and $j$ 
connected with the exchange interaction $J$, 
and $\hat{\mathcal{P}}_{ij}$ is the {\sc swap} operator
that can be written as 
\begin{equation}
  \hat{\mathcal{P}}_{ij}
  =\frac{1}{2}\left(
  \hat{X}_i \hat{X}_j +
  \hat{Y}_i \hat{Y}_j +
  \hat{Z}_i \hat{Z}_j +
  \hat{I}_i \hat{I}_j 
  \right)
  \label{SWAP}
\end{equation}
for $i\not=j$ with  
$\{ \hat{X}_{i}, \hat{Y}_{i}, \hat{Z}_{i} \}$ and $\hat{I}_i$
being the Pauli operators and the identity operator
acting on the $i$th qubit.
We consider the one-dimensional Heisenberg model 
consisting of $N=$20, 22, 24, and 28 qubits 
under periodic boundary conditions.
Although we specify the target energy $E$ and
the system size $N$ (assumed even) for our microcanonical ensemble, 
we do not specify other Hamiltonian-symmetry-related properties of the system 
such as the magnetization and the total momentum. 
Therefore, the dimension of the Hilbert space is given by $D=2^N$.

Note that 
the Hamiltonian in Eq.~(\ref{eq:Ham_SWAP}) can be written as
$\hat{\cal H} = 2J\sum_{\langle i,j\rangle} \hat{\bs{S}}_i\cdot \hat{\bs{S}}_j
+ \frac{JN_{\rm bond}}{2}$,
where $\hat{\bs{S}}_i= \frac{1}{2}(\hat{X}_i\ \hat{Y}_i \ \hat{Z}_i)$
is the vector of the spin-1/2 operators acting on the $i$th qubit and 
$N_{\rm bond}=\sum_{\langle i,j \rangle}$
is the number of bonds,
which is given by $N_{\rm bond}=N$ for the one-dimensional system under 
periodic boundary conditions. 
This relation is useful when one compares the energy
eigenvalues of $\hat{\cal H}$ in Eq.~(\ref{eq:Ham_SWAP}) with
those of $J\sum_{\langle i,j\rangle} \hat{\bs{S}}_i\cdot \hat{\bs{S}}_j$,
which is the form of a Hamiltonian conventionally used for the spin-1/2 Heisenberg model.

\subsection{Initial states}\label{sec:initial}

As an initial random state for the trace evaluation, 
we consider the random-phase state $\hat{V}_r|+\rangle$ defined in Eq.~(\ref{eq:rps2}). 
In our numerical simulation, $\hat{V}_r|+\rangle$ 
is prepared by assigning the random number
$\langle b| \hat{V}_r |+\rangle = \e^{\imag \theta_{b,r}}/\sqrt{D}$ to
the $b$th entry of the corresponding numeral vector [see Eq.~(\ref{eq:rps})]. 
In addition to $\hat{V}_r|+\rangle$, we consider 
two types of much simpler random-phase states  $\hat{v}_r|+\rangle$
using single- and two-qubit gates, as described below.

The first type of 
$\hat{v}_r|+\rangle$ is the random-phase-product state~\cite{Iitaka2020},
which would be one of the simplest examples of $\hat{v}_r|+\rangle$.
In this case, 
the unitary $\hat{v}_r = \hat{v}_r^{(1)}$ is given by 
a product of single-qubit $Z$ rotations, i.e., 
\begin{equation}
\hat{v}_r^{(1)}=
\prod_{k=0}^{N-1}\hat{R}_{Z_k}(\theta_{k,r}),
\label{eq:v1}
\end{equation}
where 
$\hat{R}_{Z_k}(\theta_{k,r})
=\e^{-\imag \theta_{k,r} \hat{Z}_k/2}$ with $\hat{Z}_k$
being Pauli-$Z$ operator acting on the $k$th qubit, 
and $N_v=N$.
The random variables $\{\theta_{k,r}\}_{k=0}^{N-1}$
are drawn uniformly from $[0,2\pi)^{N_v}$.
Note that $\hat{v}_r^{(1)}|+\rangle$ is a product state
having no entanglements between qubits.
In Appendix~\ref{app:rpps}, we show that the random-phase-product state
satisfies Eq.~(\ref{eq:mixed2}).
The random-phase-product states are thus sufficient for
taking the trace in principle. However, as we shall show explicitly in Sec.~\ref{sec:result_comp}, 
the convergence of the trace evaluation via the random sampling is rather slow.  

As the second type of $\hat{v}_r|+\rangle$, we introduce 
another random diagonal unitary circuits 
similar to those in Refs.~\cite{Nakata2017math,Nakata2017PRX},
$\hat{v}_r=\hat{v}_r^{(2)}\hat{v}_r^{(1)}$,
where $\hat{v}_r^{(1)}$ is given in Eq.~(\ref{eq:v1}) and 
$\hat{v}_r^{(2)}$ consists of all-to-all $N(N-1)/2$ two-qubit gates, i.e.,  
\begin{equation}
  \hat{v}_r^{(2)}=
  \prod_{i=0}^{N-1} \prod_{j=0}^{i-1} \e^{-\imag \hat{Z}_i \hat{Z}_j  \theta_{k(ij),r}}.
  \label{eq:v2}
\end{equation}
Here, the number of random variables in 
$\hat{v}_r=\hat{v}_r^{(2)}\hat{v}_r^{(1)}$ 
is $N_v=N+N(N-1)/2=N(N+1)/2$,
and $k(ij)$ in Eq.~(\ref{eq:v2}) maps a two-dimensional label $\{ij\}$ to
a one-dimensional label $k$ which runs from $N$ to $N(N+1)/2-1$.
It is obvious that $\hat{v}_r=\hat{v}_r^{(2)}\hat{v}_r^{(1)}$ also satisfies Eq.~(\ref{eq:mixed2})
because it contains $\hat{v}_r^{(1)}$ (see Appendix~\ref{app:rpps}), implying that 
it is also sufficient for the trace evaluation. 
However, in contrast to $\hat{v}_r^{(1)}$,  
$\hat{v}_r^{(2)}$ can apparently generate entanglements between qubits. 
The random variables $\{\theta_{k,r}\}_{k=0}^{N_v-1}$
are drawn uniformly from $[0,2\pi)^{N_v}$. 
Namely, unlike unitary $t$-designs,
the random variables are assumed to be continuous here.

\subsection{Time-evolution operator}
We approximate the time-evolution operator $U(t)$ in Eq.~(\ref{eq:ut})
by the first-order Suzuki-Trotter decomposition~\cite{Trotter1959,Suzuki1976} of the form
\begin{equation}
  \hat{U}(t)
  \approx
  \left[\hat{S}(\Delta_t)\right]^M \equiv
  \left[
  \e^{-\imag \hat{\cal H}_A \Delta_t}
  \e^{-\imag \hat{\cal H}_B \Delta_t}
  \right]^M
  \label{eq:Suzuki_Trotter}
\end{equation}
where $M$ is an integer such that $t=M\Delta_t$, 
$\hat{\cal H}_A$ ($\hat{\cal H}_B$) consists of 
the Hamiltonian terms acting on the odd (even) bonds, 
i.e., $\hat{\cal{H}}=\hat{\cal H}_A+\hat{\cal H}_B$ with 
$\hat{\cal H}_{A\,(B)}=J\sum_{i:\,{\text {odd (even)}}} \hat{\cal P}_{i,i+1}$, 
and 
$\Delta_t$ is a small time slice.  
Accordingly,
we discretize the time integrals in
Eqs.~(\ref{eq:Nr}) and (\ref{eq:Hr}) 
with the trapezoidal rule.  
In our numerical demonstration shown in Sec.~\ref{sec:results}, 
we calculate thermodynamic quantities
with the filtering time $\tau$ varying $0 < \tau J \leqslant 5$. 
It turns out that
taking the range of the time integral over  
$0 \leqslant t J \leqslant t_{\rm max} J$ (see Sec.~\ref{sec:form_summary}) with
$t_{\rm max} J=50$, instead of infinity, and $\Delta_t J = 0.01$ gives
quantitatively converged results. 
We thus set the number $M$ of time slices as large as $M = 50/\Delta_t J=5000$. 

\begin{figure*}
  \includegraphics[width=1.95\columnwidth]{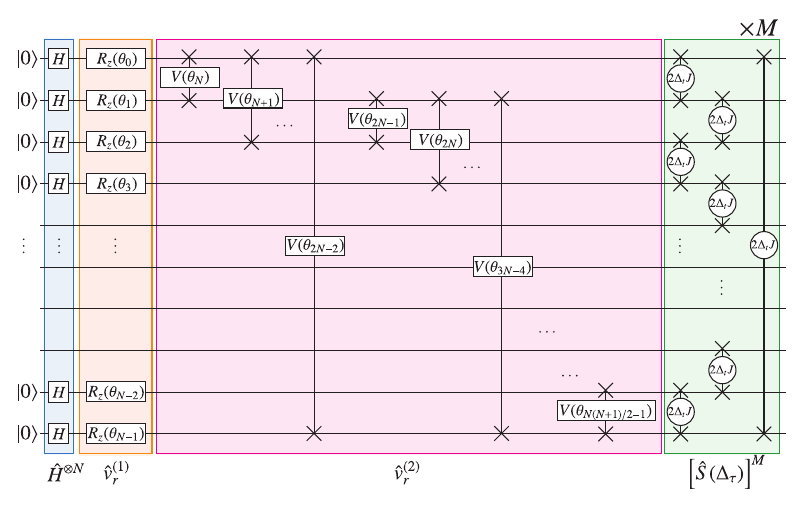}
  \caption{
    \label{fig:circuit}
    Quantum circuit for preparing the state
    $[\hat{S}(\Delta_t)]^M \hat{v}_r^{(2)} \hat{v}_r^{(1)}|+\rangle$.
    Operator expressions for
    $\hat{v}_r^{(1)}$, $\hat{v}_r^{(2)}$, and $[\hat{S}(\Delta_t)]^M$ are given in
    Eqs.~(\ref{eq:v1}), (\ref{eq:v2}), and (\ref{eq:Suzuki_Trotter}), respectively. 
    In the time-evolution operator part, $[\hat{S}(\Delta_t)]^M$, 
    we assume the one-dimensional Heisenberg model under periodic boundary conditions. 
    Note that the subscript ``$r$'' in $\theta_{k,r}$ is omitted 
    for simplicity.
  }
\end{figure*}

Figure~\ref{fig:circuit} shows schematically 
a quantum circuit for preparing the time-evolved state
$[\hat{S}(\Delta_t)]^M \hat{v}_r^{(2)}\hat{v}_r^{(1)}|+\rangle$.
Here, the ``$V$" gate with angle $\theta_k$ connecting qubits $i$ and $j$,  
appearing in the $\hat{v}_r^{(2)}$ part of the quantum circuit in the figure, denotes 
$\e^{-\imag \hat{Z}_i \hat{Z}_j  \theta_{k,r}}$.
Note that the order of quantum gates in the part of the quantum circuit 
corresponding to the diagonal unitary $\hat{v}_r^{(2)}\hat{v}_r^{(1)}$ 
is irrelevant because these quantum gates commute among each other. 
The ``$2\Delta_t J$'' gate connecting qubits $i$ and $j$ that  
appears in the $[\hat{S}(\Delta_t)]^M$ part of the quantum circuit in the figure denotes
the exponential-SWAP gate 
$\e^{-\imag \theta \hat{\cal P}_{ij}/2}$ with angle $\theta = 2\Delta_t J$~\cite{Seki2020vqe}.
The matrix element of the time-evolution operator in Eq.~(\ref{eq:Kr}),  
here approximated by the Suzuki-Trotter decomposition, i.e.,   
\begin{equation}
  K_r(t) 
  \approx
  \langle +| \hat{v}_r^{(1)\dag} \hat{v}_r^{(2)\dag} [\hat{S}(\Delta_t)]^M \hat{v}_r^{(2)}\hat{v}_r^{(1)}|+\rangle
\end{equation}
can be evaluated with the Hadamard test on quantum computers.
For example, the real part of $K_r(t)$ can be evaluated with the Hadamard test
by adding one ancillary qubit in $|+\rangle_{\rm a}=\frac{1}{\sqrt{2}}(|0\rangle_{\rm a}+|1\rangle_{\rm a})$
(where the subscript ``${\rm a}$'' indicates the ancillary qubit),
replacing the quantum gates for $[\hat{S}(\Delta_t)]^M$ by
those for the controlled-$[\hat{S}(\Delta_t)]^M$ gate with the control qubit being the ancillary qubit, 
and measuring the ancillary qubit in $|+\rangle_{\rm a}$ basis. The imaginary part can be evaluated similarly. 
In the same manner, the other matrix element in Eq.~(\ref{eq:Lr}), i.e., 
\begin{equation}
  L_r(t) 
  \approx
  J \sum_{\langle ij \rangle} \langle +| \hat{v}_r^{(1)\dag} \hat{v}_r^{(2)\dag}
  \hat{\cal P}_{ij}[\hat{S}(\Delta_t)]^M \hat{v}_r^{(2)}\hat{v}_r^{(1)}|+\rangle
\end{equation}
can also be evaluated with the Hadamard test on quantum computers. 

\section{Numerical results}\label{sec:results}

\subsection{Number of states}\label{sec:nos}
In the microcanonical ensemble,
the number of states within an energy window at a give energy $E$ plays a central role,
because its logarithm corresponds to the entropy, which is
the thermodynamic potential in the microcanonical ensemble.
We thus first examine how
our definition of the number of states within an energy window, $g_\tau(E) \delta E$,
behaves in the one-dimensional Heisenberg model for $N=$20, 22, and 24. 
The results for smaller system sizes are also provided in Appendix~\ref{app:smallN}. 

Figure~\ref{fig:nos} shows histograms of the number of states. 
Here, a histogram is made by dividing the energy range
$[E_0,E_{D-1}]$ of the whole energy eigenvalues into $N_{\rm bin}$ number of bins, and hence
the height of a bar of a histogram corresponds to 
the number of states within an energy window
$[E-\delta E_{N_{\rm bin}}/2, E+\delta E_{N_{\rm bin}}/2)$, where 
\begin{equation}
  \delta E_{N_{\rm bin}}  \equiv \frac{E_{D-1}-E_0}{N_{\rm bin}-1}, 
\end{equation}
and we allocate $N_{\rm bin}$ bins so that the minimum (maximum) energy eigenvalue $E_0$ ($E_{D-1}$) is located 
at the center of the corresponding bin. 
In Fig.~\ref{fig:nos}, we plot the results for $N_{\rm bin}=32$, 64, and 128.
Note that basically the same histogram 
but with a different $\delta E_{N_{\rm bin}}$ for $N=20$ has already been
reported in Ref.~\cite{Okamoto2018}.
For comparison, we also show in the figure the number of states calculated form our definition, i.e.,  
\begin{equation}
   \Tr\left[\hat{G}_\tau(E)\right] = g_\tau(E) \delta E,   
\end{equation}
where $\delta E$ is given in Eq.~(\ref{eq:width}) and the filtering time $\tau$ is chosen as 
\begin{equation}
  \tau = \frac{\sqrt{\pi}}{\delta E_{N_{\rm bin}}}, 
  \label{eq:tau2}
\end{equation}
hence satisfying $\delta E=\delta E_{N_{\rm bin}}$. 
More explicitly, the values of $\tau$ are 
$\tau J=$1.98, 1.80, 1.65 for $N_{\rm bin}=32$ and $N=20,22,24$,
$\tau J=$4.02, 3.65, 3.35 for $N_{\rm bin}=64$ and $N=20,22,24$, and 
$\tau J=$8.09, 7.36, 6.75 for $N_{\rm bin}=128$ and $N=20,22,24$, respectively.

\begin{figure*}
  \includegraphics[width=1.95\columnwidth]{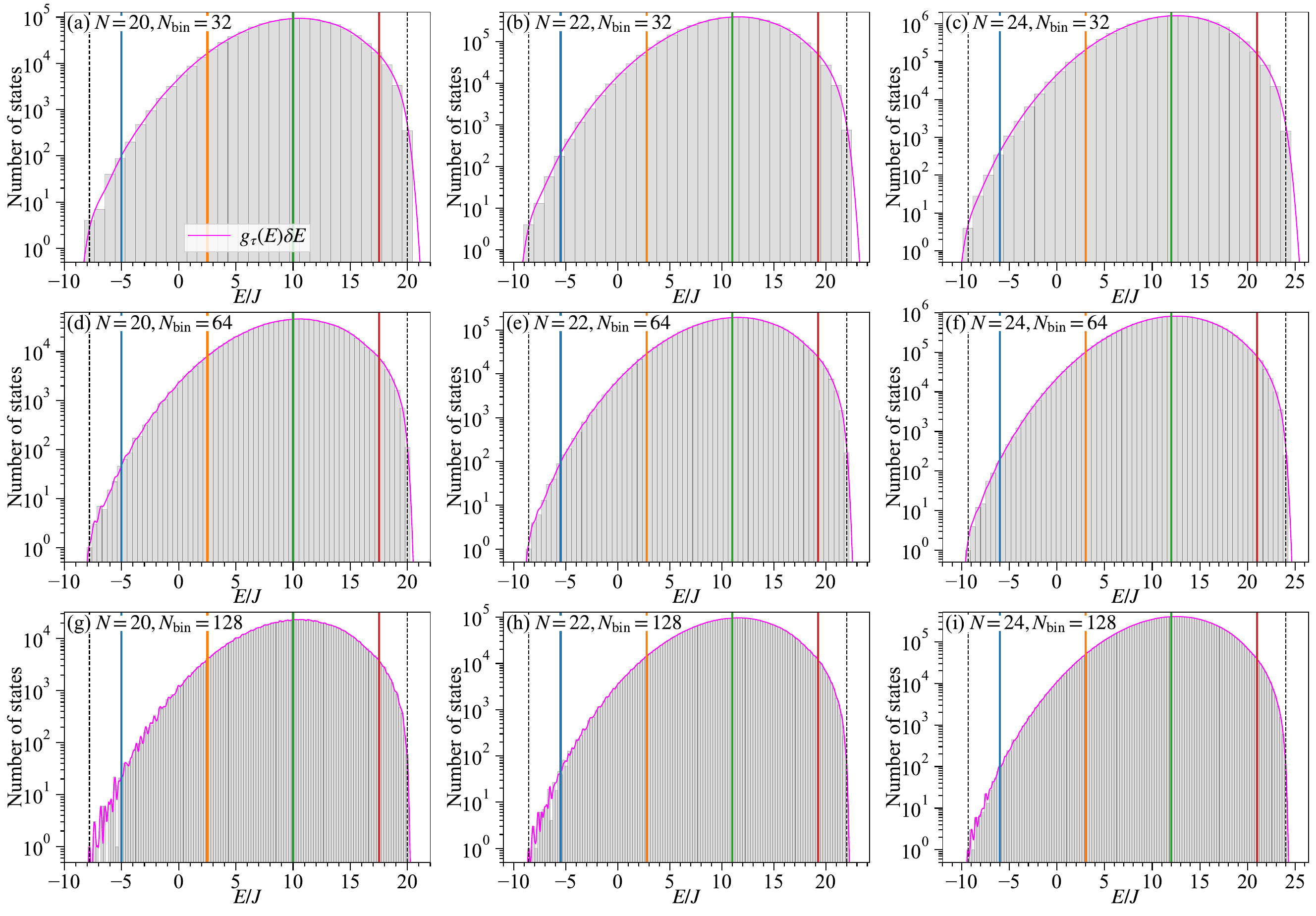}
  \caption{
    \label{fig:nos}
    Semilog plots of the histograms for the the number of states
    as a function of the target energy $E$ 
    with
    (a,b,c) $N_{\rm bin}=32$,
    (d,e,f) $N_{\rm bin}=64$, and 
    (g,h,i) $N_{\rm bin}=128$ 
    for 
    (a,d,g) $N=20$,
    (b,e,h) $N=22$, and 
    (c,f,i) $N=24$.
    For comparison, 
    the number of states calculated from 
    $\Tr[\hat{G}_\tau(E)]=g_\tau(E)\delta E$ with the filtering time $\tau$ chosen as in Eq.~(\ref{eq:tau2}) 
    is also shown in each panel by the magenta line. 
    The thin dashed vertical lines at the edges of the
    histograms indicate the minimum and maximum energy
    eigenvalues $E_0$ and $E_{D-1}$.
    The thick vertical lines at $E/NJ=-0.25,0.125,0.5,$ and $0.875$
    (indicated by blue, orange, green, and red, respectively) denote the target energies used in the results shown in
    Figs.~\ref{fig:ESbeta}, \ref{fig:Efluct}, \ref{fig:rand}, and \ref{fig:rand3}.
    Notice that the different panels employ the different axes scales. 
  }
\end{figure*}

It is found in Fig.~\ref{fig:nos} that $\Tr[\hat{G}_\tau(E)]=g_\tau(E)\delta E$ closely follows the corresponding
histogram of the number of states, indicating that our definitions of the density of states
$g_\tau(E)$ and the width of the energy window $\delta E$ are reasonable.
We note that, in contrast to the histograms,
$\Tr[\hat{G}_\tau(E)]=g_\tau(E)\delta E$ is a continuous function of $E$ and hence
its derivative with respect to $E$ is well defined even for finite-size systems.  
However, when the energy eigenvalues are distributed 
sparsely as compared to a given energy window, $g_{\tau}(E) \delta E$
behaves snaky [for example, see a low-energy region in 
Figs.~\ref{fig:nos}(g)--\ref{fig:nos}(i)] and such a behavior becomes more prominent for
the smaller $N$ and the larger $N_{\rm bin}$ (see Fig.~\ref{fig:nos_small} in Appendix~\ref{app:smallN}).
On the other hand, thermodynamically, 
the entropy $S_\tau(E)=\ln [g_\tau(E)\delta E]$ should be 
a concave function of $E$ so that the inverse temperature $\beta_\tau(E)=\partial S_\tau(E)/\partial E$
decreases monotonically with $E$.
This implies that if the filtering time $\tau$ is so large that $\delta E$ is smaller 
than the energy-eigenvalue spacing at energy around $E$, the statistical mechanical treatment of these quantum states
becomes irrelevant and hence loses connections to thermodynamics,
as it is usually the case in statistical mechanics.
In this sense, the proposed method is expected to be most effective 
for larger systems where the distribution of the energy eigenvalues is  
dense.

\subsection{Filtering-time dependence}
In the thermodynamic limit, thermodynamic quantities
should not depend significantly on the width of the energy window $\delta E=\sqrt{\pi}/\tau$, 
as long as a large number of energy eigenstates are contained within the energy window 
$[E-\delta E/2,E+\delta E/2)$, 
rationalizing the treatment of a microcanonical ensemble. 
However, for finite-size systems, the width of the
energy window affects the results in general.  
Therefore, here we numerically examine how 
the energy expectation value ${\cal E}_{\tau}(E)$ in Eq.~(\ref{eq:ETr}),
the entropy $S_\tau(E)$ in Eq.~(\ref{eq:STr}), 
the inverse temperature $\beta_\tau(E)$ in Eq.~(\ref{eq:betaTr}), and
the energy fluctuation
depend on the filtering time $\tau$ and thus the width of the energy window $\delta E=\sqrt{\pi}/\tau$. 
For this purpose, we employ the full diagonalization method to evaluate these quantities numerically exactly 
without introducing any random sampling. 
For each system size $N$, four target energies $E/NJ=-0.25,0.125,0.5$, and $0.875$ are selected
(see the vertical lines in Fig.~\ref{fig:nos}). 
Notice that we choose $E$ proportional to $N$ in order 
to compare the results for different system sizes. 
The results for smaller system sizes are also provided in Appendix~\ref{app:smallN}.

\subsubsection{Energy expectation value}
\begin{figure*}
  \includegraphics[width=1.95\columnwidth]{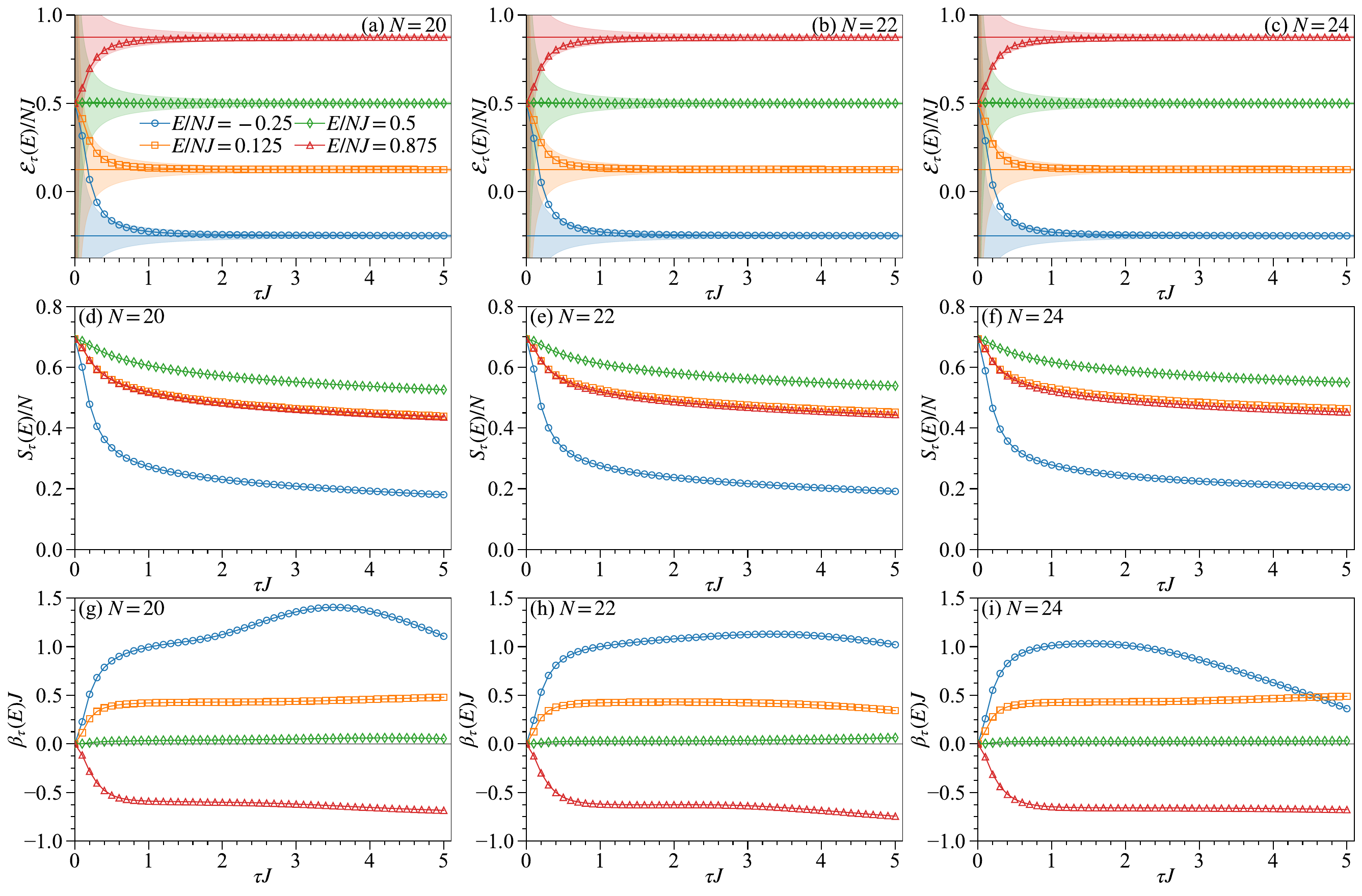}
  \caption{
    \label{fig:ESbeta}
    (a,b,c) The energy expectation value ${\cal E}_{\tau}(E)$,
    (d,e,f) the entropy $S_\tau(E)$, and 
    (g,h,i) the inverse temperature $\beta_\tau(E)$ 
    as a function of the filtering time $\tau$
    for 
    (a,d,g) $N=20$,
    (b,e,h) $N=22$, and
    (c,f,i) $N=24$ 
    calculated using the full diagonalization method.
    In (a,b,c), the horizontal blue, orange, green, and red lines indicate 
    the target energies $E/NJ=-0.25$, $0.125$, $0.5$, and $0.875$, respectively (also see Fig.~\ref{fig:nos}), and 
    the shaded areas indicate
    the corresponding energy windows $[E-\delta E/2,E+\delta E/2)$. 
    In (d,e,f), the results for $E/NJ=0.125$ (orange) and $E/NJ=0.875$ (red) are almost on top of each other in this scale. 
  }
\end{figure*}

Figures~\ref{fig:ESbeta}(a)-\ref{fig:ESbeta}(c) show the $\tau$ dependence of
the energy expectation value ${\cal E}_{\tau}(E)$, 
which may approach the target energy $E$
with increasing the filtering time $\tau$,
as long as there exist energy eigenvalues $\{E_n\}$
in the range of Eq.~(\ref{eq:range}). 
The energy expectation value
starts from its value in the zero-filtering-time limit,
\begin{equation}
  \lim_{\tau \to 0} {\cal E}_{\tau}(E)=\frac{1}{D}\Tr\left[\hat{\cal H}\right], 
\end{equation}
which is in our case reduced 
simply to $\lim_{\tau \to 0} {\cal E}_{\tau}(E)=JN_{\rm bond}/2$
because $\Tr[\hat{\bs{S}}_i \cdot \hat{\bs{S}}_j]=0$ for $i\ne j $.
As expected,  ${\cal E}_{\tau}(E)$
deviates from its zero-filtering-time limit and
approaches the target energy $E$ with increasing $\tau$.
Moreover, it is interesting to find that ${\cal E}_\tau(E)$ as a function of $\tau$ is almost always within the
energy window $[E-\delta E/2,E+\delta E/2)$ (indicated with the shaded areas in the figures) with 
no significant difference for different system sizes 
[also see Figs.~\ref{fig:ESbeta_small}(a)--\ref{fig:ESbeta_small}(c) in Appendix~\ref{app:smallN}]. 
Although we already have the target energy $E$ as an input parameter,
the energy expectation value ${\cal E}_{\tau}(E)$ is a useful quantity 
to check if the filtering time $\tau$ is so large that
${\cal E}_{\tau}(E)$ is consistent with the target energy $E$.

\subsubsection{Entropy}
Figures~\ref{fig:ESbeta}(d)-\ref{fig:ESbeta}(f) show the $\tau$ dependence of
the entropy $S_{\tau}(E)$. 
The entropy 
in the zero-filtering-time limit is given by 
\begin{equation}
  \lim_{\tau \to 0} S_{\tau}(E)=\ln D,
  \label{eq:Slimit}
\end{equation}
as explained in Sec.~\ref{sec:window} [see Eq.~(\ref{eq:nos_tau0})].
It is also readily shown that the entropy $S_{\tau}(E)$ is
a decreasing function of $\tau$, i.e., 
\begin{equation}
  \frac{\partial S_{\tau}(E)}{\partial \tau}
  =
  \frac{1}{\Tr\left[\hat{G}_\tau(E)\right]}
  \frac{\partial \Tr\left[\hat{G}_\tau(E)\right]}
       {\partial \tau}
       \leqslant 0,
  \label{eq:dSdtau}
\end{equation}
because
$\partial_\tau \Tr[\hat{G}_\tau(E)] =-2\tau{\sum_{n=0}^{D-1} (E_n-E)^2\e^{-(E_n-E)^2\tau^2}}\leqslant 0$ and
$\Tr[\hat{G}_\tau(E)] >0$.
The inequality in Eq.~(\ref{eq:dSdtau}) can be understood intuitively
because the number of states in the energy window
$[E-\delta E/2,E+\delta E/2)$, which is given by $\Tr[\hat{G}_\tau(E)]$,  
never increases with decreasing $\delta E$ (also see Fig.~\ref{fig:nos}).
The equality in Eq.~(\ref{eq:dSdtau}) is achieved when $\tau=0$, and
hence the maximum value of the entropy is in the
zero-filtering-time limit given in Eq.~(\ref{eq:Slimit}).

Among the four target energies, 
the difference in the entropy among the different system sizes $N$
is somewhat significant for $E/NJ=-0.25$,  
for which the larger $N$ has the larger entropy 
[also see Figs.~\ref{fig:ESbeta_small}(d)--\ref{fig:ESbeta_small}(f) in Appendix~\ref{app:smallN}].
This behavior is consistent with the results for 
the number of states shown in Fig.~\ref{fig:nos},
where the energy eigenvalues are distributed rather sparsely around 
$E/NJ=-0.25$.

\subsubsection{Inverse temperature}
Figures~\ref{fig:ESbeta}(g)-\ref{fig:ESbeta}(i) show the $\tau$ dependence of
the inverse temperature $\beta_\tau(E)$.
The inverse temperature starts with its value 
in the zero-filtering-time limit, 
\begin{equation}
  \lim_{\tau \to 0} \beta_\tau(E)=0, 
\end{equation}
corresponding to the infinite temperature.
At finite $\tau$, $\beta_\tau(E)$ can be either positive or negative
as it is obvious from Eq.~(\ref{eq:betaTr}). 
The positive (negative) temperature can be obtained 
when the energy expectation value ${\cal E}_{\tau}(E)$
is larger (smaller) than the target energy $E$. 
Indeed, the inverse temperature for $E/NJ=0.5$ 
remains around 0 even when $\tau$ is increased, 
and the inverse temperature for the target energy smaller (larger) than $E/NJ=0.5$ 
(corresponding to the infinite temperature) is positive (negative). 
This is expected because the slope of the magenta line in Fig.~\ref{fig:nos} is 
related to the inverse temperature at the corresponding target energy, i.e.,  
$\beta_\tau(E)=\partial_E\ln{\Tr[\hat{G}_{\tau}(E)]}$ from Eqs.~(\ref{eq:beta}) and (\ref{eq:STr}). 

We also find that for target energies $E/NJ=0.125$, 0.5, and 0.875, 
around which the distribution of energy eigenstates is dense,  
the inverse temperature $\beta_\tau(E)$ becomes almost independent of $\tau$, i.e., 
\begin{equation}
  \frac{\partial \beta_\tau(E)}{\partial \tau} \approx 0,
  \label{eq:partial_beta}
\end{equation}
once $\tau$ becomes sufficiently large, i.e., $\tau J \sim 1.0$,
where the energy expectation value ${\cal E}_\tau(E)$ essentially reaches
the target energy $E$, and this almost $\tau$ independent behavior 
of $\beta_\tau(E)$ continues up to $\tau J \alt 3$. 
Moreover, the $\tau$ dependence becomes less significant for larger $N$ 
[also see Figs.~\ref{fig:ESbeta_small}(g)--\ref{fig:ESbeta_small}(i) in Appendix~\ref{app:smallN}].
The $\tau$-independent $\beta_\tau (E)$ implies 
that the energy expectation value ${\cal E}_{\tau}(E)$ approaches the target energy $E$
in $\tau$ as $1/\tau^2$, i.e., ${\cal E}_\tau(E)-E = \beta(E)/2\tau^2$,
where $\beta(E)$ is a representative value of $\beta_\tau(E)$
satisfying the condition in Eq.~(\ref{eq:partial_beta}).

On the other hand, the inverse temperature $\beta_\tau(E)$ depends rather wildly on $\tau$ for $E/NJ=-0.25$. 
In particular, 
the inverse temperature for $N=14$, 16 and 18 becomes negative for $\tau J \gtrsim 4$ 
[see Figs.~\ref{fig:ESbeta_small}(g)--\ref{fig:ESbeta_small}(i) in Appendix~\ref{app:smallN}], 
although a positive value is expected for the target energy smaller than $E/NJ=0.5$.
Such a dependence is due to the sparse distribution
of energy eigenstates around $E/NJ=-0.25$ (see Figs.~\ref{fig:nos} and~\ref{fig:nos_small}).
Namely, if $\tau$ is so large that the number of energy eigenstates within the
energy window is insufficiently small for considering the microcanonical ensemble,
the calculation of thermodynamic quantities fails, as it is well known in statistical mechanics. 
Finally, we note that except for $E/NJ=-0.25$, the $\tau$-dependence of the inverse temperature
is more moderate for $N=24$ than for $N=20$ and 22 
because of the denser distribution of energy eigenstates around the target energy $E$.  
As discussed in Sec.~\ref{sec:result_comp}, this will be more evident for $N=28$ 
(see Fig.~\ref{fig:rand2}), where the full-diagonalization results are not available.

\subsubsection{Energy fluctuation}\label{sec:fluct}
\begin{figure*}
  \includegraphics[width=1.9\columnwidth]{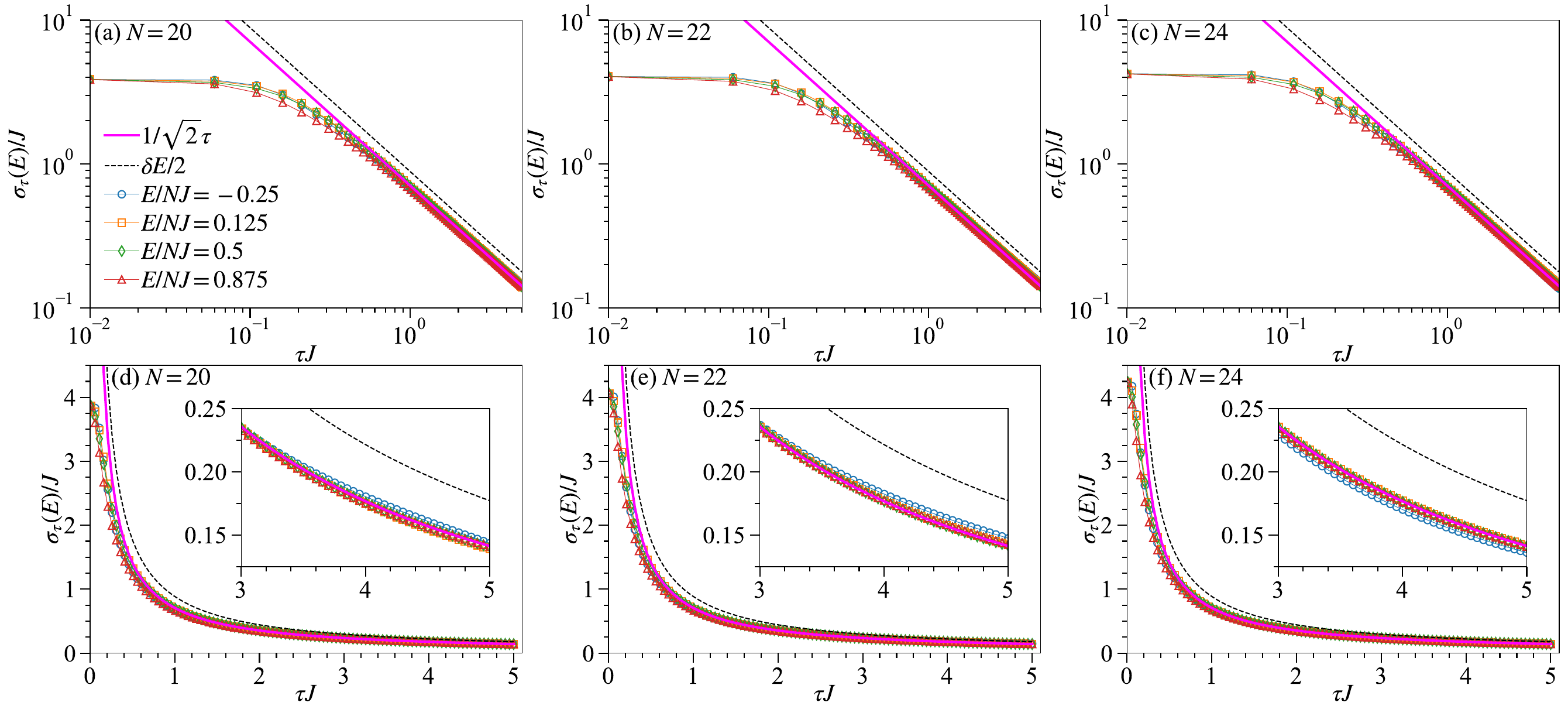}
  \caption{
    \label{fig:Efluct}
    The energy fluctuation $\sigma_\tau(E)$
    as a function of the filtering time $\tau$ for
    different target energies $E/NJ=-0.25$, $0.125$, $0.5$, and $0.875$ 
    denoted by blue circles, orange squares, green diamonds, and red triangles, respectively, 
    for     
    (a) $N=20$,
    (b) $N=22$, and 
    (c) $N=24$ 
    calculated using the full diagonalization method. 
    For comparison, $1/\sqrt{2}\tau$ and $\delta E/2$ are also indicated
    by thick magenta line and thin black dashed line, respectively. 
    (d-f) show the same results as in (a-c) but plotted in a linear scale. 
    The insets in (d-f) show the magnifications for $3 \leqslant \tau J \leqslant 5$. 
  }
\end{figure*}

Figure~\ref{fig:Efluct} shows the
$\tau$ dependence of the energy fluctuation $\sigma_\tau(E)$ defined as 
\begin{equation}
  \sigma_\tau(E) \equiv \sqrt{
    \frac{\Tr\left[\hat{\cal H}^2\hat{G}_\tau(E)\right]}{\Tr\left[\hat{G}_{\tau}(E)\right]}
    -{\cal E}_\tau(E)^2
  }.
\end{equation}
We first notice in the figure that the system-size dependence of $\sigma_\tau(E)$ is not significant
(also see Fig.~\ref{fig:Efluct_small} in Appendix~\ref{app:smallN}). 
As expected from the Gaussian form of $\hat{G}_\tau(E)$ in Eq.~(\ref{eq:Gop}), we also find that 
the energy fluctuation for large $\tau$ behaves as
\begin{equation}
  \sigma_\tau(E) \sim \frac{1}{\sqrt{2}\tau}.
  \label{eq:sigma_tau}
\end{equation}
This is because $1/\sqrt{2}\tau$ is the ``sigma'' of the Gaussian in Eq.~(\ref{eq:Gop}).
However, as opposed to what is expected from the form of $1/\sqrt{2}\tau$, 
$\sigma_\tau(E)$ never diverges for $\tau \to 0$ 
simply because the energy eigenvalues are distributed 
only in the finite range of $[E_0,E_{D-1}]$, and hence 
$\sigma_\tau(E)$ deviates from $1/\sqrt{2}\tau$ for small $\tau$. 
Finally, we note that the half width of the energy window, 
$\delta E/2=\sqrt{\pi}/2\tau$, is less relevant for
describing the energy fluctuation for large $\tau$.

\subsection{Comparison with random sampling for trace evaluation} \label{sec:result_comp}
The numerical results shown so far have been calculated by fully diagonalizing the Hamiltonian and thus 
they are free from statistical errors. Now we shall compare these results with those obtained on the basis 
of the energy-filtered random-phase states with the random sampling 
for the trace evaluations as in Eqs.~(\ref{eq:TrG_R}) and (\ref{eq:TrHG_R}). 
Note that the latter results with the number $R$ of random samples being infinity 
are essentially identical to the former results obtained by the full diagonalization method. 
One technical advantage of the random sampling approach is that one can treat larger systems, as we shall 
also demonstrate here.

\begin{figure}
  \includegraphics[width=1.0\columnwidth]{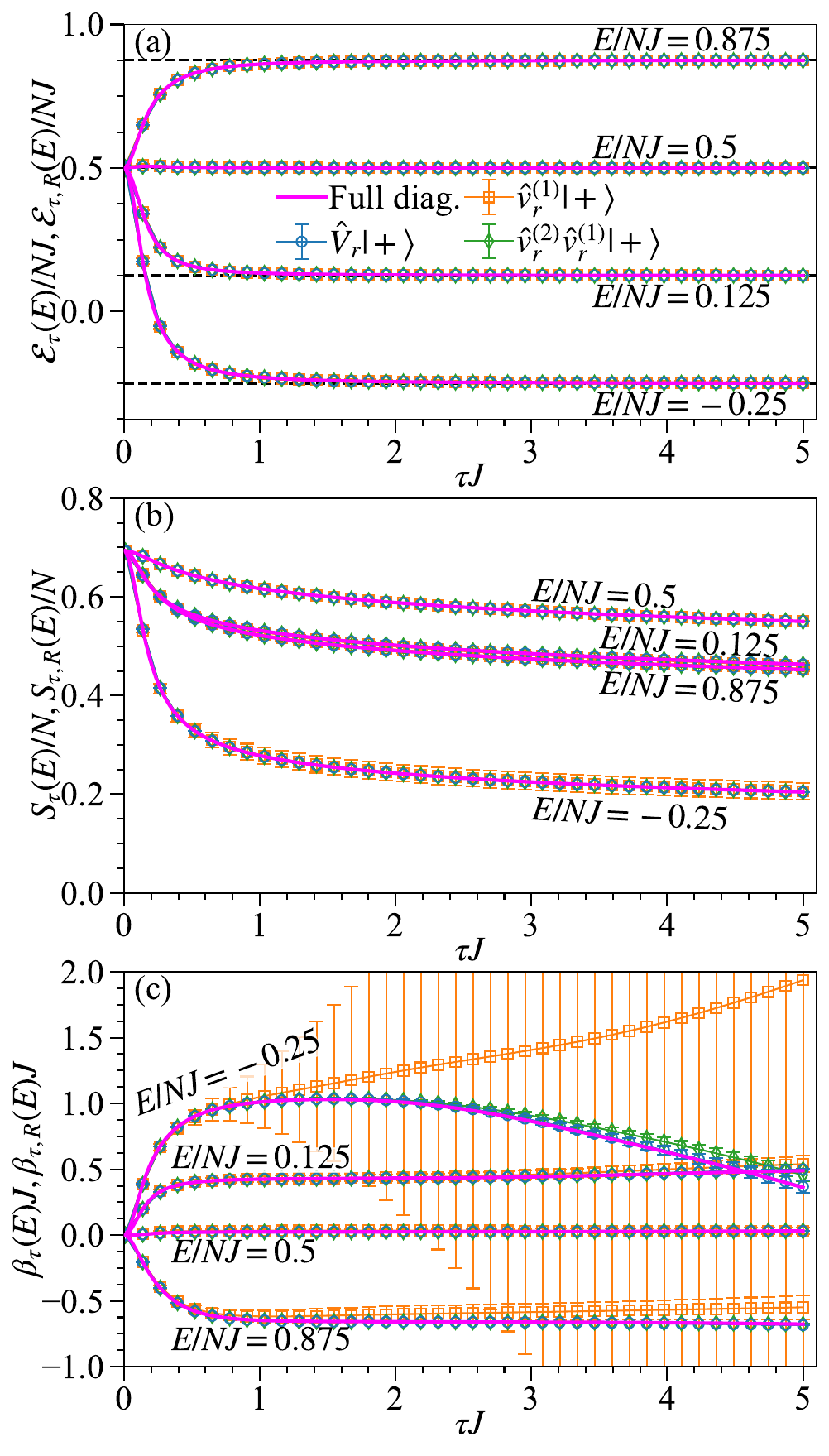}
  \caption{
    \label{fig:rand}
    (a) The energy expectation value ${\cal E}_{\tau,R}(E)$,
    (b) the entropy $S_{\tau,R}(E)$, and 
    (c) the inverse temperature $\beta_{\tau,R}(E)$ 
    as a function of the filtering time $\tau$
    for different target energies $E/NJ=-0.25$, $0.125$, $0.5$, and 0.875
    calculated with the random sampling of the trace evaluations 
    using three types of the random-phase states, i.e., $\hat{V}_r|+\rangle$ (blue circles),
    $\hat{v}_r^{(1)}|+\rangle$ (orange squares), and
    $\hat{v}_r^{(2)}\hat{v}_r^{(1)}|+\rangle$ (green diamonds). 
    The system size is fixed at $N=24$ for all the calculations and 
    the number of samples is $R=64$. 
    For comparison, the results for ${\cal E}_{\tau}(E)$, $S_{\tau}(E)$, and $\beta_{\tau}(E)$ 
    calculated by the full diagonalization method are also shown by magenta solid lines. 
    Note that these full-diagonalization results are the same as those in
    Figs.~\ref{fig:ESbeta}(c), \ref{fig:ESbeta}(f), and \ref{fig:ESbeta}(i).
    %
    %
    The black dashed horizontal lines in (a) indicate the target energies.
    %
    %
    For most of the cases, the results, including the statistical errors, for $\hat{V}_r|+\rangle$ and 
    $\hat{v}_r^{(2)}\hat{v}_r^{(1)}|+\rangle$ are almost indistinguishable.
    Moreover, the results in (b) for $E/NJ=0.125$ and $0.875$ are on top of 
    each other in this scale. 
    %
  }
\end{figure}

Figure~\ref{fig:rand} shows the results for 
the energy expectation value ${\cal E}_{\tau,R}(E)$,
the entropy $S_{\tau,R}(E)$, and 
the inverse temperature $\beta_{\tau,R}(E)$
of the $N=24$ system evaluated 
using the random-phase states 
$\hat{V}_r|+\rangle$,  
$\hat{v}_r^{(1)}|+\rangle$, and
$\hat{v}_r^{(2)}\hat{v}_r^{(1)}|+\rangle$, 
which are also compared with those for 
${\cal E}_{\tau}(E)$, $S_{\tau}(E)$, and $\beta_{\tau}(E)$ obtained by the 
full diagonalization method. 
For the trace evaluation with the random-phase states, 
we set the number of random samples at $R=64$.
We indeed confirm in Fig.~\ref{fig:rand} (also see Fig.~\ref{fig:rand3}) that generally 
${\cal E}_{\tau,R}(E)$, $S_{\tau,R}(E)$, and $\beta_{\tau,R}(E)$
evaluated by these different random-phase states agree with 
those obtained by the full diagonalization method for all four target energies 
within a few error bars. Here, the error bar is due to the statistical error of a finite number of samples 
and is estimated form the standard error of the mean.  

\begin{figure}
  \includegraphics[width=1.0\columnwidth]{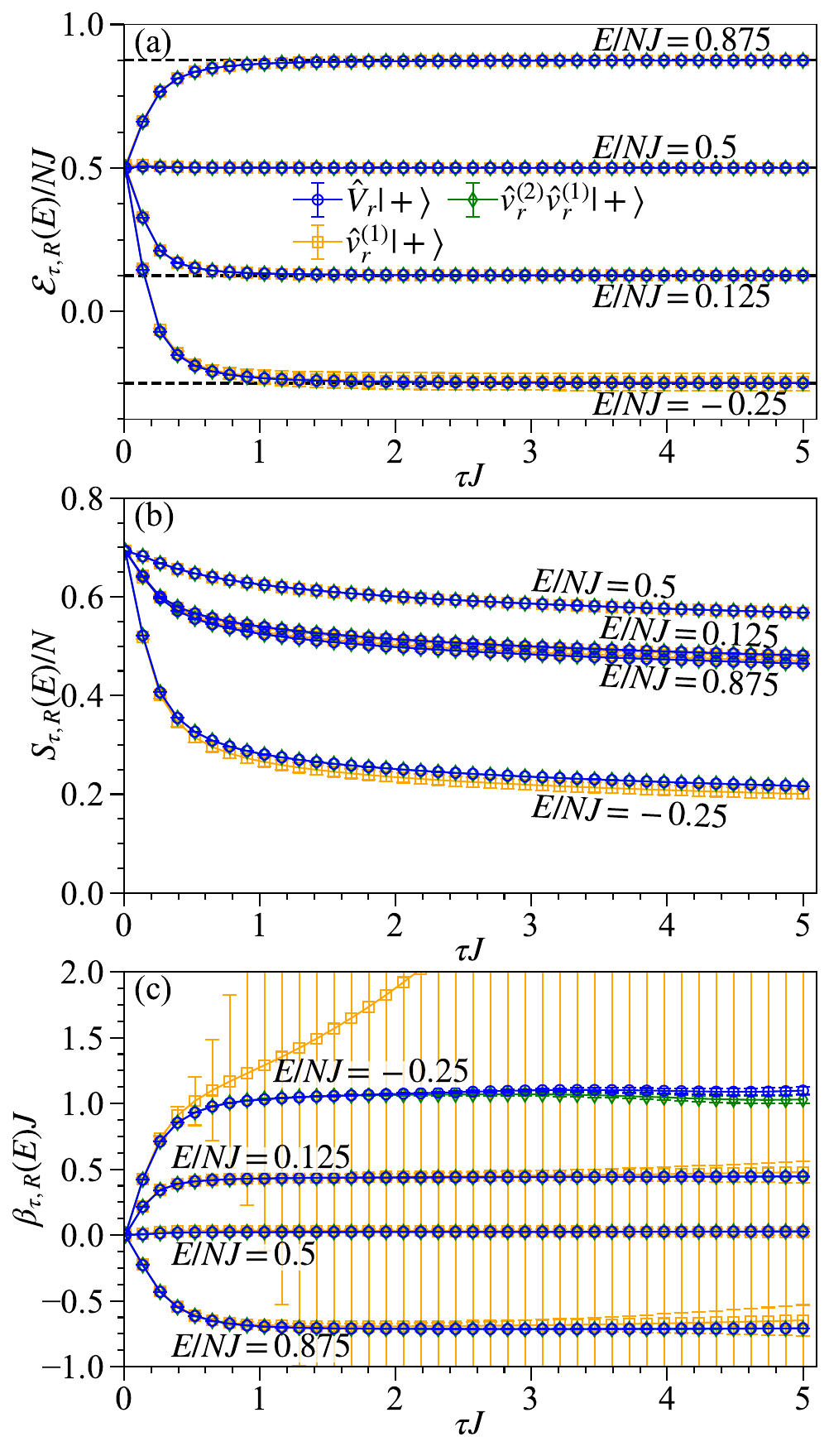}
  \caption{
    \label{fig:rand2}
    The same as Fig.~\ref{fig:rand} but for $N=28$. 
    The full-diagonalization results are not shown because the huge computational resource required is not 
    currently available. 
  }
\end{figure}

However, more interestingly, among the results for the three distinct types of the random-phase states, 
the statistical errors for the random-phase-product states $\hat{v}_r^{(1)}|+\rangle$ 
is significantly larger than those for the others, 
especially for the inverse temperature at the target energy $E/NJ=-0.25$,  
as shown in Fig.~\ref{fig:rand}(c) [also see Fig.~\ref{fig:rand3}(l)], while
the difference between the results for
$\hat{v}_r^{(2)}\hat{v}_r^{(1)}|+\rangle$ and $\hat{V}_r|+\rangle$
is not significant. 
As shown in Fig.~\ref{fig:rand2} (also see Fig.~\ref{fig:rand4}), 
the same trend but with larger statistical errors
for $\hat{v}_r^{(1)}|+\rangle$ 
is also found for the larger system of $N=28$ with $R=64$, where 
the full-diagonalization results are not available because 
of the huge computational resource required. 
It is also interesting to notice in Fig.~\ref{fig:rand2}(c) 
[also see Figs.~\ref{fig:rand4}(c), \ref{fig:rand4}(f), \ref{fig:rand4}(i), and \ref{fig:rand4}(l)]
that the inverse temperature evaluated for 
$\hat{v}_r^{(2)}\hat{v}_r^{(1)}|+\rangle$ as well as $\hat{V}_r|+\rangle$
is almost independent of $\tau$ for a wider region of $\tau$ even at the target energy $E/NJ=-0.25$. 
Although the simple circuit structure of 
$\hat{v}_r^{(1)}|+\rangle$ for the random-phase states, 
which involve only the single-qubit rotations, 
is quite appealing, 
our results reveal that the random-phase-product states $\hat{v}_r^{(1)}|+\rangle$ 
require the larger number $R$ of samples  
than the other random-phase states $\hat{v}_r^{(2)}\hat{v}_r^{(1)}|+\rangle$ and $\hat{V}_r|+\rangle$
to achieve a desired statistical accuracy. 
We should note here that the 
efficiency of the sampling with
the random-phase-product states $\hat{v}_r^{(1)}|+\rangle$ is under recent debate,
and several studies on improving~\cite{Iwaki2021,Goto2021} and
quantifying~\cite{Iwaki2022} the efficiency of the sampling
beyond the level of the random-phase-product states 
have been recently reported.

Let us now discuss the variance due to the random samplings considered here. 
In the case of the random sampling for the trace evaluations using the random-phase states $\{\hat{V}_r|+\rangle\}_r$,
the covariance between 
$A_{r}=\langle +|\hat{V}_r^\dag \hat{A}\hat{V}_r|+\rangle$ and 
$B_{r}^*=\langle +|\hat{V}_r^\dag \hat{B}^\dag\hat{V}_r|+\rangle$ 
decreases as in Eq.~(\ref{eq:cov}).
Here, $\hat{\cal H}$, $\hat{G}_\tau(E)$, and $\hat{\cal H}\hat{G}_\tau(E)$ are
relevant for operators $\hat{A}$ and $\hat{B}$. 
Since
$|\langle b|\hat{G}_\tau(E)|b\rangle|=  O(1)$, 
$|\langle b|\hat{\cal H}\hat{G}_\tau(E)|b\rangle|= O(N)$,
$\Tr[\hat{G}_\tau^2(E)] = \Tr[\hat{G}_{\sqrt{2}\tau}(E)]=O(D)$, and 
$\Tr[\hat{\cal H}^2\hat{G}_\tau^2(E)] = O(DN^2)$, 
the corresponding (co)variance decreases as $1/D$, 
i.e., exponentially in the system size $N$ 
(see Sec.~\ref{sec:trace}). 
Since the diagonal-unitary $t$-designs can simulate
up to $t$-th statistical moment of the random-phase states~\cite{Nakata2014}, 
diagonal-unitary $t$-designs with $t\geqslant 2$
also possess the property of the exponentially small (co)variance as in Eq.~(\ref{eq:cov}),
but those with $t=1$ do not, 
as $A_{r}^2$ and $A_{r}B_{r}$ involve two $\hat{V}_r$'s and two $\hat{V}_r^\dag$'s.
This is thus consistent with the numerical observation that
the statistical errors for $\hat{v}_r^{(1)}|+\rangle$
are qualitatively distinct from those for 
$\hat{v}_r^{(2)}\hat{v}_r^{(1)}|+\rangle$ and $\hat{V}_r|+\rangle$. 
We should note that a similar discussion
for the canonical TPQ state using random Clifford circuits 
(i.e., not for diagonal-unitary designs) is found in the recent 
study~\cite{Coopmans2022}.
Considering quantum computation, it is impractical to 
prepare $\hat{V}_r|+\rangle$ because 
it requires an exponentially large number
of single- and two-qubit gates. 
Therefore, as our numerical results also clearly demonstrated, 
$\hat{v}_r^{(2)}\hat{v}_r^{(1)}|+\rangle$ might be an appealing choice 
among the random-phase states 
as it requires only a polynomial number of single- and two-qubit gates.

\section{Conclusions and discussions}\label{sec:conclusions}
We have proposed a method to calculate thermodynamic quantities of 
quantum many-body systems based on microcanonical TPQ states,
i.e., energy-filtered random-phase states. 
In this formalism, a microcanonical ensemble is specified by a target energy $E$ as well as its 
associated energy window of width $\delta E$, and the density of states is thus expressed by a sum of 
Gaussians centered at the target energy $E$ with its spread corresponding to the width of the 
energy window $\delta E$. 
Since the density of states is a continuous function of $E$ in this formalism, 
we can derive analytical expressions of thermodynamic quantities such as  
the entropy and the inverse temperature. 
This formalism also allows us to estimate these thermodynamic quantities 
by evaluating the trace of the time-evolution operator 
and the trace of the time-evolution operator multiplied by the Hamiltonian of the system, 
which is thus suitable for quantum computation. 
By introducing the random sampling for the trace evaluations using the random-phase states, 
we can recognize that our formalism is a microcanonical counterpart of the canonical TPQ state, and 
the corresponding TPQ state is now an energy-filtered random-phase state 
with the target energy $E$ and the filtering time $\tau$, the latter being related
to the width of the energy window via $\delta E = \sqrt{\pi}/\tau$. 

We have then numerically validated the proposed method 
by calculating thermodynamic quantities of
the one-dimensional spin-1/2 Heisenberg model  
with the full diagonalization method up to $N=24$ qubits and
the random sampling method up to $N=28$ qubits. 
It is found that the proposed method is most effective for
a target energy around which a large number of energy eigenstates exist and
hence the statistical-mechanical treatment of thermodynamic quantities
is reasonable. 
We have also numerically examined three different types of the random-phase states
$\hat{V}_r|+\rangle$, $\hat{v}_r^{(1)}|+\rangle$, and $\hat{v}_r^{(2)}\hat{v}_r^{(1)}|+\rangle$, 
the latter two being prepared by the diagonal-unitary circuits, 
for the random sampling of the trace evaluations. 
We have shown that for a fixed number $R$ of samples, the random-phase states
$\hat{v}_r^{(2)}\hat{v}_r^{(1)}|+\rangle$ provide the results as
accurate as $\hat{V}_r|+\rangle$, where they can be prepared 
with a polynomial number $O(N^2)$ of one- and two-qubit gates.
It is worth noting that
the method for the trace evaluations 
using the random-phase states
$\hat{V}_r|+\rangle_r$,
$\hat{v}_r^{(1)}|+\rangle_r$, or
$\hat{v}_r^{(2)}\hat{v}_r^{(1)}|+\rangle_r$ 
is independent of the spatial dimensions of the system,
and hence the same method is expected to be applicable for computing
thermodynamic quantities in higher spatial dimensions.

Considering quantum computation, the number of quantum gates
required is $O(N^2)+O(MN)$ in the present setting, where
$O(N^2)$ is for preparation of the state $\hat{v}_r^{(2)}\hat{v}_r^{(1)}|+\rangle$,
and $O(MN)$ is for the time-evolution operator. If a $k$-local Hamiltonian is
considered, the number of quantum gates
will be $O(N^2)+O(kMN)$, which is still polynomial in $N$. 
The quantum circuit corresponding to $\hat{v}_r^{(2)}$
consists of $O(N^2)$ all-to-all two-qubit operations. 
Therefore, implementing $\hat{v}_r^{(2)}\hat{v}_r^{(1)}|+\rangle$ for large systems (e.g., $N\sim 10^2$) 
on, for example, superconducting quantum computers would be challenging.  
Thus implementing $\hat{v}_r^{(1)}|+\rangle$ instead would be a 
good compromise for demonstrating the proposed method using a real quantum computer 
as it requires only $N$ Hadamard and $N$ single-qubit-rotation gates.
On the other hand, the time evolution of a quantum state by a quantum
many-body Hamiltonian has been demonstrated recently~\cite{arute2020observation}, 
suggesting that the time-evolution part of the circuit 
would be feasible in near-term quantum computers.

If the number of states around a target energy is 
so small that only a few energy eigenstates are found in the energy window, 
which often occurs especially for small systems, 
then the statistical-mechanical treatment of thermodynamic quantities is irrelevant.
Rather, in such a case, one might be interested in 
these individual energy eigenvalues and eigenstates separately.  
While there are several methods to compute 
a few eigenstates of the Hamiltonian, 
those that are related to the present formalism are   
the filter-diagonalization method~\cite{Wall1995} in classical computation, and 
the quantum filter-diagonalization method~\cite{parrish2019quantum} and 
the related Krylov-subspace method~\cite{Stair2020} in quantum computation.
More recently, a Gaussian-filter method for exploring ground-state properties has been
proposed~\cite{He2022}, where the initial state is prepared so as to
have a large overlap with the target ground state, instead of being prepared as a random state. 
The quantum subspace expansion method such as one~\cite{seki2021spatial} 
based on the quantum power method~\cite{seki2021} might be also useful.
We also note that in a similar spirit 
but for the canonical ensemble in classical computation,
an improved version of the finite-temperature Lanczos method 
by explicitly calculating low-lying eigenpairs has recently been proposed~\cite{Morita2020}
to reduce the number of samples at low temperatures
(see also Ref.~\cite{Aichhorn2003} for another approach to an efficient sampling at low temperatures).

In the formalism developed here, we have introduced the width of the energy window $\delta E\sim 1/\tau$ 
as a spread of the Gaussian $\hat{G}_\tau(E)$ in Eq.~(\ref{eq:Gop}). 
However, this is in principle not necessary. 
For example, we could also employ a more conventional  
rectangular function centered at $E$ with its width $\delta E$, 
as in the histograms in Fig.~\ref{fig:nos}. 
A rectangular cut of the energy window is appealing 
for its conceptual simplicity in counting the number of states within this window. 
A drawback of this is that the entropy is no longer a continuous function of $E$ 
and hence the derivative of it with respect to $E$ is not well defined. 
In addition, considering the numerical simulation based on
the Fourier transform of $\hat{G}_\tau(E)$ but with a sum of rectangular functions, 
there appears a cardinal sine (${\rm sinc}$) function 
in place of the Gaussian in the integrand of Eq.~(\ref{eq:GFT}). 
The appearance of the ${\rm sinc}$ function in the time domain
might be less appealing than the Gaussian
from a numerical point of view 
because ${\rm sinc}{x}=\sin{x}/{x}$
decays slowly as $1/x$ with oscillation.
Therefore, using a Gaussian instead of a rectangular function 
much simplifies the analytical and numerical
treatments of the microcanonical ensemble. 
Although details of the choice for the energy window should not affect
thermodynamic quantities in the thermodynamic limit, 
it is still interesting to explore how functional forms for $\hat{G}_\tau(E)$,
other than the Gaussian studied here, could improve  
the numerical efficiency of the calculations for finite systems.

Let us now briefly discuss a possible generalization of the present method
to other ensembles. To this end, let us first recall  
the following properties of the random-phase state $|\phi_r\rangle$:  
\begin{alignat}{1}
  D \lim_{R\to \infty}\langle \langle |\phi_r \rangle \langle \phi_r| \rangle \rangle_R &= \hat{I} 
\end{alignat}
and
\begin{alignat}{1}
  D \langle \langle\, \langle\phi_r|\phi_r\rangle\, \rangle \rangle_R &= \Tr\left[\hat{I}\right]=D. 
\end{alignat}
Along with the analysis in Sec.~\ref{sec:microTPQ}, 
$|\phi_r\rangle$ can be considered as a pure quantum state corresponding to
the uniform ensemble~\cite{Tolman}, which is represented by a density matrix 
of the maximally mixed state, i.e., $\hat{\rho}_{\rm uni}\equiv \hat{I}/D$.  
The uniform ensemble is 
an ensemble used when no prior knowledge of the system is available,
in the sense that all the pure states appear in
$\hat{\rho}_{\rm uni}$ with the equal probability $1/D$ 
because the lack of knowledge can be quantified by, for example, the von Neumann entropy 
having the maximum value $\ln D$. 
The density matrix $\hat{\rho}_\tau(E)=\hat{G}_\tau(E)/\Tr[\hat{G}_\tau(E)]$ in Eq.~(\ref{eq:rho_E})
specifies the energy $E$ of the system and the width of the energy window $\sqrt{\pi}/\tau$,
and the acquirement of such knowledge 
is incorporated in the pure-state formalism by multiplying
$[\hat{G}_\tau(E)]^{\frac{1}{2}}$ to the random-phase state $|\phi_r\rangle$,
i.e., by filtering the energy. 

We can push forward the above strategy to 
priorly specify several physical quantities, not limited to the energy,  
of the system by using multiple filters. 
Let us introduce an operator (also see Ref.~\cite{Tolman})
\begin{equation}
  \hat{\cal G}_{\hat{Q},\sigma}(Q) = \exp \left[ -\frac{(\hat{Q}-Q)^2}{2\sigma^2} \right], 
\end{equation}
where $\hat{Q}$ is a Hermitian operator in the
Hilbert space of the system and $Q$ and $\sigma$ are real parameters, 
hence $\hat{\cal G}_{\hat{Q},\sigma}(Q)$ being Hermitian and positive definite.  
We can also introduce $N_q$ products of such operators
$\hat{\cal G}_{\vec{\hat{Q}},\vec{\sigma}}(\vec{Q}) 
\equiv \prod_{q=1}^{N_q}\hat{\cal G}_{\hat{Q}_q,\sigma_q}(Q_q)$
to consider a density matrix
\begin{equation}
  \hat{\rho}_{\vec{\hat{Q}},\vec{\sigma}}(\vec{Q}) \equiv
  \frac{
    \hat{\cal G}_{\vec{\hat{Q}},\vec{\sigma}}(\vec{Q})
  }
       {
         \Tr\left[\hat{\cal G}_{\vec{\hat{Q}},\vec{\sigma}}(\vec{Q})\right]
       },
       \label{eq:rhoGen}
\end{equation} 
in which the eigenvalues of $\hat{Q}_1, \hat{Q}_2, \cdots, \hat{Q}_{N_q}$ 
are concentrated around $Q_1, Q_2, \cdots, Q_{N_q}$
with their variances approximately $\sigma_1^2, \sigma_2^2, \cdots, \sigma_{N_q}^2$.
Here, we assume that 
$[\hat{Q}_q,\hat{Q}_{q^\prime}]=0$ and 
$[\hat{Q}_q,\hat{\cal H}]=0$ so that 
$\hat{\rho}_{\vec{\hat{Q}},\vec{\sigma}}(\vec{Q})$ is Hermitian and does not evolve in time.
Examples of $\hat{Q}_q$
include the Hamiltonian itself, as already described above with
$\hat{\cal G}_{\hat{\cal H},\frac{1}{\sqrt{2}\tau}}(E)=\hat{G}_\tau(E)$, 
the magnetization operator,
the particle-number operator (for fermions or bosons), and 
the total-spin operator, assuming that the system preserves the corresponding symmetries.  
The pure state associated with $\hat{\rho}_{\vec{\hat{Q}},\vec{\sigma}}(\vec{Q})$
is given by $\hat{\cal G}_{\vec{\hat{Q}},\sqrt{2}\vec{\sigma}}(\vec{Q})|\phi_r\rangle$, 
and the Fourier representation of
$\hat{\cal G}_{\vec{\hat{Q}},\vec{\sigma}}(\vec{Q})$
is given by 
\begin{equation}
\hat{\cal G}_{\vec{\hat{Q}},\vec{\sigma}}(\vec{Q})=
\prod_{q=1}^{N_{q}} \left[ \frac{1}{\sqrt{2\pi}\sigma_q}\int_{-\infty}^{\infty}
\dd s_q\e^{-\frac{s_q^2 \sigma_q^2}{2}}\e^{-\imag (\hat{Q}_q-Q_q) s_q} \right],
\end{equation}
where the integration variable $s_q$ has 
the dimension of $1/Q_q$ [also see Eq.~(\ref{eq:GFT})].

It is worth noting that the above generalization can be further extended
to a set of noncommuting operators $[\hat{Q}_q,\hat{Q}_{q^\prime}]\not{=}0$ 
by replacing $\hat{\cal G}_{\vec{\hat{Q}}, \vec{\sigma}}(\vec{Q})$ in Eq.~(\ref{eq:rhoGen}) with
\begin{equation}
  {\cal G}_{\vec{\hat{Q}},\bs{\Sigma}}(\vec{Q})
=\exp\left[
  -\frac{1}{2}\left(\vec{\hat{Q}}-\vec{Q}\right)^T 
  \bs{\Sigma}^{-1}
  \left(\vec{\hat{Q}}-\vec{Q}\right)
  \right],
\end{equation}
where $\bs{\Sigma}$ is an $N_q \times N_q$ covariance matrix whose $(i,j)$ entry 
approximately gives the covariance between the expectations of $\hat{Q}_i$ and $\hat{Q}_j$. 
${\cal G}_{\vec{\hat{Q}},\bs{\Sigma}}(\vec{Q})$ will be
relevant to generalized thermodynamic ensembles with
noncommuting operators studied in the context of generalized thermodynamic
resource theories~\cite{Halpern2016,Halpern2018,Halpern2020,Halpern2022},
which are recently examined experimentally with a trapped-ion system~\cite{Kranzl2022}.
When 
$\bs{\Sigma}={\rm diag}(\sigma_1^2, \sigma_2^2, \cdots,\sigma_{N_q}^2)$ and
$[\hat{Q}_q,\hat{Q}_{q^\prime}]=0$,
${\cal G}_{\vec{\hat{Q}},\bs{\Sigma}}(\vec{Q})$ is reduced to
$\hat{\cal G}_{\vec{\hat{Q}}, \vec{\sigma}}(\vec{Q})$. 
We expect that it is useful to consider a variety of ensembles along these lines 
when we can priorly specify several physical quantities of the system  
associated with Hermitian operators $\hat{Q}_1, \hat{Q}_2, \cdots, \hat{Q}_{N_{q}}$.  
Finally, we note that 
our formalism developed here is
related to the previous work in Ref.~\cite{Sirui2021}
(also see related works in Refs.~\cite{Yang2022,Schuckert2022}).
Aside from technical details,
the main conceptual differences 
can be summarized as follows: 
(i)
we derive the analytical expression for the inverse temperature that is not a parameter but is evaluated, 
covering both positive- and negative-temperature sides, 
and
(ii)
our formalism reveals a connection to a TPQ-state formalism of the microcanonical ensemble 
via the energy-filtered random-phase states that are closely related to the states introduced in a different context 
to extract eigenvalues and eigenstates of $\hat{H}$ at any desired energy range, known as 
the filter-diagonalization method~\cite{Wall1995}.

\acknowledgements
We are grateful to Nicole Yunger Halpern for valuable comment. 
We are also grateful to Atsushi Iwaki for valuable discussion and comment,   
and for bringing Refs.~\cite{Challa1988,Yoneta2019} to our attention.
A part of the numerical simulations has been performed
using the HOKUSAI supercomputer at RIKEN (Project IDs: Q22525 and C22007).
This work is supported by
Grant-in-Aid for Research Activity Start-up (No.~JP19K23433),
Grant-in-Aid for Scientific Research (C) (No.~JP22K03520),
Grant-in-Aid for Scientific Research (B) (No.~JP18H01183), and 
Grant-in-Aid for Scientific Research (A) (No.~JP21H04446) from MEXT, Japan. 
This work is also supported in part by the COE research grant in computational science from 
Hyogo Prefecture and Kobe City through Foundation for Computational Science.

\appendix

\section{Cumulative number of states}\label{app:cumulative}
In this Appendix, we discuss the cumulative number of states.
The cumulative number of states $W_\tau(E)$ below energy $E$ is given by 
\begin{equation}
  W_\tau(E) = \int_{-\infty}^E \dd E^\prime g_\tau(E^\prime)
  = \frac{1}{2}\sum_{n=0}^{D-1} {\rm erfc}[(E_n-E)\tau], 
\end{equation}
where ${\rm erfc}(x)=\frac{2}{\sqrt{\pi}}\int_x^{\infty}\dd y\ \e^{-y^2}$ 
is the complementary error function, satisfying that 
${\rm erfc}(-\infty)=2$ and ${\rm erfc}(\infty)=0$.
Hence, the total number of states is given by $W_\tau(\infty)=D$. 
Note also that
$\frac{1}{2}{\rm erfc}(-x\tau)$ is reduced to the step function $\theta(x)$, 
i.e., $\theta(x)=1$ for $x>0$ and $\theta(x)=0$ for $x<0$, 
in the limit of $\tau\to\infty$.
In other words,  for finite $\tau$, the cumulative number of states $W_\tau(E)$ 
is a continuous function of $E$ and in general not integer because
$W_\tau(E)$ is not a sum of step functions. 
The number of states within the range of
$[E-\delta E/2, E+\delta E/2)$ for small $\delta E$ is given by
\begin{alignat}{1}
  W_\tau\left(E+\frac{\delta E}{2}\right)-
  W_\tau\left(E-\frac{\delta E}{2}\right)
  &= \int_{E-\frac{\delta E}{2}}^{E+\frac{\delta E}{2}} \dd E^\prime g_\tau (E^\prime)
  \notag \\
  &\sim g_\tau (E)\delta E,
  \label{eq:nos}
\end{alignat}
where the integrand $g_\tau(E^\prime)$ is
approximated by its representative value at the mid point,  
$g_\tau(E)$.
The argument in the logarithm of Eq.~(\ref{eq:S}) 
is exactly the quantity in
Eq.~(\ref{eq:nos}) with $\delta E$ given in Eq.~(\ref{eq:width}). 

\section{Equivalence between Eq.~(\ref{eq:Nr}) and Eq.~(\ref{eq:norm})}\label{app:equiv}
In this Appendix, we prove 
the equivalence between Eq.~(\ref{eq:Nr}) and Eq.~(\ref{eq:norm}).
From Eq.~(\ref{eq:filterTPQ}), the squared norm of
the energy-filtered random-phase state $|\psi_{\tau,r}(E)\rangle$ 
can be expressed as
\begin{alignat}{1}
  &\langle \psi_{\tau,r}(E)|\psi_{\tau,r}(E)\rangle \notag \\
 =&\frac{1}{2\pi\tau^2}
  \int_{-\infty}^{\infty} \dd s \int_{-\infty}^{\infty} \dd t\ 
  \e^{-\frac{s^2+t^2}{2\tau^2}} \e^{\imag E(t-s)} \langle \phi_r|\hat{U}(t-s)|\phi_r\rangle \notag \\
  =&
  \frac{1}{4\pi\tau^2}
  \int_{-\infty}^{\infty} \dd t_+ \int_{-\infty}^{\infty} \dd t_-\ 
  \e^{-\frac{t_+^2+t_-^2}{4\tau^2}} \e^{\imag Et_-} \langle \phi_r|\hat{U}(t_-)|\phi_r\rangle \notag \\
  =&
  \frac{1}{2\sqrt{\pi}\tau}
  \int_{-\infty}^{\infty} \dd t_-\ 
  \e^{-\frac{t_-^2}{4\tau^2}} \e^{\imag Et_-} \langle \phi_r|\hat{U}(t_-)|\phi_r\rangle.
  \label{eq:equiv}
\end{alignat}
where the variables are changed from
$s$ and $t$ to
$t_+ \equiv t+s$ and 
$t_- \equiv t-s$
in the second equality,
and the integral over $t_+$ is performed in the third equality.
Note that by the change of variables in the second equality,
the volume element is transformed as
$\dd s \dd t=
\left|\frac{\partial(s,t)}{\partial(t_+,t_-)}\right|\dd t_+ \dd t_-$
with the  Jacobian determinant 
$\frac{\partial(s,t)}{\partial(t_+,t_-)}
=\frac{1}{2}$.
By substituting 
$K_r(t_-)=\langle \phi_r|\hat{U}(t_-)|\phi_r\rangle$
into Eq.~(\ref{eq:equiv}), we obtain Eq.~(\ref{eq:Nr}), and hence 
the equivalence between Eq.~(\ref{eq:Nr}) and Eq.~(\ref{eq:norm}) is proved. 
The equivalence between Eq.~(\ref{eq:Hr}) and Eq.~(\ref{eq:H}) can also be proved similarly.

\section{Translation to canonical ensemble}\label{app:canonical}
In this Appendix, we discuss how the present formalism
of the microcanonical ensemble can be translated into the canonical ensemble. 
Although thermodynamic quantities calculated with
the microcanonical ensemble should be identical to those with 
the canonical ensemble in the thermodynamic limit,
this does not hold generally for finite systems. 
On one hand, there may be a case where 
one wishes to specify the inverse temperature $\beta$
rather than the target energy $E$ as a parameter, and hence wishes to
use the canonical ensemble rather than the microcanonical ensemble.
On the other hand, it is well known that the partition function $Z_{\rm can}(\beta)$ of the canonical ensemble
can be formally written as
\begin{alignat}{1}
  Z_{\rm can}(\beta) \equiv \Tr\left[\e^{-\beta \hat{\cal H}}\right]
  = \sum_{n=0}^{D-1} \int_{-\infty}^{\infty} \dd E \e^{-\beta E} \delta(E_n-E). 
\end{alignat}
From the partition function $Z_{\rm can}(\beta)$, the canonical energy and the canonical entropy can be calculated as
$E_{\rm can}(\beta)=-\partial_\beta \ln Z_{\rm can}(\beta)$ and
$S_{\rm can}(\beta)=\beta E_{\rm can}(\beta)+\ln Z_{\rm can}(\beta)$, respectively.

\begin{figure}
  \includegraphics[width=0.95\columnwidth]{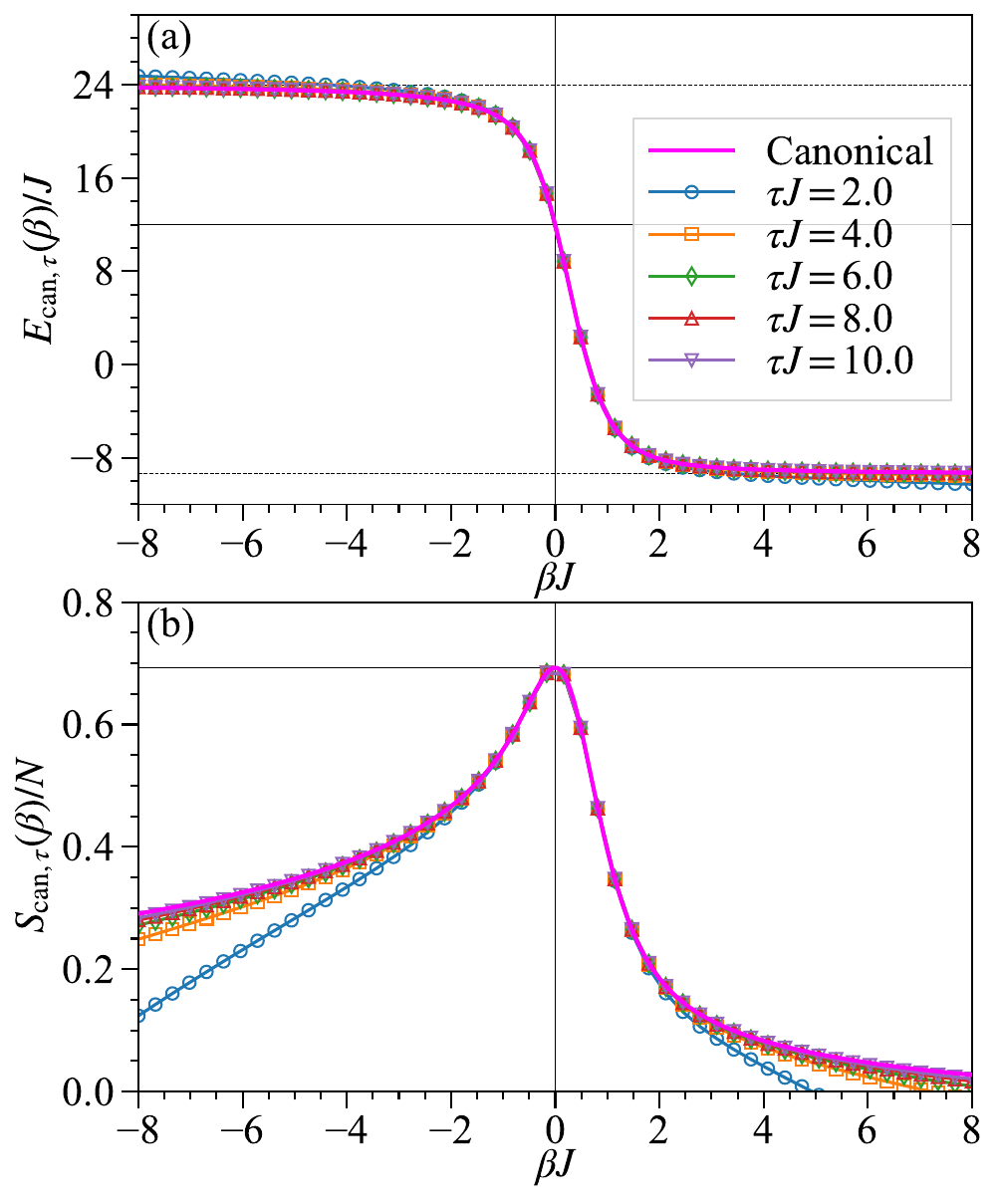}
  \caption{
    \label{fig:can}
    (a) The $\tau$-dependent canonical energy $E_{{\rm can},\tau}(\beta)$ and
    (b) the $\tau$-dependent canonical entropy $S_{{\rm can},\tau}(\beta)$
    as a function of the inverse temperature $\beta$ for
    several $\tau$ values for $N=24$.
    The solid magenta lines in (a) and (b) indicate
    the canonical energy $E_{{\rm can}}(\beta)$ and
    the canonical entropy $S_{{\rm can}}(\beta)$, respectively. 
    In (a), the dashed horizontal lines indicate $E_0$ and $E_{D-1}$,
    and the solid horizontal line indicates $J N_{\rm bond}/2=12J$,
    corresponding to the energy at $\beta=0$.
    In (b), the horizontal line indicates $\ln 2$, corresponding to
    the entropy per qubit at $\beta=0$. 
  }
\end{figure}

Now, in order to be compatible with the formalism of the microcanonical ensemble considered here,
let us define a $\tau$-dependent partition function 
\begin{alignat}{1}
  Z_{{\rm can},\tau}(\beta) \equiv \int_{-\infty}^{\infty} \dd E  \e^{-\beta E} g_\tau(E).
  \label{eq:Ztau}
\end{alignat}
Because $g_\tau(E)$ is a sum of Gaussians, the integral can be performed analytically
and it results in 
\begin{alignat}{1}
  Z_{{\rm can},\tau}(\beta) = \sum_{n=0}^{D-1} \e^{-\beta E_n
    + \frac{\beta^2}{4\tau^2}} =
  \e^{
    \frac{\beta^2}{4\tau^2}}
    Z_{\rm can}(\beta). 
\end{alignat}
Accordingly, we define 
a $\tau$-dependent canonical energy and
a $\tau$-dependent canonical entropy as 
\begin{alignat}{1}
  E_{{\rm can},\tau}(\beta)&\equiv-\frac{\partial}{\partial \beta} \ln Z_{{\rm can},\tau}(\beta)
  = E_{\rm can}(\beta)
  -\frac{\beta}{2\tau^2}
  \label{eq:Ecantau}
\end{alignat}
and
\begin{alignat}{1}
  S_{{\rm can},\tau}(\beta)&\equiv\beta E_{{\rm can},\tau}(\beta)+\ln Z_{{\rm can},\tau}(\beta)
  = S_{\rm can}(\beta)
  - \frac{\beta^2}{4\tau^2},
  \label{eq:Scantau}
\end{alignat}
respectively. 
Note that $\beta$ here is not calculated as in the microcanonical ensemble but it is simply a parameter.  
Figure~\ref{fig:can} shows
$E_{{\rm can},\tau}(\beta)$ and
$S_{{\rm can},\tau}(\beta)$ for the one-dimensional spin-1/2 Heisenberg model 
$\hat{\mathcal{H}}$ in Eq.~(\ref{eq:Ham_SWAP}) 
with $N=24$ qubits calculated by the full diagonalization method.
As expected from Eqs.~(\ref{eq:Ecantau}) and (\ref{eq:Scantau}),
the results converge to those of the canonical ensemble,
$E_{{\rm can}}(\beta)$ and
$S_{{\rm can}}(\beta)$, with increasing $\tau$. 

Finally, when the random sampling method for the trace evaluation is employed,
the $\tau$-dependent partition function $Z_{{\rm can},\tau}(\beta)$ should be
calculated by replacing $g_\tau(E)$ with $g_{\tau,R}(E)$ in Eq.~(\ref{eq:Ztau}) 
[see Eqs.~(\ref{eq:Nr}) and (\ref{eq:g_filterTPQ})]. 
Accordingly, 
the $\tau$-dependent partition function $Z_{{\rm can},\tau}(\beta)$ as well as  
the $\tau$-dependent canonical energy $E_{{\rm can},\tau}(\beta)$ and
the $\tau$-dependent canonical entropy $S_{{\rm can},\tau}(\beta)$
acquires statistical errors.

\section{Average of the random-phase-product states}\label{app:rpps}
In this Appendix, we show that
the random-phase-product states of the form
$v_r^{(1)}|+\rangle$ satisfy Eq.~(\ref{eq:mixed2}). 
The ideal random average for
the states $v_r^{(1)}|+\rangle$ parametrized as in Eq.~(\ref{eq:v1}), where $N_v=N$, is given by
$
\mathbb{E}_N[A_r]=
\frac{1}{(2\pi)^{N}}
\int_{0}^{2\pi}\dd \theta_{0,r}
\int_{0}^{2\pi}\dd \theta_{1,r} \cdots
\int_{0}^{2\pi}\dd \theta_{N-1,r} A_r.
$
We can then easily show that  
\begin{equation}
  \mathbb{E}_N
  \left[\prod_{k=0}^{N-1} \e^{\imag \theta_{k,r}
      \left\{(-1)^{b_k}-(-1)^{b_k^\prime}\right\}/2}
    \right]
  =\prod_{k=0}^{N-1} \delta_{b_k b_k^\prime}=\delta_{bb^\prime}
  \label{eq:Kron2}
\end{equation}
and thus $\hat{v}_r^{(1)}|+\rangle$ satisfy Eq.~(\ref{eq:mixed2}). 
Here, $b_k$ and $b_k^\prime$ are binary ($0$ or $1$) variables at the $k$th qubit, 
and $b$ and $b^\prime$ are the corresponding bit strings of length $N$, $\{b_k \}_{k=0}^{N-1}$ 
and $\{b'_k \}_{k=0}^{N-1}$, respectively. 

Instead of the continuous random variables,
we can consider $\{\theta_{k,r}\}_{k=0}^{N-1}$ 
as a set of discrete random variables drawn
uniformly from $\{0,\pi\}^N$, $\{0,\frac{2\pi}{3},\frac{4\pi}{3}\}^N$, or in general
$\{0, \frac{2\pi}{l}, \frac{2\pi\times 2}{l}, \cdots \frac{2\pi(l-1)}{l}\}^N$ with  $l>0$ integer.   
In this case,
the ideal random average is given by 
$\mathbb{E}_N[A_r]=
\frac{1}{l^N}
\sum_{\theta_{0,r}=0}^{2\pi(l-1)/l}
\sum_{\theta_{1,r}=0}^{2\pi(l-1)/l}
\cdots 
\sum_{\theta_{N-1,r}=0}^{2\pi(l-1)/l} A_r$,
and one can also readily confirm that 
Eq.~(\ref{eq:Kron2}) and hence Eq.~(\ref{eq:mixed2})
are still satisfied. 
A finite set of random diagonal unitaries $\{\hat{v}_r^{(1)}\}$ 
with discrete random variables corresponds to the diagonal-unitary 1-design. 

\section{Additional numerical results}\label{app:smallN}

\subsection{Numerical results for smaller systems}
Figure~\ref{fig:nos_small} shows exactly the same results as in Fig.~\ref{fig:nos} 
but for $N=14$, 16, and 18. 
The values of $\tau$ used for evaluating $\Tr\left[\hat{G}_\tau(E)\right] = g_\tau(E) \delta E$ are
$\tau J=$2.81, 2.46, 2.19 for $N_{\rm bin}=32$ and $N=14,16,18$,
$\tau J=$5.72, 5.01, 4.46 for $N_{\rm bin}=64$ and $N=14,16,18$, and
$\tau J=$11.5, 10.1, 8.99 for $N_{\rm bin}=128$ and $N=14,16,18$, respectively.
As discussed in Sec.~\ref{sec:nos}, 
$g_\tau(E)\delta E$ becomes more snaky for the smaller $N$ and the larger $N_{\rm bin}$.

\begin{figure*}
  \includegraphics[width=1.95\columnwidth]{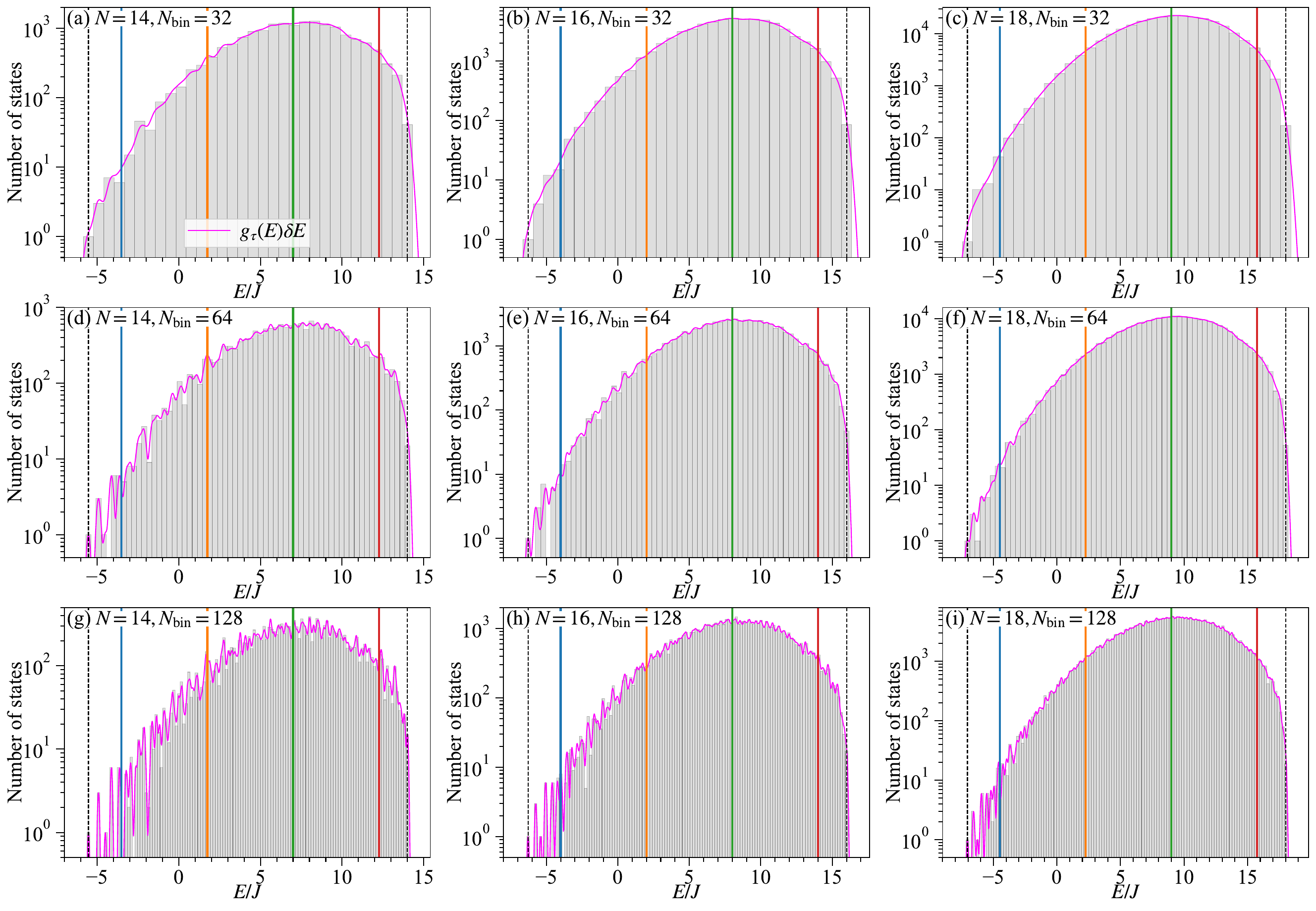}
  \caption{
    \label{fig:nos_small}
    The same as Fig.~\ref{fig:nos} but for 
    $N=14$, 16, and 18. 
        The thick vertical lines at $E/NJ=-0.25,0.125,0.5,$ and $0.875$
    (indicated by blue, orange, green, and red, respectively) denote the target energies used in the results shown in
    Figs.~\ref{fig:ESbeta_small} and \ref{fig:Efluct_small}. 
  }\label{fig:nos_small}
\end{figure*}

\begin{figure*}
  \includegraphics[width=1.95\columnwidth]{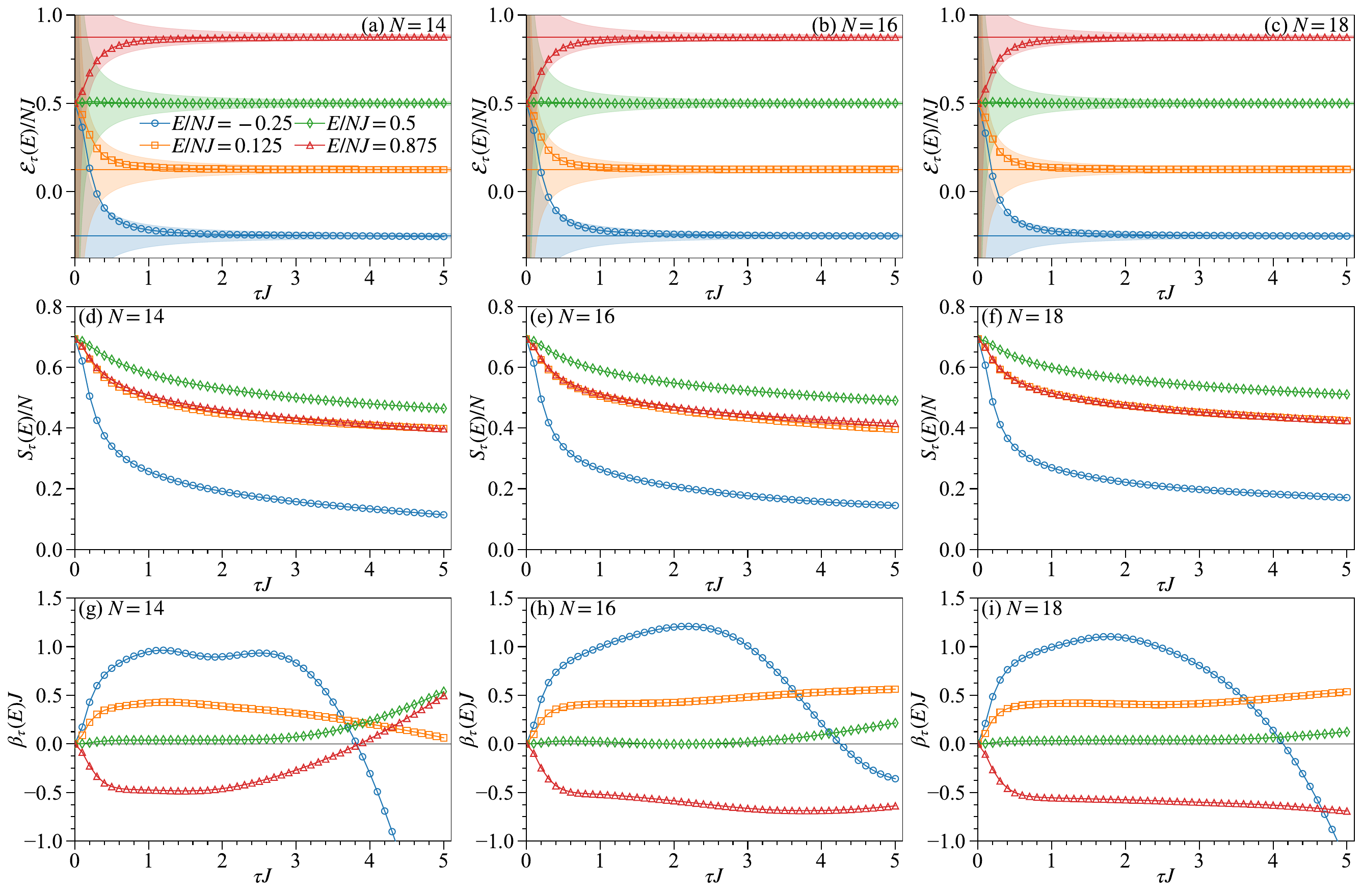}
  \caption{
    \label{fig:ESbeta_small}
    The same as Fig.~\ref{fig:ESbeta} but for 
    $N=14$, 16, and 18.
  }
\end{figure*}

Figure~\ref{fig:ESbeta_small} shows exactly the same results as in Fig.~\ref{fig:ESbeta} 
but for $N=14$, 16, and 18. 
As compared to ${\cal E}_{\tau}(E)$ and $S_\tau(E)$, the inverse temperature 
$\beta_\tau(E)$ depends rather wildly on the filtering time $\tau$.
In particular, 
the inverse temperature for $E/NJ=-0.25$
becomes negative for $\tau J \gtrsim 4$.  
However, as shown in Figs.~\ref{fig:ESbeta} and~\ref{fig:rand2} (also see Fig.~\ref{fig:rand4}), 
such a wild dependence on $\tau$ is moderated for the larger systems.

\begin{figure*}
  \includegraphics[width=1.9\columnwidth]{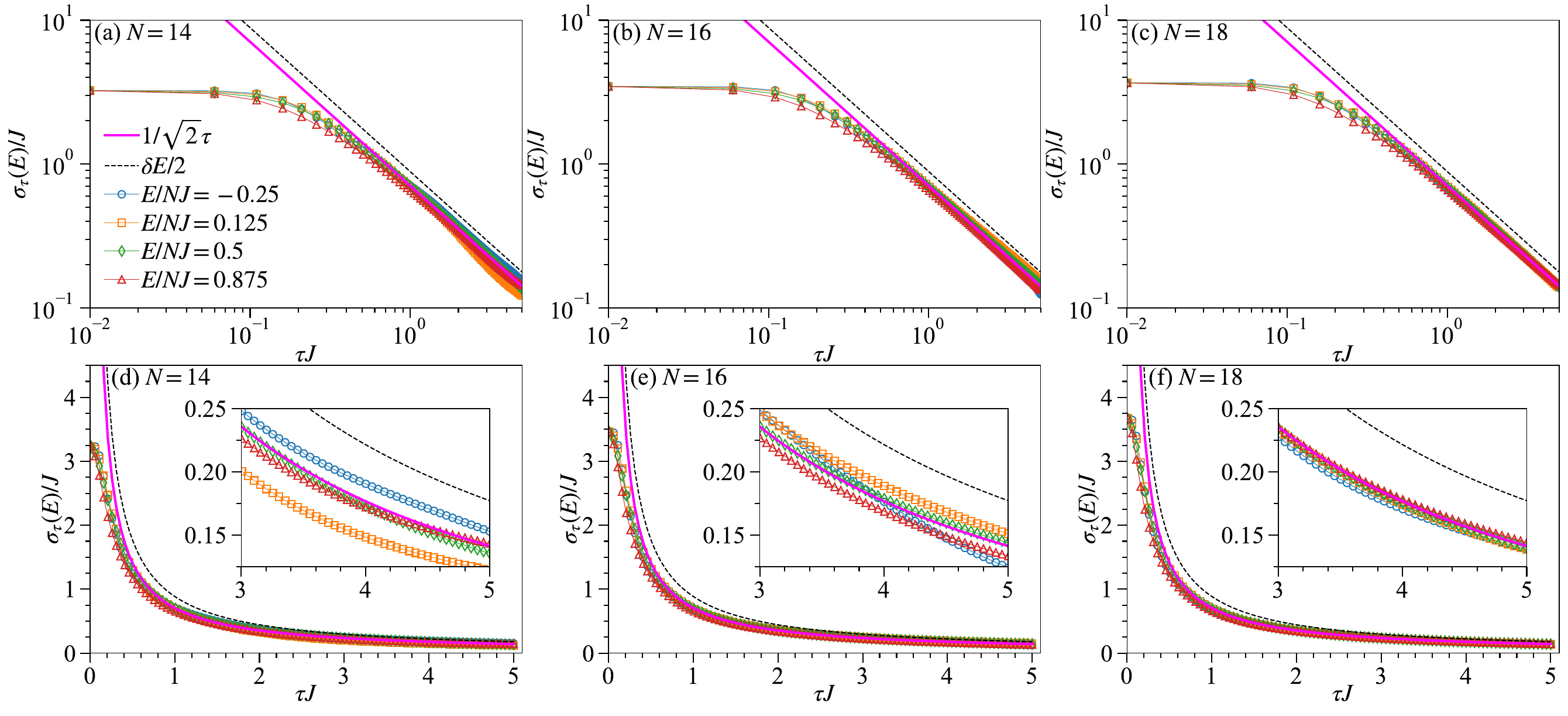}
  \caption{
    \label{fig:Efluct_small}
    The same as Fig.~\ref{fig:Efluct} but for
    $N=14$, 16, and 18.
  }
\end{figure*}

Figure~\ref{fig:Efluct_small} shows exactly the same results as in Fig.~\ref{fig:Efluct} 
but for $N=14$, 16, and 18.
As expected, 
the energy fluctuation $\sigma_\tau(E)$ for large $\tau$ behaves as
$\sigma_\tau(E) \sim 1/\sqrt{2}\tau$, independently of the system size $N$. 
It is also found that 
the dependence of $\sigma_\tau(E)$ on the target energy $E$ becomes less significant
and matches better with $1/\sqrt{2}\tau$ 
for the larger systems [see insets in Figs.~\ref{fig:Efluct_small}(d)--\ref{fig:Efluct_small}(f)].

\subsection{Detailed results on random sampling for trace evaluation }
\begin{figure*}
  \includegraphics[width=1.95\columnwidth]{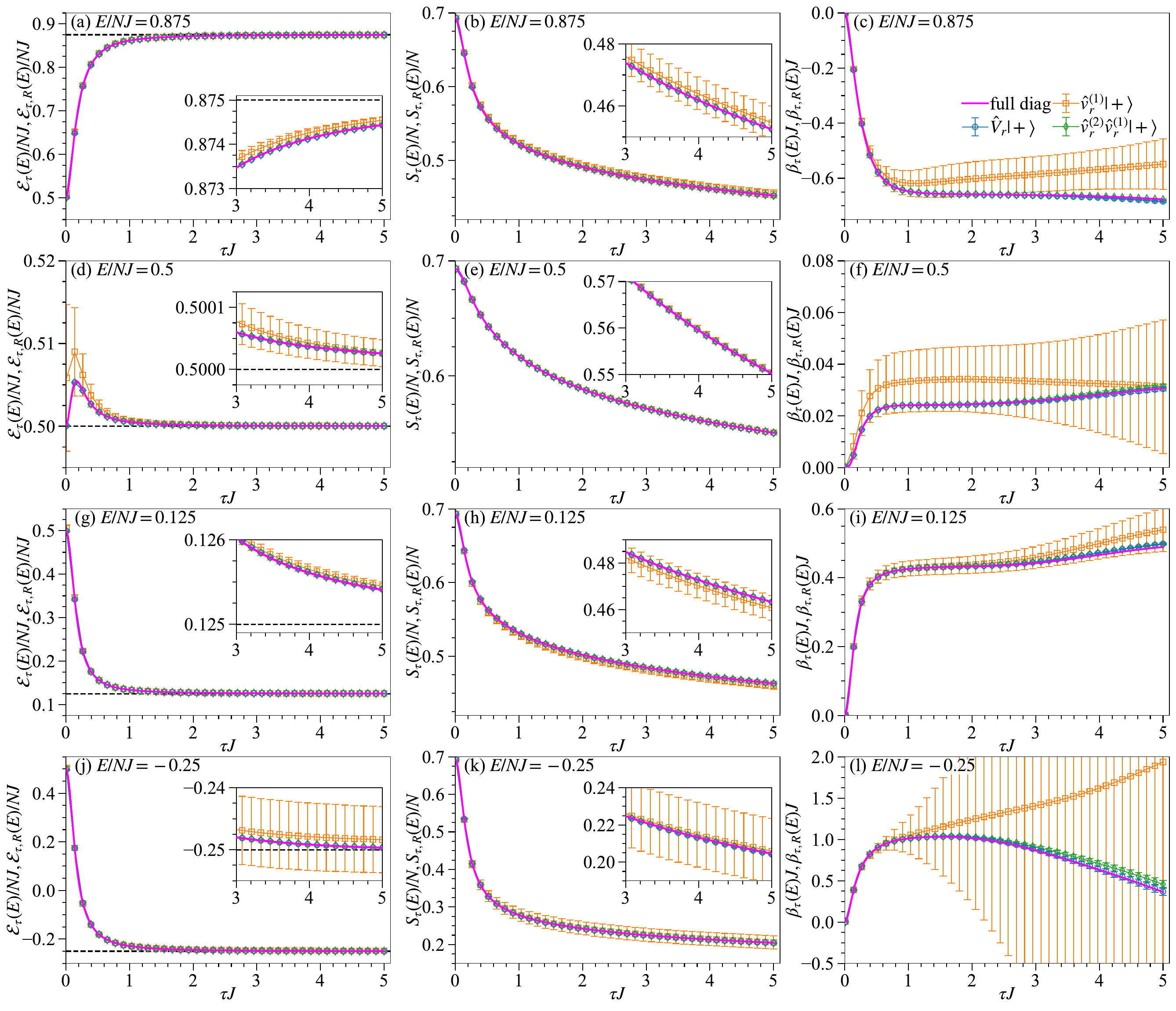}
  \caption{
    \label{fig:rand3}
    The same results as in Fig.~\ref{fig:rand}
    but plotted separately for
    the different target energies $E/NJ=0.875$, $0.5$, $0.125$, and $-0.25$.
    (a,d,g,j) The energy expectation value ${\cal E}_{\tau,R}(E)$,
    (b,e,h,k) the entropy $S_{\tau,R}(E)$, and 
    (c,f,i,l) the inverse temperature $\beta_{\tau,R}(E)$ 
    as a function of the filtering time $\tau$ for  $N=24$.  
    The black dashed horizontal lines in (a), (d), (g), and (j) indicate the target energies.
    The insets show the magnifications for $3 \leqslant \tau J \leqslant 5$.
    Notice that the different panels employ the different scales in the $y$ axis. 
  }
\end{figure*}

\begin{figure*}
  \includegraphics[width=1.95\columnwidth]{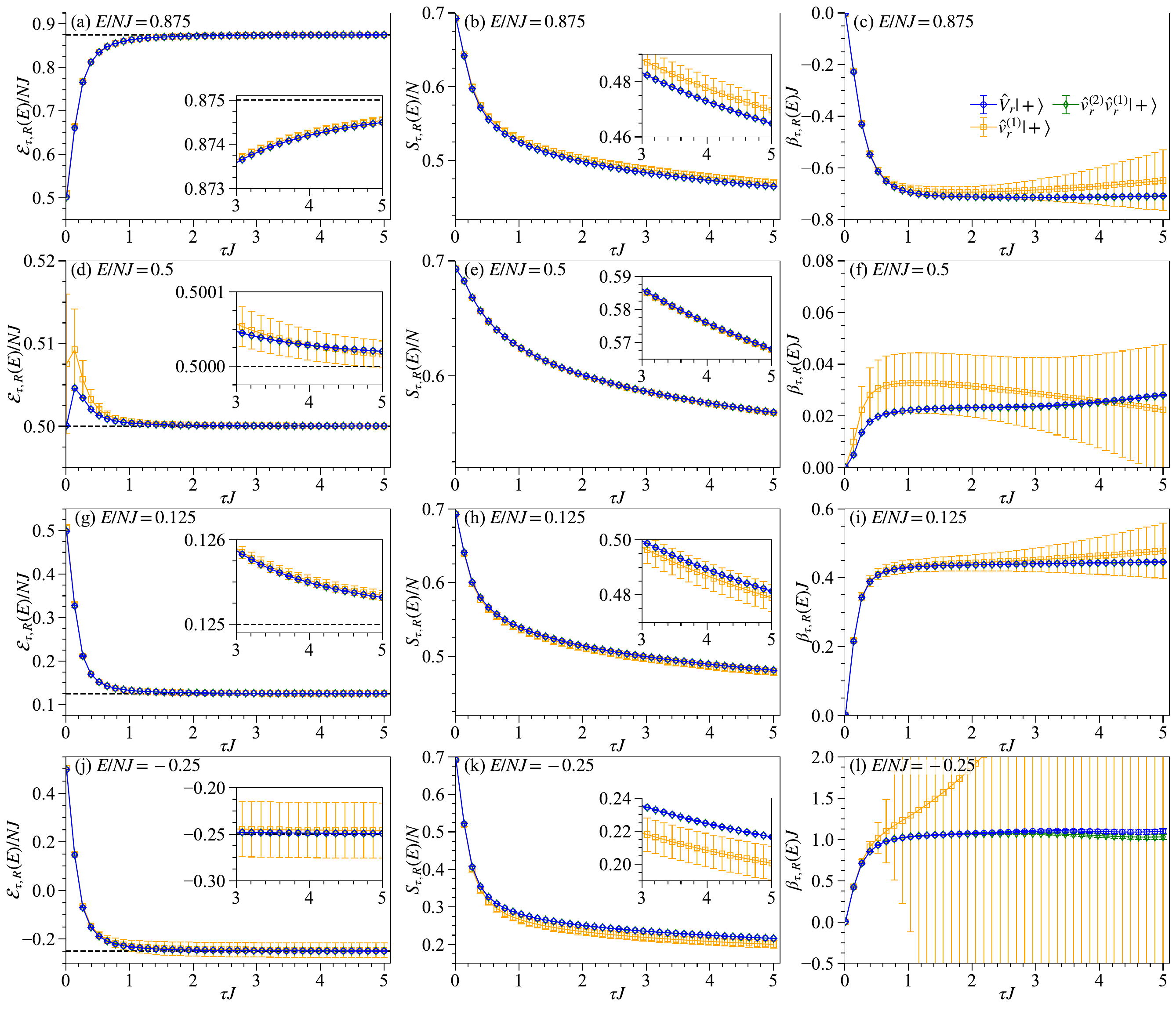}
  \caption{
    \label{fig:rand4}
    The same as in Fig.~\ref{fig:rand3} but for $N=28$.
    The full-diagonalization results are not shown because the huge computational resource required is not 
    currently available.     
  }
\end{figure*}

Here we provide detailed results on the random sampling method for the trace evaluations.
Figure~\ref{fig:rand3} shows the same results as in Fig.~\ref{fig:rand} for $N=24$ 
but plotted separately for the different target energies $E/NJ=0.875$, $0.5$, $0.125$, and $-0.25$.
It is clearly observed that the statistical errors with the 
the random-phase states $\hat{v}_r^{(1)}|+\rangle$ behave qualitatively differently
from those with the other random-phase states
$\hat{v}_r^{(2)}\hat{v}_r^{(1)}|+\rangle$ and $\hat{V}_r|+\rangle$.
The same trend is found for the larger system of $N=28$, as shown in Fig.~\ref{fig:rand4}.

\bibliography{biball}

\begin{thebibliography}{56}%
\makeatletter
\providecommand \@ifxundefined [1]{%
 \@ifx{#1\undefined}
}%
\providecommand \@ifnum [1]{%
 \ifnum #1\expandafter \@firstoftwo
 \else \expandafter \@secondoftwo
 \fi
}%
\providecommand \@ifx [1]{%
 \ifx #1\expandafter \@firstoftwo
 \else \expandafter \@secondoftwo
 \fi
}%
\providecommand \natexlab [1]{#1}%
\providecommand \enquote  [1]{``#1''}%
\providecommand \bibnamefont  [1]{#1}%
\providecommand \bibfnamefont [1]{#1}%
\providecommand \citenamefont [1]{#1}%
\providecommand \href@noop [0]{\@secondoftwo}%
\providecommand \href [0]{\begingroup \@sanitize@url \@href}%
\providecommand \@href[1]{\@@startlink{#1}\@@href}%
\providecommand \@@href[1]{\endgroup#1\@@endlink}%
\providecommand \@sanitize@url [0]{\catcode `\\12\catcode `\$12\catcode
  `\&12\catcode `\#12\catcode `\^12\catcode `\_12\catcode `\%12\relax}%
\providecommand \@@startlink[1]{}%
\providecommand \@@endlink[0]{}%
\providecommand \url  [0]{\begingroup\@sanitize@url \@url }%
\providecommand \@url [1]{\endgroup\@href {#1}{\urlprefix }}%
\providecommand \urlprefix  [0]{URL }%
\providecommand \Eprint [0]{\href }%
\providecommand \doibase [0]{http://dx.doi.org/}%
\providecommand \selectlanguage [0]{\@gobble}%
\providecommand \bibinfo  [0]{\@secondoftwo}%
\providecommand \bibfield  [0]{\@secondoftwo}%
\providecommand \translation [1]{[#1]}%
\providecommand \BibitemOpen [0]{}%
\providecommand \bibitemStop [0]{}%
\providecommand \bibitemNoStop [0]{.\EOS\space}%
\providecommand \EOS [0]{\spacefactor3000\relax}%
\providecommand \BibitemShut  [1]{\csname bibitem#1\endcsname}%
\let\auto@bib@innerbib\@empty
\bibitem [{\citenamefont {Fano}(1957)}]{Fano1957}%
  \BibitemOpen
  \bibfield  {author} {\bibinfo {author} {\bibfnamefont {U.}~\bibnamefont
  {Fano}},\ }\bibfield  {title} {\enquote {\bibinfo {title} {{Description of
  States in Quantum Mechanics by Density Matrix and Operator Techniques}},}\
  }\href {\doibase 10.1103/RevModPhys.29.74} {\bibfield  {journal} {\bibinfo
  {journal} {Rev. Mod. Phys.}\ }\textbf {\bibinfo {volume} {29}},\ \bibinfo
  {pages} {74--93} (\bibinfo {year} {1957})}\BibitemShut {NoStop}%
\bibitem [{\citenamefont {Suzuki}(1985)}]{Suzuki1985}%
  \BibitemOpen
  \bibfield  {author} {\bibinfo {author} {\bibfnamefont {Masuo}\ \bibnamefont
  {Suzuki}},\ }\bibfield  {title} {\enquote {\bibinfo {title} {Thermo field
  dynamics in equilibrium and non-equilibrium interacting quantum systems},}\
  }\href {\doibase 10.1143/JPSJ.54.4483} {\bibfield  {journal} {\bibinfo
  {journal} {Journal of the Physical Society of Japan}\ }\textbf {\bibinfo
  {volume} {54}},\ \bibinfo {pages} {4483--4485} (\bibinfo {year}
  {1985})}\BibitemShut {NoStop}%
\bibitem [{\citenamefont {Feiguin}\ and\ \citenamefont
  {White}(2005)}]{Feiguin2005}%
  \BibitemOpen
  \bibfield  {author} {\bibinfo {author} {\bibfnamefont {Adrian~E.}\
  \bibnamefont {Feiguin}}\ and\ \bibinfo {author} {\bibfnamefont {Steven~R.}\
  \bibnamefont {White}},\ }\bibfield  {title} {\enquote {\bibinfo {title}
  {{Finite-temperature density matrix renormalization using an enlarged Hilbert
  space}},}\ }\href {\doibase 10.1103/PhysRevB.72.220401} {\bibfield  {journal}
  {\bibinfo  {journal} {Phys. Rev. B}\ }\textbf {\bibinfo {volume} {72}},\
  \bibinfo {pages} {220401} (\bibinfo {year} {2005})}\BibitemShut {NoStop}%
\bibitem [{\citenamefont {Nomura}\ \emph {et~al.}(2021)\citenamefont {Nomura},
  \citenamefont {Yoshioka},\ and\ \citenamefont {Nori}}]{Nomura2021}%
  \BibitemOpen
  \bibfield  {author} {\bibinfo {author} {\bibfnamefont {Yusuke}\ \bibnamefont
  {Nomura}}, \bibinfo {author} {\bibfnamefont {Nobuyuki}\ \bibnamefont
  {Yoshioka}}, \ and\ \bibinfo {author} {\bibfnamefont {Franco}\ \bibnamefont
  {Nori}},\ }\bibfield  {title} {\enquote {\bibinfo {title} {{Purifying Deep
  Boltzmann Machines for Thermal Quantum States}},}\ }\href {\doibase
  10.1103/PhysRevLett.127.060601} {\bibfield  {journal} {\bibinfo  {journal}
  {Phys. Rev. Lett.}\ }\textbf {\bibinfo {volume} {127}},\ \bibinfo {pages}
  {060601} (\bibinfo {year} {2021})}\BibitemShut {NoStop}%
\bibitem [{\citenamefont {Li}\ and\ \citenamefont {Haldane}(2008)}]{Li2008}%
  \BibitemOpen
  \bibfield  {author} {\bibinfo {author} {\bibfnamefont {Hui}\ \bibnamefont
  {Li}}\ and\ \bibinfo {author} {\bibfnamefont {F.~D.~M.}\ \bibnamefont
  {Haldane}},\ }\bibfield  {title} {\enquote {\bibinfo {title} {{Entanglement
  Spectrum as a Generalization of Entanglement Entropy: Identification of
  Topological Order in Non-Abelian Fractional Quantum Hall Effect States}},}\
  }\href {\doibase 10.1103/PhysRevLett.101.010504} {\bibfield  {journal}
  {\bibinfo  {journal} {Phys. Rev. Lett.}\ }\textbf {\bibinfo {volume} {101}},\
  \bibinfo {pages} {010504} (\bibinfo {year} {2008})}\BibitemShut {NoStop}%
\bibitem [{\citenamefont {Poilblanc}(2010)}]{Poilblanc2010}%
  \BibitemOpen
  \bibfield  {author} {\bibinfo {author} {\bibfnamefont {Didier}\ \bibnamefont
  {Poilblanc}},\ }\bibfield  {title} {\enquote {\bibinfo {title} {{Entanglement
  Spectra of Quantum Heisenberg Ladders}},}\ }\href {\doibase
  10.1103/PhysRevLett.105.077202} {\bibfield  {journal} {\bibinfo  {journal}
  {Phys. Rev. Lett.}\ }\textbf {\bibinfo {volume} {105}},\ \bibinfo {pages}
  {077202} (\bibinfo {year} {2010})}\BibitemShut {NoStop}%
\bibitem [{\citenamefont {Qi}\ \emph {et~al.}(2012)\citenamefont {Qi},
  \citenamefont {Katsura},\ and\ \citenamefont {Ludwig}}]{Qi2012}%
  \BibitemOpen
  \bibfield  {author} {\bibinfo {author} {\bibfnamefont {Xiao-Liang}\
  \bibnamefont {Qi}}, \bibinfo {author} {\bibfnamefont {Hosho}\ \bibnamefont
  {Katsura}}, \ and\ \bibinfo {author} {\bibfnamefont {Andreas W.~W.}\
  \bibnamefont {Ludwig}},\ }\bibfield  {title} {\enquote {\bibinfo {title}
  {{General Relationship between the Entanglement Spectrum and the Edge State
  Spectrum of Topological Quantum States}},}\ }\href {\doibase
  10.1103/PhysRevLett.108.196402} {\bibfield  {journal} {\bibinfo  {journal}
  {Phys. Rev. Lett.}\ }\textbf {\bibinfo {volume} {108}},\ \bibinfo {pages}
  {196402} (\bibinfo {year} {2012})}\BibitemShut {NoStop}%
\bibitem [{\citenamefont {Dalmonte}\ \emph {et~al.}(2018)\citenamefont
  {Dalmonte}, \citenamefont {Vermersch},\ and\ \citenamefont
  {Zoller}}]{Dalmonte2018}%
  \BibitemOpen
  \bibfield  {author} {\bibinfo {author} {\bibfnamefont {M.}~\bibnamefont
  {Dalmonte}}, \bibinfo {author} {\bibfnamefont {B.}~\bibnamefont {Vermersch}},
  \ and\ \bibinfo {author} {\bibfnamefont {P.}~\bibnamefont {Zoller}},\
  }\bibfield  {title} {\enquote {\bibinfo {title} {{Quantum simulation and
  spectroscopy of entanglement Hamiltonians}},}\ }\href {\doibase
  10.1038/s41567-018-0151-7} {\bibfield  {journal} {\bibinfo  {journal} {Nature
  Physics}\ }\textbf {\bibinfo {volume} {14}},\ \bibinfo {pages} {827--831}
  (\bibinfo {year} {2018})}\BibitemShut {NoStop}%
\bibitem [{\citenamefont {Giudici}\ \emph {et~al.}(2018)\citenamefont
  {Giudici}, \citenamefont {Mendes-Santos}, \citenamefont {Calabrese},\ and\
  \citenamefont {Dalmonte}}]{Giudici2018}%
  \BibitemOpen
  \bibfield  {author} {\bibinfo {author} {\bibfnamefont {G.}~\bibnamefont
  {Giudici}}, \bibinfo {author} {\bibfnamefont {T.}~\bibnamefont
  {Mendes-Santos}}, \bibinfo {author} {\bibfnamefont {P.}~\bibnamefont
  {Calabrese}}, \ and\ \bibinfo {author} {\bibfnamefont {M.}~\bibnamefont
  {Dalmonte}},\ }\bibfield  {title} {\enquote {\bibinfo {title} {{Entanglement
  Hamiltonians of lattice models via the Bisognano-Wichmann theorem}},}\ }\href
  {\doibase 10.1103/PhysRevB.98.134403} {\bibfield  {journal} {\bibinfo
  {journal} {Phys. Rev. B}\ }\textbf {\bibinfo {volume} {98}},\ \bibinfo
  {pages} {134403} (\bibinfo {year} {2018})}\BibitemShut {NoStop}%
\bibitem [{\citenamefont {Mendes-Santos}\ \emph {et~al.}(2020)\citenamefont
  {Mendes-Santos}, \citenamefont {Giudici}, \citenamefont {Fazio},\ and\
  \citenamefont {Dalmonte}}]{MendesSantos2020}%
  \BibitemOpen
  \bibfield  {author} {\bibinfo {author} {\bibfnamefont {T.}~\bibnamefont
  {Mendes-Santos}}, \bibinfo {author} {\bibfnamefont {G.}~\bibnamefont
  {Giudici}}, \bibinfo {author} {\bibfnamefont {R.}~\bibnamefont {Fazio}}, \
  and\ \bibinfo {author} {\bibfnamefont {M.}~\bibnamefont {Dalmonte}},\
  }\bibfield  {title} {\enquote {\bibinfo {title} {{Measuring von Neumann
  entanglement entropies without wave functions}},}\ }\href {\doibase
  10.1088/1367-2630/ab6875} {\bibfield  {journal} {\bibinfo  {journal} {New
  Journal of Physics}\ }\textbf {\bibinfo {volume} {22}},\ \bibinfo {pages}
  {013044} (\bibinfo {year} {2020})}\BibitemShut {NoStop}%
\bibitem [{\citenamefont {Seki}\ and\ \citenamefont
  {Yunoki}(2020{\natexlab{a}})}]{Seki2020thermal}%
  \BibitemOpen
  \bibfield  {author} {\bibinfo {author} {\bibfnamefont {Kazuhiro}\
  \bibnamefont {Seki}}\ and\ \bibinfo {author} {\bibfnamefont {Seiji}\
  \bibnamefont {Yunoki}},\ }\bibfield  {title} {\enquote {\bibinfo {title}
  {Emergence of a thermal equilibrium in a subsystem of a pure ground state by
  quantum entanglement},}\ }\href {\doibase 10.1103/PhysRevResearch.2.043087}
  {\bibfield  {journal} {\bibinfo  {journal} {Phys. Rev. Research}\ }\textbf
  {\bibinfo {volume} {2}},\ \bibinfo {pages} {043087} (\bibinfo {year}
  {2020}{\natexlab{a}})}\BibitemShut {NoStop}%
\bibitem [{\citenamefont {Tolman}(1980)}]{Tolman}%
  \BibitemOpen
  \bibfield  {author} {\bibinfo {author} {\bibfnamefont {Richard~C.}\
  \bibnamefont {Tolman}},\ }\href@noop {} {\emph {\bibinfo {title} {{The
  Principles of Statistical Mechanics}}}}\ (\bibinfo  {publisher} {Dover},\
  \bibinfo {address} {New York},\ \bibinfo {year} {1980})\ Chap.~\bibinfo
  {chapter} {IX}\BibitemShut {NoStop}%
\bibitem [{\citenamefont {Drabold}\ and\ \citenamefont
  {Sankey}(1993)}]{Drabold1993}%
  \BibitemOpen
  \bibfield  {author} {\bibinfo {author} {\bibfnamefont {David~A.}\
  \bibnamefont {Drabold}}\ and\ \bibinfo {author} {\bibfnamefont {Otto~F.}\
  \bibnamefont {Sankey}},\ }\bibfield  {title} {\enquote {\bibinfo {title}
  {{Maximum entropy approach for linear scaling in the electronic structure
  problem}},}\ }\href {\doibase 10.1103/PhysRevLett.70.3631} {\bibfield
  {journal} {\bibinfo  {journal} {Phys. Rev. Lett.}\ }\textbf {\bibinfo
  {volume} {70}},\ \bibinfo {pages} {3631--3634} (\bibinfo {year}
  {1993})}\BibitemShut {NoStop}%
\bibitem [{\citenamefont {Iitaka}\ and\ \citenamefont
  {Ebisuzaki}(2004)}]{Iitaka2004}%
  \BibitemOpen
  \bibfield  {author} {\bibinfo {author} {\bibfnamefont {Toshiaki}\
  \bibnamefont {Iitaka}}\ and\ \bibinfo {author} {\bibfnamefont {Toshikazu}\
  \bibnamefont {Ebisuzaki}},\ }\bibfield  {title} {\enquote {\bibinfo {title}
  {Random phase vector for calculating the trace of a large matrix},}\ }\href
  {\doibase 10.1103/PhysRevE.69.057701} {\bibfield  {journal} {\bibinfo
  {journal} {Phys. Rev. E}\ }\textbf {\bibinfo {volume} {69}},\ \bibinfo
  {pages} {057701} (\bibinfo {year} {2004})}\BibitemShut {NoStop}%
\bibitem [{\citenamefont {Wei\ss{}e}\ \emph {et~al.}(2006)\citenamefont
  {Wei\ss{}e}, \citenamefont {Wellein}, \citenamefont {Alvermann},\ and\
  \citenamefont {Fehske}}]{Weisse2006}%
  \BibitemOpen
  \bibfield  {author} {\bibinfo {author} {\bibfnamefont {Alexander}\
  \bibnamefont {Wei\ss{}e}}, \bibinfo {author} {\bibfnamefont {Gerhard}\
  \bibnamefont {Wellein}}, \bibinfo {author} {\bibfnamefont {Andreas}\
  \bibnamefont {Alvermann}}, \ and\ \bibinfo {author} {\bibfnamefont {Holger}\
  \bibnamefont {Fehske}},\ }\bibfield  {title} {\enquote {\bibinfo {title} {The
  kernel polynomial method},}\ }\href {\doibase 10.1103/RevModPhys.78.275}
  {\bibfield  {journal} {\bibinfo  {journal} {Rev. Mod. Phys.}\ }\textbf
  {\bibinfo {volume} {78}},\ \bibinfo {pages} {275--306} (\bibinfo {year}
  {2006})}\BibitemShut {NoStop}%
\bibitem [{\citenamefont {Jin}\ \emph {et~al.}(2021)\citenamefont {Jin},
  \citenamefont {Willsch}, \citenamefont {Willsch}, \citenamefont {Lagemann},
  \citenamefont {Michielsen},\ and\ \citenamefont {De~Raedt}}]{Jin2021}%
  \BibitemOpen
  \bibfield  {author} {\bibinfo {author} {\bibfnamefont {Fengpin}\ \bibnamefont
  {Jin}}, \bibinfo {author} {\bibfnamefont {Dennis}\ \bibnamefont {Willsch}},
  \bibinfo {author} {\bibfnamefont {Madita}\ \bibnamefont {Willsch}}, \bibinfo
  {author} {\bibfnamefont {Hannes}\ \bibnamefont {Lagemann}}, \bibinfo {author}
  {\bibfnamefont {Kristel}\ \bibnamefont {Michielsen}}, \ and\ \bibinfo
  {author} {\bibfnamefont {Hans}\ \bibnamefont {De~Raedt}},\ }\bibfield
  {title} {\enquote {\bibinfo {title} {{Random State Technology}},}\ }\href
  {\doibase 10.7566/JPSJ.90.012001} {\bibfield  {journal} {\bibinfo  {journal}
  {Journal of the Physical Society of Japan}\ }\textbf {\bibinfo {volume}
  {90}},\ \bibinfo {pages} {012001} (\bibinfo {year} {2021})}\BibitemShut
  {NoStop}%
\bibitem [{\citenamefont {Jakli\v{c}}\ and\ \citenamefont
  {Prelov\v{s}ek}(1994)}]{Jaklic1994}%
  \BibitemOpen
  \bibfield  {author} {\bibinfo {author} {\bibfnamefont {J.}~\bibnamefont
  {Jakli\v{c}}}\ and\ \bibinfo {author} {\bibfnamefont {P.}~\bibnamefont
  {Prelov\v{s}ek}},\ }\bibfield  {title} {\enquote {\bibinfo {title} {{Lanczos
  method for the calculation of finite-temperature quantities in correlated
  systems}},}\ }\href {\doibase 10.1103/PhysRevB.49.5065} {\bibfield  {journal}
  {\bibinfo  {journal} {Phys. Rev. B}\ }\textbf {\bibinfo {volume} {49}},\
  \bibinfo {pages} {5065--5068} (\bibinfo {year} {1994})}\BibitemShut {NoStop}%
\bibitem [{\citenamefont {Prelov{\v{s}}ek}\ and\ \citenamefont
  {Bon{\v{c}}a}(2013)}]{Prelovsek}%
  \BibitemOpen
  \bibfield  {author} {\bibinfo {author} {\bibfnamefont {P.}~\bibnamefont
  {Prelov{\v{s}}ek}}\ and\ \bibinfo {author} {\bibfnamefont {J.}~\bibnamefont
  {Bon{\v{c}}a}},\ }\enquote {\bibinfo {title} {{Ground State and Finite
  Temperature Lanczos Methods}},}\ in\ \href {\doibase
  10.1007/978-3-642-35106-8_1} {\emph {\bibinfo {booktitle} {Strongly
  Correlated Systems: Numerical Methods}}},\ \bibinfo {editor} {edited by\
  \bibinfo {editor} {\bibfnamefont {Adolfo}\ \bibnamefont {Avella}}\ and\
  \bibinfo {editor} {\bibfnamefont {Ferdinando}\ \bibnamefont {Mancini}}}\
  (\bibinfo  {publisher} {Springer Berlin Heidelberg},\ \bibinfo {address}
  {Berlin, Heidelberg},\ \bibinfo {year} {2013})\ pp.\ \bibinfo {pages}
  {1--30}\BibitemShut {NoStop}%
\bibitem [{\citenamefont {Sugiura}\ and\ \citenamefont
  {Shimizu}(2013)}]{Sugiura2013}%
  \BibitemOpen
  \bibfield  {author} {\bibinfo {author} {\bibfnamefont {Sho}\ \bibnamefont
  {Sugiura}}\ and\ \bibinfo {author} {\bibfnamefont {Akira}\ \bibnamefont
  {Shimizu}},\ }\bibfield  {title} {\enquote {\bibinfo {title} {{Canonical
  Thermal Pure Quantum State}},}\ }\href {\doibase
  10.1103/PhysRevLett.111.010401} {\bibfield  {journal} {\bibinfo  {journal}
  {Phys. Rev. Lett.}\ }\textbf {\bibinfo {volume} {111}},\ \bibinfo {pages}
  {010401} (\bibinfo {year} {2013})}\BibitemShut {NoStop}%
\bibitem [{\citenamefont {Long}\ \emph {et~al.}(2003)\citenamefont {Long},
  \citenamefont {Prelov\ifmmode~\check{s}\else \v{s}\fi{}ek}, \citenamefont
  {El~Shawish}, \citenamefont {Karadamoglou},\ and\ \citenamefont
  {Zotos}}]{Long2003}%
  \BibitemOpen
  \bibfield  {author} {\bibinfo {author} {\bibfnamefont {M.~W.}\ \bibnamefont
  {Long}}, \bibinfo {author} {\bibfnamefont {P.}~\bibnamefont
  {Prelov\ifmmode~\check{s}\else \v{s}\fi{}ek}}, \bibinfo {author}
  {\bibfnamefont {S.}~\bibnamefont {El~Shawish}}, \bibinfo {author}
  {\bibfnamefont {J.}~\bibnamefont {Karadamoglou}}, \ and\ \bibinfo {author}
  {\bibfnamefont {X.}~\bibnamefont {Zotos}},\ }\bibfield  {title} {\enquote
  {\bibinfo {title} {{Finite-temperature dynamical correlations using the
  microcanonical ensemble and the Lanczos algorithm}},}\ }\href {\doibase
  10.1103/PhysRevB.68.235106} {\bibfield  {journal} {\bibinfo  {journal} {Phys.
  Rev. B}\ }\textbf {\bibinfo {volume} {68}},\ \bibinfo {pages} {235106}
  (\bibinfo {year} {2003})}\BibitemShut {NoStop}%
\bibitem [{\citenamefont {Okamoto}\ \emph {et~al.}(2018)\citenamefont
  {Okamoto}, \citenamefont {Alvarez}, \citenamefont {Dagotto},\ and\
  \citenamefont {Tohyama}}]{Okamoto2018}%
  \BibitemOpen
  \bibfield  {author} {\bibinfo {author} {\bibfnamefont {Satoshi}\ \bibnamefont
  {Okamoto}}, \bibinfo {author} {\bibfnamefont {Gonzalo}\ \bibnamefont
  {Alvarez}}, \bibinfo {author} {\bibfnamefont {Elbio}\ \bibnamefont
  {Dagotto}}, \ and\ \bibinfo {author} {\bibfnamefont {Takami}\ \bibnamefont
  {Tohyama}},\ }\bibfield  {title} {\enquote {\bibinfo {title} {{Accuracy of
  the microcanonical Lanczos method to compute real-frequency dynamical
  spectral functions of quantum models at finite temperatures}},}\ }\href
  {\doibase 10.1103/PhysRevE.97.043308} {\bibfield  {journal} {\bibinfo
  {journal} {Phys. Rev. E}\ }\textbf {\bibinfo {volume} {97}},\ \bibinfo
  {pages} {043308} (\bibinfo {year} {2018})}\BibitemShut {NoStop}%
\bibitem [{\citenamefont {Sugiura}\ and\ \citenamefont
  {Shimizu}(2012)}]{Sugiura2012}%
  \BibitemOpen
  \bibfield  {author} {\bibinfo {author} {\bibfnamefont {Sho}\ \bibnamefont
  {Sugiura}}\ and\ \bibinfo {author} {\bibfnamefont {Akira}\ \bibnamefont
  {Shimizu}},\ }\bibfield  {title} {\enquote {\bibinfo {title} {{Thermal Pure
  Quantum States at Finite Temperature}},}\ }\href {\doibase
  10.1103/PhysRevLett.108.240401} {\bibfield  {journal} {\bibinfo  {journal}
  {Phys. Rev. Lett.}\ }\textbf {\bibinfo {volume} {108}},\ \bibinfo {pages}
  {240401} (\bibinfo {year} {2012})}\BibitemShut {NoStop}%
\bibitem [{\citenamefont {Seki}\ and\ \citenamefont
  {Yunoki}(2020{\natexlab{b}})}]{Seki2020ftlm}%
  \BibitemOpen
  \bibfield  {author} {\bibinfo {author} {\bibfnamefont {Kazuhiro}\
  \bibnamefont {Seki}}\ and\ \bibinfo {author} {\bibfnamefont {Seiji}\
  \bibnamefont {Yunoki}},\ }\bibfield  {title} {\enquote {\bibinfo {title}
  {{Thermodynamic properties of an $S=\frac{1}{2}$ ring-exchange model on the
  triangular lattice}},}\ }\href {\doibase 10.1103/PhysRevB.101.235115}
  {\bibfield  {journal} {\bibinfo  {journal} {Phys. Rev. B}\ }\textbf {\bibinfo
  {volume} {101}},\ \bibinfo {pages} {235115} (\bibinfo {year}
  {2020}{\natexlab{b}})}\BibitemShut {NoStop}%
\bibitem [{\citenamefont {Suzuki}(1990)}]{Suzuki1990}%
  \BibitemOpen
  \bibfield  {author} {\bibinfo {author} {\bibfnamefont {Masuo}\ \bibnamefont
  {Suzuki}},\ }\bibfield  {title} {\enquote {\bibinfo {title} {{Fractal
  decomposition of exponential operators with applications to many-body
  theories and Monte Carlo simulations}},}\ }\href {\doibase
  https://doi.org/10.1016/0375-9601(90)90962-N} {\bibfield  {journal} {\bibinfo
   {journal} {Physics Letters A}\ }\textbf {\bibinfo {volume} {146}},\ \bibinfo
  {pages} {319 -- 323} (\bibinfo {year} {1990})}\BibitemShut {NoStop}%
\bibitem [{\citenamefont {Wall}\ and\ \citenamefont
  {Neuhauser}(1995)}]{Wall1995}%
  \BibitemOpen
  \bibfield  {author} {\bibinfo {author} {\bibfnamefont {Michael~R.}\
  \bibnamefont {Wall}}\ and\ \bibinfo {author} {\bibfnamefont {Daniel}\
  \bibnamefont {Neuhauser}},\ }\bibfield  {title} {\enquote {\bibinfo {title}
  {{Extraction, through filter-diagonalization, of general quantum eigenvalues
  or classical normal mode frequencies from a small number of residues or a
  short-time segment of a signal. I. Theory and application to a
  quantum-dynamics model}},}\ }\href {\doibase 10.1063/1.468999} {\bibfield
  {journal} {\bibinfo  {journal} {The Journal of Chemical Physics}\ }\textbf
  {\bibinfo {volume} {102}},\ \bibinfo {pages} {8011--8022} (\bibinfo {year}
  {1995})}\BibitemShut {NoStop}%
\bibitem [{\citenamefont {Challa}\ and\ \citenamefont
  {Hetherington}(1988)}]{Challa1988}%
  \BibitemOpen
  \bibfield  {author} {\bibinfo {author} {\bibfnamefont {Murty S.~S.}\
  \bibnamefont {Challa}}\ and\ \bibinfo {author} {\bibfnamefont {J.~H.}\
  \bibnamefont {Hetherington}},\ }\bibfield  {title} {\enquote {\bibinfo
  {title} {{Gaussian ensemble: An alternate Monte Carlo scheme}},}\ }\href
  {\doibase 10.1103/PhysRevA.38.6324} {\bibfield  {journal} {\bibinfo
  {journal} {Phys. Rev. A}\ }\textbf {\bibinfo {volume} {38}},\ \bibinfo
  {pages} {6324--6337} (\bibinfo {year} {1988})}\BibitemShut {NoStop}%
\bibitem [{\citenamefont {Yoneta}\ and\ \citenamefont
  {Shimizu}(2019)}]{Yoneta2019}%
  \BibitemOpen
  \bibfield  {author} {\bibinfo {author} {\bibfnamefont {Yasushi}\ \bibnamefont
  {Yoneta}}\ and\ \bibinfo {author} {\bibfnamefont {Akira}\ \bibnamefont
  {Shimizu}},\ }\bibfield  {title} {\enquote {\bibinfo {title} {{Squeezed
  ensemble for systems with first-order phase transitions}},}\ }\href {\doibase
  10.1103/PhysRevB.99.144105} {\bibfield  {journal} {\bibinfo  {journal} {Phys.
  Rev. B}\ }\textbf {\bibinfo {volume} {99}},\ \bibinfo {pages} {144105}
  (\bibinfo {year} {2019})}\BibitemShut {NoStop}%
\bibitem [{\citenamefont {Nakata}\ \emph {et~al.}(2014)\citenamefont {Nakata},
  \citenamefont {Koashi},\ and\ \citenamefont {Murao}}]{Nakata2014}%
  \BibitemOpen
  \bibfield  {author} {\bibinfo {author} {\bibfnamefont {Yoshifumi}\
  \bibnamefont {Nakata}}, \bibinfo {author} {\bibfnamefont {Masato}\
  \bibnamefont {Koashi}}, \ and\ \bibinfo {author} {\bibfnamefont {Mio}\
  \bibnamefont {Murao}},\ }\bibfield  {title} {\enquote {\bibinfo {title}
  {{Generating a state $t$-design by diagonal quantum circuits}},}\ }\href
  {\doibase 10.1088/1367-2630/16/5/053043} {\bibfield  {journal} {\bibinfo
  {journal} {New Journal of Physics}\ }\textbf {\bibinfo {volume} {16}},\
  \bibinfo {pages} {053043} (\bibinfo {year} {2014})}\BibitemShut {NoStop}%
\bibitem [{\citenamefont {Nakata}\ \emph {et~al.}(2012)\citenamefont {Nakata},
  \citenamefont {Turner},\ and\ \citenamefont {Murao}}]{Nakata2012}%
  \BibitemOpen
  \bibfield  {author} {\bibinfo {author} {\bibfnamefont {Yoshifumi}\
  \bibnamefont {Nakata}}, \bibinfo {author} {\bibfnamefont {Peter~S.}\
  \bibnamefont {Turner}}, \ and\ \bibinfo {author} {\bibfnamefont {Mio}\
  \bibnamefont {Murao}},\ }\bibfield  {title} {\enquote {\bibinfo {title}
  {{Phase-random states: Ensembles of states with fixed amplitudes and
  uniformly distributed phases in a fixed basis}},}\ }\href {\doibase
  10.1103/PhysRevA.86.012301} {\bibfield  {journal} {\bibinfo  {journal} {Phys.
  Rev. A}\ }\textbf {\bibinfo {volume} {86}},\ \bibinfo {pages} {012301}
  (\bibinfo {year} {2012})}\BibitemShut {NoStop}%
\bibitem [{\citenamefont {Iwaki}\ and\ \citenamefont
  {Hotta}(2022)}]{Iwaki2022}%
  \BibitemOpen
  \bibfield  {author} {\bibinfo {author} {\bibfnamefont {Atsushi}\ \bibnamefont
  {Iwaki}}\ and\ \bibinfo {author} {\bibfnamefont {Chisa}\ \bibnamefont
  {Hotta}},\ }\bibfield  {title} {\enquote {\bibinfo {title} {Purity of thermal
  mixed quantum states},}\ }\href {\doibase 10.1103/PhysRevB.106.094409}
  {\bibfield  {journal} {\bibinfo  {journal} {Phys. Rev. B}\ }\textbf {\bibinfo
  {volume} {106}},\ \bibinfo {pages} {094409} (\bibinfo {year}
  {2022})}\BibitemShut {NoStop}%
\bibitem [{\citenamefont {Dankert}\ \emph {et~al.}(2009)\citenamefont
  {Dankert}, \citenamefont {Cleve}, \citenamefont {Emerson},\ and\
  \citenamefont {Livine}}]{Dankert2009}%
  \BibitemOpen
  \bibfield  {author} {\bibinfo {author} {\bibfnamefont {Christoph}\
  \bibnamefont {Dankert}}, \bibinfo {author} {\bibfnamefont {Richard}\
  \bibnamefont {Cleve}}, \bibinfo {author} {\bibfnamefont {Joseph}\
  \bibnamefont {Emerson}}, \ and\ \bibinfo {author} {\bibfnamefont {Etera}\
  \bibnamefont {Livine}},\ }\bibfield  {title} {\enquote {\bibinfo {title}
  {Exact and approximate unitary 2-designs and their application to fidelity
  estimation},}\ }\href {\doibase 10.1103/PhysRevA.80.012304} {\bibfield
  {journal} {\bibinfo  {journal} {Phys. Rev. A}\ }\textbf {\bibinfo {volume}
  {80}},\ \bibinfo {pages} {012304} (\bibinfo {year} {2009})}\BibitemShut
  {NoStop}%
\bibitem [{\citenamefont {Iitaka}(2020)}]{Iitaka2020}%
  \BibitemOpen
  \bibfield  {author} {\bibinfo {author} {\bibfnamefont {Toshiaki}\
  \bibnamefont {Iitaka}},\ }\href@noop {} {\enquote {\bibinfo {title} {{Random
  Phase Product State for Canonical Ensemble}},}\ } (\bibinfo {year} {2020}),\
  \Eprint {http://arxiv.org/abs/2006.14459} {arXiv:2006.14459
  [cond-mat.str-el]} \BibitemShut {NoStop}%
\bibitem [{\citenamefont {Nakata}\ \emph
  {et~al.}(2017{\natexlab{a}})\citenamefont {Nakata}, \citenamefont {Hirche},
  \citenamefont {Morgan},\ and\ \citenamefont {Winter}}]{Nakata2017math}%
  \BibitemOpen
  \bibfield  {author} {\bibinfo {author} {\bibfnamefont {Yoshifumi}\
  \bibnamefont {Nakata}}, \bibinfo {author} {\bibfnamefont {Christoph}\
  \bibnamefont {Hirche}}, \bibinfo {author} {\bibfnamefont {Ciara}\
  \bibnamefont {Morgan}}, \ and\ \bibinfo {author} {\bibfnamefont {Andreas}\
  \bibnamefont {Winter}},\ }\bibfield  {title} {\enquote {\bibinfo {title}
  {{Unitary 2-designs from random $X$- and $Z$-diagonal unitaries}},}\ }\href
  {\doibase 10.1063/1.4983266} {\bibfield  {journal} {\bibinfo  {journal}
  {Journal of Mathematical Physics}\ }\textbf {\bibinfo {volume} {58}},\
  \bibinfo {pages} {052203} (\bibinfo {year} {2017}{\natexlab{a}})}\BibitemShut
  {NoStop}%
\bibitem [{\citenamefont {Nakata}\ \emph
  {et~al.}(2017{\natexlab{b}})\citenamefont {Nakata}, \citenamefont {Hirche},
  \citenamefont {Koashi},\ and\ \citenamefont {Winter}}]{Nakata2017PRX}%
  \BibitemOpen
  \bibfield  {author} {\bibinfo {author} {\bibfnamefont {Yoshifumi}\
  \bibnamefont {Nakata}}, \bibinfo {author} {\bibfnamefont {Christoph}\
  \bibnamefont {Hirche}}, \bibinfo {author} {\bibfnamefont {Masato}\
  \bibnamefont {Koashi}}, \ and\ \bibinfo {author} {\bibfnamefont {Andreas}\
  \bibnamefont {Winter}},\ }\bibfield  {title} {\enquote {\bibinfo {title}
  {{Efficient Quantum Pseudorandomness with Nearly Time-Independent Hamiltonian
  Dynamics}},}\ }\href {\doibase 10.1103/PhysRevX.7.021006} {\bibfield
  {journal} {\bibinfo  {journal} {Phys. Rev. X}\ }\textbf {\bibinfo {volume}
  {7}},\ \bibinfo {pages} {021006} (\bibinfo {year}
  {2017}{\natexlab{b}})}\BibitemShut {NoStop}%
\bibitem [{\citenamefont {Trotter}(1959)}]{Trotter1959}%
  \BibitemOpen
  \bibfield  {author} {\bibinfo {author} {\bibfnamefont {H.~F.}\ \bibnamefont
  {Trotter}},\ }\bibfield  {title} {\enquote {\bibinfo {title} {On the product
  of semi-groups of operators},}\ }\href
  {https://doi.org/10.1090/S0002-9939-1959-0108732-6} {\bibfield  {journal}
  {\bibinfo  {journal} {Proc. Am. Math. Soc.}\ }\textbf {\bibinfo {volume}
  {10}},\ \bibinfo {pages} {545} (\bibinfo {year} {1959})}\BibitemShut
  {NoStop}%
\bibitem [{\citenamefont {Suzuki}(1976)}]{Suzuki1976}%
  \BibitemOpen
  \bibfield  {author} {\bibinfo {author} {\bibfnamefont {Masuo}\ \bibnamefont
  {Suzuki}},\ }\bibfield  {title} {\enquote {\bibinfo {title} {{Generalized
  Trotter's formula and systematic approximants of exponential operators and
  inner derivations with applications to many-body problems}},}\ }\href
  {http://projecteuclid.org/euclid.cmp/1103900351} {\bibfield  {journal}
  {\bibinfo  {journal} {Comm. Math. Phys.}\ }\textbf {\bibinfo {volume} {51}},\
  \bibinfo {pages} {183--190} (\bibinfo {year} {1976})}\BibitemShut {NoStop}%
\bibitem [{\citenamefont {Seki}\ \emph {et~al.}(2020)\citenamefont {Seki},
  \citenamefont {Shirakawa},\ and\ \citenamefont {Yunoki}}]{Seki2020vqe}%
  \BibitemOpen
  \bibfield  {author} {\bibinfo {author} {\bibfnamefont {Kazuhiro}\
  \bibnamefont {Seki}}, \bibinfo {author} {\bibfnamefont {Tomonori}\
  \bibnamefont {Shirakawa}}, \ and\ \bibinfo {author} {\bibfnamefont {Seiji}\
  \bibnamefont {Yunoki}},\ }\bibfield  {title} {\enquote {\bibinfo {title}
  {Symmetry-adapted variational quantum eigensolver},}\ }\href {\doibase
  10.1103/PhysRevA.101.052340} {\bibfield  {journal} {\bibinfo  {journal}
  {Phys. Rev. A}\ }\textbf {\bibinfo {volume} {101}},\ \bibinfo {pages}
  {052340} (\bibinfo {year} {2020})}\BibitemShut {NoStop}%
\bibitem [{\citenamefont {Iwaki}\ \emph {et~al.}(2021)\citenamefont {Iwaki},
  \citenamefont {Shimizu},\ and\ \citenamefont {Hotta}}]{Iwaki2021}%
  \BibitemOpen
  \bibfield  {author} {\bibinfo {author} {\bibfnamefont {Atsushi}\ \bibnamefont
  {Iwaki}}, \bibinfo {author} {\bibfnamefont {Akira}\ \bibnamefont {Shimizu}},
  \ and\ \bibinfo {author} {\bibfnamefont {Chisa}\ \bibnamefont {Hotta}},\
  }\bibfield  {title} {\enquote {\bibinfo {title} {{Thermal pure quantum matrix
  product states recovering a volume law entanglement}},}\ }\href {\doibase
  10.1103/PhysRevResearch.3.L022015} {\bibfield  {journal} {\bibinfo  {journal}
  {Phys. Rev. Research}\ }\textbf {\bibinfo {volume} {3}},\ \bibinfo {pages}
  {L022015} (\bibinfo {year} {2021})}\BibitemShut {NoStop}%
\bibitem [{\citenamefont {Goto}\ \emph {et~al.}(2021)\citenamefont {Goto},
  \citenamefont {Kaneko},\ and\ \citenamefont {Danshita}}]{Goto2021}%
  \BibitemOpen
  \bibfield  {author} {\bibinfo {author} {\bibfnamefont {Shimpei}\ \bibnamefont
  {Goto}}, \bibinfo {author} {\bibfnamefont {Ryui}\ \bibnamefont {Kaneko}}, \
  and\ \bibinfo {author} {\bibfnamefont {Ippei}\ \bibnamefont {Danshita}},\
  }\bibfield  {title} {\enquote {\bibinfo {title} {{Matrix product state
  approach for a quantum system at finite temperatures using random phases and
  Trotter gates}},}\ }\href {\doibase 10.1103/PhysRevB.104.045133} {\bibfield
  {journal} {\bibinfo  {journal} {Phys. Rev. B}\ }\textbf {\bibinfo {volume}
  {104}},\ \bibinfo {pages} {045133} (\bibinfo {year} {2021})}\BibitemShut
  {NoStop}%
\bibitem [{\citenamefont {Coopmans}\ \emph {et~al.}(2022)\citenamefont
  {Coopmans}, \citenamefont {Kikuchi},\ and\ \citenamefont
  {Benedetti}}]{Coopmans2022}%
  \BibitemOpen
  \bibfield  {author} {\bibinfo {author} {\bibfnamefont {Luuk}\ \bibnamefont
  {Coopmans}}, \bibinfo {author} {\bibfnamefont {Yuta}\ \bibnamefont
  {Kikuchi}}, \ and\ \bibinfo {author} {\bibfnamefont {Marcello}\ \bibnamefont
  {Benedetti}},\ }\href {\doibase 10.48550/ARXIV.2206.05302} {\enquote
  {\bibinfo {title} {{Predicting Gibbs State Expectation Values with Pure
  Thermal Shadows}},}\ } (\bibinfo {year} {2022}),\ \Eprint
  {http://arxiv.org/abs/2206.05302} {arXiv:2206.05302 [quant-ph]} \BibitemShut
  {NoStop}%
\bibitem [{\citenamefont {Arute}\ \emph {et~al.}(2020)\citenamefont {Arute},
  \citenamefont {Arya}, \citenamefont {Babbush}, \citenamefont {Bacon},
  \citenamefont {Bardin}, \citenamefont {Barends}, \citenamefont {Bengtsson},
  \citenamefont {Boixo}, \citenamefont {Broughton}, \citenamefont {Buckley},
  \citenamefont {Buell}, \citenamefont {Burkett}, \citenamefont {Bushnell},
  \citenamefont {Chen}, \citenamefont {Chen}, \citenamefont {Chen},
  \citenamefont {Chiaro}, \citenamefont {Collins}, \citenamefont {Cotton},
  \citenamefont {Courtney}, \citenamefont {Demura}, \citenamefont {Derk},
  \citenamefont {Dunsworth}, \citenamefont {Eppens}, \citenamefont {Eckl},
  \citenamefont {Erickson}, \citenamefont {Farhi}, \citenamefont {Fowler},
  \citenamefont {Foxen}, \citenamefont {Gidney}, \citenamefont {Giustina},
  \citenamefont {Graff}, \citenamefont {Gross}, \citenamefont {Habegger},
  \citenamefont {Harrigan}, \citenamefont {Ho}, \citenamefont {Hong},
  \citenamefont {Huang}, \citenamefont {Huggins}, \citenamefont {Ioffe},
  \citenamefont {Isakov}, \citenamefont {Jeffrey}, \citenamefont {Jiang},
  \citenamefont {Jones}, \citenamefont {Kafri}, \citenamefont {Kechedzhi},
  \citenamefont {Kelly}, \citenamefont {Kim}, \citenamefont {Klimov},
  \citenamefont {Korotkov}, \citenamefont {Kostritsa}, \citenamefont
  {Landhuis}, \citenamefont {Laptev}, \citenamefont {Lindmark}, \citenamefont
  {Lucero}, \citenamefont {Marthaler}, \citenamefont {Martin}, \citenamefont
  {Martinis}, \citenamefont {Marusczyk}, \citenamefont {McArdle}, \citenamefont
  {McClean}, \citenamefont {McCourt}, \citenamefont {McEwen}, \citenamefont
  {Megrant}, \citenamefont {Mejuto-Zaera}, \citenamefont {Mi}, \citenamefont
  {Mohseni}, \citenamefont {Mruczkiewicz}, \citenamefont {Mutus}, \citenamefont
  {Naaman}, \citenamefont {Neeley}, \citenamefont {Neill}, \citenamefont
  {Neven}, \citenamefont {Newman}, \citenamefont {Niu}, \citenamefont
  {O'Brien}, \citenamefont {Ostby}, \citenamefont {Pat{\'o}}, \citenamefont
  {Petukhov}, \citenamefont {Putterman}, \citenamefont {Quintana},
  \citenamefont {Reiner}, \citenamefont {Roushan}, \citenamefont {Rubin},
  \citenamefont {Sank}, \citenamefont {Satzinger}, \citenamefont {Smelyanskiy},
  \citenamefont {Strain}, \citenamefont {Sung}, \citenamefont {Schmitteckert},
  \citenamefont {Szalay}, \citenamefont {Tubman}, \citenamefont {Vainsencher},
  \citenamefont {White}, \citenamefont {Vogt}, \citenamefont {Yao},
  \citenamefont {Yeh}, \citenamefont {Zalcman},\ and\ \citenamefont
  {Zanker}}]{arute2020observation}%
  \BibitemOpen
  \bibfield  {author} {\bibinfo {author} {\bibfnamefont {Frank}\ \bibnamefont
  {Arute}}, \bibinfo {author} {\bibfnamefont {Kunal}\ \bibnamefont {Arya}},
  \bibinfo {author} {\bibfnamefont {Ryan}\ \bibnamefont {Babbush}}, \bibinfo
  {author} {\bibfnamefont {Dave}\ \bibnamefont {Bacon}}, \bibinfo {author}
  {\bibfnamefont {Joseph~C.}\ \bibnamefont {Bardin}}, \bibinfo {author}
  {\bibfnamefont {Rami}\ \bibnamefont {Barends}}, \bibinfo {author}
  {\bibfnamefont {Andreas}\ \bibnamefont {Bengtsson}}, \bibinfo {author}
  {\bibfnamefont {Sergio}\ \bibnamefont {Boixo}}, \bibinfo {author}
  {\bibfnamefont {Michael}\ \bibnamefont {Broughton}}, \bibinfo {author}
  {\bibfnamefont {Bob~B.}\ \bibnamefont {Buckley}}, \bibinfo {author}
  {\bibfnamefont {David~A.}\ \bibnamefont {Buell}}, \bibinfo {author}
  {\bibfnamefont {Brian}\ \bibnamefont {Burkett}}, \bibinfo {author}
  {\bibfnamefont {Nicholas}\ \bibnamefont {Bushnell}}, \bibinfo {author}
  {\bibfnamefont {Yu}~\bibnamefont {Chen}}, \bibinfo {author} {\bibfnamefont
  {Zijun}\ \bibnamefont {Chen}}, \bibinfo {author} {\bibfnamefont {Yu-An}\
  \bibnamefont {Chen}}, \bibinfo {author} {\bibfnamefont {Ben}\ \bibnamefont
  {Chiaro}}, \bibinfo {author} {\bibfnamefont {Roberto}\ \bibnamefont
  {Collins}}, \bibinfo {author} {\bibfnamefont {Stephen~J.}\ \bibnamefont
  {Cotton}}, \bibinfo {author} {\bibfnamefont {William}\ \bibnamefont
  {Courtney}}, \bibinfo {author} {\bibfnamefont {Sean}\ \bibnamefont {Demura}},
  \bibinfo {author} {\bibfnamefont {Alan}\ \bibnamefont {Derk}}, \bibinfo
  {author} {\bibfnamefont {Andrew}\ \bibnamefont {Dunsworth}}, \bibinfo
  {author} {\bibfnamefont {Daniel}\ \bibnamefont {Eppens}}, \bibinfo {author}
  {\bibfnamefont {Thomas}\ \bibnamefont {Eckl}}, \bibinfo {author}
  {\bibfnamefont {Catherine}\ \bibnamefont {Erickson}}, \bibinfo {author}
  {\bibfnamefont {Edward}\ \bibnamefont {Farhi}}, \bibinfo {author}
  {\bibfnamefont {Austin}\ \bibnamefont {Fowler}}, \bibinfo {author}
  {\bibfnamefont {Brooks}\ \bibnamefont {Foxen}}, \bibinfo {author}
  {\bibfnamefont {Craig}\ \bibnamefont {Gidney}}, \bibinfo {author}
  {\bibfnamefont {Marissa}\ \bibnamefont {Giustina}}, \bibinfo {author}
  {\bibfnamefont {Rob}\ \bibnamefont {Graff}}, \bibinfo {author} {\bibfnamefont
  {Jonathan~A.}\ \bibnamefont {Gross}}, \bibinfo {author} {\bibfnamefont
  {Steve}\ \bibnamefont {Habegger}}, \bibinfo {author} {\bibfnamefont
  {Matthew~P.}\ \bibnamefont {Harrigan}}, \bibinfo {author} {\bibfnamefont
  {Alan}\ \bibnamefont {Ho}}, \bibinfo {author} {\bibfnamefont {Sabrina}\
  \bibnamefont {Hong}}, \bibinfo {author} {\bibfnamefont {Trent}\ \bibnamefont
  {Huang}}, \bibinfo {author} {\bibfnamefont {William}\ \bibnamefont
  {Huggins}}, \bibinfo {author} {\bibfnamefont {Lev~B.}\ \bibnamefont {Ioffe}},
  \bibinfo {author} {\bibfnamefont {Sergei~V.}\ \bibnamefont {Isakov}},
  \bibinfo {author} {\bibfnamefont {Evan}\ \bibnamefont {Jeffrey}}, \bibinfo
  {author} {\bibfnamefont {Zhang}\ \bibnamefont {Jiang}}, \bibinfo {author}
  {\bibfnamefont {Cody}\ \bibnamefont {Jones}}, \bibinfo {author}
  {\bibfnamefont {Dvir}\ \bibnamefont {Kafri}}, \bibinfo {author}
  {\bibfnamefont {Kostyantyn}\ \bibnamefont {Kechedzhi}}, \bibinfo {author}
  {\bibfnamefont {Julian}\ \bibnamefont {Kelly}}, \bibinfo {author}
  {\bibfnamefont {Seon}\ \bibnamefont {Kim}}, \bibinfo {author} {\bibfnamefont
  {Paul~V.}\ \bibnamefont {Klimov}}, \bibinfo {author} {\bibfnamefont
  {Alexander~N.}\ \bibnamefont {Korotkov}}, \bibinfo {author} {\bibfnamefont
  {Fedor}\ \bibnamefont {Kostritsa}}, \bibinfo {author} {\bibfnamefont {David}\
  \bibnamefont {Landhuis}}, \bibinfo {author} {\bibfnamefont {Pavel}\
  \bibnamefont {Laptev}}, \bibinfo {author} {\bibfnamefont {Mike}\ \bibnamefont
  {Lindmark}}, \bibinfo {author} {\bibfnamefont {Erik}\ \bibnamefont {Lucero}},
  \bibinfo {author} {\bibfnamefont {Michael}\ \bibnamefont {Marthaler}},
  \bibinfo {author} {\bibfnamefont {Orion}\ \bibnamefont {Martin}}, \bibinfo
  {author} {\bibfnamefont {John~M.}\ \bibnamefont {Martinis}}, \bibinfo
  {author} {\bibfnamefont {Anika}\ \bibnamefont {Marusczyk}}, \bibinfo {author}
  {\bibfnamefont {Sam}\ \bibnamefont {McArdle}}, \bibinfo {author}
  {\bibfnamefont {Jarrod~R.}\ \bibnamefont {McClean}}, \bibinfo {author}
  {\bibfnamefont {Trevor}\ \bibnamefont {McCourt}}, \bibinfo {author}
  {\bibfnamefont {Matt}\ \bibnamefont {McEwen}}, \bibinfo {author}
  {\bibfnamefont {Anthony}\ \bibnamefont {Megrant}}, \bibinfo {author}
  {\bibfnamefont {Carlos}\ \bibnamefont {Mejuto-Zaera}}, \bibinfo {author}
  {\bibfnamefont {Xiao}\ \bibnamefont {Mi}}, \bibinfo {author} {\bibfnamefont
  {Masoud}\ \bibnamefont {Mohseni}}, \bibinfo {author} {\bibfnamefont
  {Wojciech}\ \bibnamefont {Mruczkiewicz}}, \bibinfo {author} {\bibfnamefont
  {Josh}\ \bibnamefont {Mutus}}, \bibinfo {author} {\bibfnamefont {Ofer}\
  \bibnamefont {Naaman}}, \bibinfo {author} {\bibfnamefont {Matthew}\
  \bibnamefont {Neeley}}, \bibinfo {author} {\bibfnamefont {Charles}\
  \bibnamefont {Neill}}, \bibinfo {author} {\bibfnamefont {Hartmut}\
  \bibnamefont {Neven}}, \bibinfo {author} {\bibfnamefont {Michael}\
  \bibnamefont {Newman}}, \bibinfo {author} {\bibfnamefont {Murphy~Yuezhen}\
  \bibnamefont {Niu}}, \bibinfo {author} {\bibfnamefont {Thomas~E.}\
  \bibnamefont {O'Brien}}, \bibinfo {author} {\bibfnamefont {Eric}\
  \bibnamefont {Ostby}}, \bibinfo {author} {\bibfnamefont {B{\'a}lint}\
  \bibnamefont {Pat{\'o}}}, \bibinfo {author} {\bibfnamefont {Andre}\
  \bibnamefont {Petukhov}}, \bibinfo {author} {\bibfnamefont {Harald}\
  \bibnamefont {Putterman}}, \bibinfo {author} {\bibfnamefont {Chris}\
  \bibnamefont {Quintana}}, \bibinfo {author} {\bibfnamefont {Jan-Michael}\
  \bibnamefont {Reiner}}, \bibinfo {author} {\bibfnamefont {Pedram}\
  \bibnamefont {Roushan}}, \bibinfo {author} {\bibfnamefont {Nicholas~C.}\
  \bibnamefont {Rubin}}, \bibinfo {author} {\bibfnamefont {Daniel}\
  \bibnamefont {Sank}}, \bibinfo {author} {\bibfnamefont {Kevin~J.}\
  \bibnamefont {Satzinger}}, \bibinfo {author} {\bibfnamefont {Vadim}\
  \bibnamefont {Smelyanskiy}}, \bibinfo {author} {\bibfnamefont {Doug}\
  \bibnamefont {Strain}}, \bibinfo {author} {\bibfnamefont {Kevin~J.}\
  \bibnamefont {Sung}}, \bibinfo {author} {\bibfnamefont {Peter}\ \bibnamefont
  {Schmitteckert}}, \bibinfo {author} {\bibfnamefont {Marco}\ \bibnamefont
  {Szalay}}, \bibinfo {author} {\bibfnamefont {Norm~M.}\ \bibnamefont
  {Tubman}}, \bibinfo {author} {\bibfnamefont {Amit}\ \bibnamefont
  {Vainsencher}}, \bibinfo {author} {\bibfnamefont {Theodore}\ \bibnamefont
  {White}}, \bibinfo {author} {\bibfnamefont {Nicolas}\ \bibnamefont {Vogt}},
  \bibinfo {author} {\bibfnamefont {Z.~Jamie}\ \bibnamefont {Yao}}, \bibinfo
  {author} {\bibfnamefont {Ping}\ \bibnamefont {Yeh}}, \bibinfo {author}
  {\bibfnamefont {Adam}\ \bibnamefont {Zalcman}}, \ and\ \bibinfo {author}
  {\bibfnamefont {Sebastian}\ \bibnamefont {Zanker}},\ }\href@noop {} {\enquote
  {\bibinfo {title} {{Observation of separated dynamics of charge and spin in
  the Fermi-Hubbard model}},}\ } (\bibinfo {year} {2020}),\ \Eprint
  {http://arxiv.org/abs/2010.07965} {arXiv:2010.07965 [quant-ph]} \BibitemShut
  {NoStop}%
\bibitem [{\citenamefont {Parrish}\ and\ \citenamefont
  {McMahon}(2019)}]{parrish2019quantum}%
  \BibitemOpen
  \bibfield  {author} {\bibinfo {author} {\bibfnamefont {Robert~M.}\
  \bibnamefont {Parrish}}\ and\ \bibinfo {author} {\bibfnamefont {Peter~L.}\
  \bibnamefont {McMahon}},\ }\href@noop {} {\enquote {\bibinfo {title}
  {{Quantum Filter Diagonalization: Quantum Eigendecomposition without Full
  Quantum Phase Estimation}},}\ } (\bibinfo {year} {2019}),\ \Eprint
  {http://arxiv.org/abs/1909.08925} {arXiv:1909.08925 [quant-ph]} \BibitemShut
  {NoStop}%
\bibitem [{\citenamefont {Stair}\ \emph {et~al.}(2020)\citenamefont {Stair},
  \citenamefont {Huang},\ and\ \citenamefont {Evangelista}}]{Stair2020}%
  \BibitemOpen
  \bibfield  {author} {\bibinfo {author} {\bibfnamefont {Nicholas~H.}\
  \bibnamefont {Stair}}, \bibinfo {author} {\bibfnamefont {Renke}\ \bibnamefont
  {Huang}}, \ and\ \bibinfo {author} {\bibfnamefont {Francesco~A.}\
  \bibnamefont {Evangelista}},\ }\bibfield  {title} {\enquote {\bibinfo {title}
  {{A Multireference Quantum Krylov Algorithm for Strongly Correlated
  Electrons}},}\ }\href {\doibase 10.1021/acs.jctc.9b01125} {\bibfield
  {journal} {\bibinfo  {journal} {Journal of Chemical Theory and Computation}\
  }\textbf {\bibinfo {volume} {16}},\ \bibinfo {pages} {2236--2245} (\bibinfo
  {year} {2020})}\BibitemShut {NoStop}%
\bibitem [{\citenamefont {He}\ \emph {et~al.}(2022)\citenamefont {He},
  \citenamefont {Zhang},\ and\ \citenamefont {Wang}}]{He2022}%
  \BibitemOpen
  \bibfield  {author} {\bibinfo {author} {\bibfnamefont {Min-Quan}\
  \bibnamefont {He}}, \bibinfo {author} {\bibfnamefont {Dan-Bo}\ \bibnamefont
  {Zhang}}, \ and\ \bibinfo {author} {\bibfnamefont {Z.~D.}\ \bibnamefont
  {Wang}},\ }\bibfield  {title} {\enquote {\bibinfo {title} {{Quantum Gaussian
  filter for exploring ground-state properties}},}\ }\href {\doibase
  10.1103/PhysRevA.106.032420} {\bibfield  {journal} {\bibinfo  {journal}
  {Phys. Rev. A}\ }\textbf {\bibinfo {volume} {106}},\ \bibinfo {pages}
  {032420} (\bibinfo {year} {2022})}\BibitemShut {NoStop}%
\bibitem [{\citenamefont {Seki}\ and\ \citenamefont
  {Yunoki}(2022)}]{seki2021spatial}%
  \BibitemOpen
  \bibfield  {author} {\bibinfo {author} {\bibfnamefont {Kazuhiro}\
  \bibnamefont {Seki}}\ and\ \bibinfo {author} {\bibfnamefont {Seiji}\
  \bibnamefont {Yunoki}},\ }\bibfield  {title} {\enquote {\bibinfo {title}
  {{Spatial, spin, and charge symmetry projections for a Fermi-Hubbard model on
  a quantum computer}},}\ }\href {\doibase 10.1103/PhysRevA.105.032419}
  {\bibfield  {journal} {\bibinfo  {journal} {Phys. Rev. A}\ }\textbf {\bibinfo
  {volume} {105}},\ \bibinfo {pages} {032419} (\bibinfo {year}
  {2022})}\BibitemShut {NoStop}%
\bibitem [{\citenamefont {Seki}\ and\ \citenamefont {Yunoki}(2021)}]{seki2021}%
  \BibitemOpen
  \bibfield  {author} {\bibinfo {author} {\bibfnamefont {Kazuhiro}\
  \bibnamefont {Seki}}\ and\ \bibinfo {author} {\bibfnamefont {Seiji}\
  \bibnamefont {Yunoki}},\ }\bibfield  {title} {\enquote {\bibinfo {title}
  {{Quantum Power Method by a Superposition of Time-Evolved States}},}\ }\href
  {\doibase 10.1103/PRXQuantum.2.010333} {\bibfield  {journal} {\bibinfo
  {journal} {PRX Quantum}\ }\textbf {\bibinfo {volume} {2}},\ \bibinfo {pages}
  {010333} (\bibinfo {year} {2021})}\BibitemShut {NoStop}%
\bibitem [{\citenamefont {Morita}\ and\ \citenamefont
  {Tohyama}(2020)}]{Morita2020}%
  \BibitemOpen
  \bibfield  {author} {\bibinfo {author} {\bibfnamefont {Katsuhiro}\
  \bibnamefont {Morita}}\ and\ \bibinfo {author} {\bibfnamefont {Takami}\
  \bibnamefont {Tohyama}},\ }\bibfield  {title} {\enquote {\bibinfo {title}
  {{Finite-temperature properties of the Kitaev-Heisenberg models on kagome and
  triangular lattices studied by improved finite-temperature Lanczos
  methods}},}\ }\href {\doibase 10.1103/PhysRevResearch.2.013205} {\bibfield
  {journal} {\bibinfo  {journal} {Phys. Rev. Research}\ }\textbf {\bibinfo
  {volume} {2}},\ \bibinfo {pages} {013205} (\bibinfo {year}
  {2020})}\BibitemShut {NoStop}%
\bibitem [{\citenamefont {Aichhorn}\ \emph {et~al.}(2003)\citenamefont
  {Aichhorn}, \citenamefont {Daghofer}, \citenamefont {Evertz},\ and\
  \citenamefont {von~der Linden}}]{Aichhorn2003}%
  \BibitemOpen
  \bibfield  {author} {\bibinfo {author} {\bibfnamefont {Markus}\ \bibnamefont
  {Aichhorn}}, \bibinfo {author} {\bibfnamefont {Maria}\ \bibnamefont
  {Daghofer}}, \bibinfo {author} {\bibfnamefont {Hans~Gerd}\ \bibnamefont
  {Evertz}}, \ and\ \bibinfo {author} {\bibfnamefont {Wolfgang}\ \bibnamefont
  {von~der Linden}},\ }\bibfield  {title} {\enquote {\bibinfo {title}
  {{Low-temperature Lanczos method for strongly correlated systems}},}\ }\href
  {\doibase 10.1103/PhysRevB.67.161103} {\bibfield  {journal} {\bibinfo
  {journal} {Phys. Rev. B}\ }\textbf {\bibinfo {volume} {67}},\ \bibinfo
  {pages} {161103} (\bibinfo {year} {2003})}\BibitemShut {NoStop}%
\bibitem [{\citenamefont {Yunger~Halpern}\ \emph {et~al.}(2016)\citenamefont
  {Yunger~Halpern}, \citenamefont {Faist}, \citenamefont {Oppenheim},\ and\
  \citenamefont {Winter}}]{Halpern2016}%
  \BibitemOpen
  \bibfield  {author} {\bibinfo {author} {\bibfnamefont {Nicole}\ \bibnamefont
  {Yunger~Halpern}}, \bibinfo {author} {\bibfnamefont {Philippe}\ \bibnamefont
  {Faist}}, \bibinfo {author} {\bibfnamefont {Jonathan}\ \bibnamefont
  {Oppenheim}}, \ and\ \bibinfo {author} {\bibfnamefont {Andreas}\ \bibnamefont
  {Winter}},\ }\bibfield  {title} {\enquote {\bibinfo {title} {Microcanonical
  and resource-theoretic derivations of the thermal state of a quantum system
  with noncommuting charges},}\ }\href {\doibase 10.1038/ncomms12051}
  {\bibfield  {journal} {\bibinfo  {journal} {Nature Communications}\ }\textbf
  {\bibinfo {volume} {7}},\ \bibinfo {pages} {12051} (\bibinfo {year}
  {2016})}\BibitemShut {NoStop}%
\bibitem [{\citenamefont {Halpern}(2018)}]{Halpern2018}%
  \BibitemOpen
  \bibfield  {author} {\bibinfo {author} {\bibfnamefont {Nicole~Yunger}\
  \bibnamefont {Halpern}},\ }\bibfield  {title} {\enquote {\bibinfo {title}
  {{Beyond heat baths {II}: framework for generalized thermodynamic resource
  theories}},}\ }\href {\doibase 10.1088/1751-8121/aaa62f} {\bibfield
  {journal} {\bibinfo  {journal} {Journal of Physics A: Mathematical and
  Theoretical}\ }\textbf {\bibinfo {volume} {51}},\ \bibinfo {pages} {094001}
  (\bibinfo {year} {2018})}\BibitemShut {NoStop}%
\bibitem [{\citenamefont {Yunger~Halpern}\ \emph {et~al.}(2020)\citenamefont
  {Yunger~Halpern}, \citenamefont {Beverland},\ and\ \citenamefont
  {Kalev}}]{Halpern2020}%
  \BibitemOpen
  \bibfield  {author} {\bibinfo {author} {\bibfnamefont {Nicole}\ \bibnamefont
  {Yunger~Halpern}}, \bibinfo {author} {\bibfnamefont {Michael~E.}\
  \bibnamefont {Beverland}}, \ and\ \bibinfo {author} {\bibfnamefont {Amir}\
  \bibnamefont {Kalev}},\ }\bibfield  {title} {\enquote {\bibinfo {title}
  {Noncommuting conserved charges in quantum many-body thermalization},}\
  }\href {\doibase 10.1103/PhysRevE.101.042117} {\bibfield  {journal} {\bibinfo
   {journal} {Phys. Rev. E}\ }\textbf {\bibinfo {volume} {101}},\ \bibinfo
  {pages} {042117} (\bibinfo {year} {2020})}\BibitemShut {NoStop}%
\bibitem [{\citenamefont {Yunger~Halpern}\ and\ \citenamefont
  {Majidy}(2022)}]{Halpern2022}%
  \BibitemOpen
  \bibfield  {author} {\bibinfo {author} {\bibfnamefont {Nicole}\ \bibnamefont
  {Yunger~Halpern}}\ and\ \bibinfo {author} {\bibfnamefont {Shayan}\
  \bibnamefont {Majidy}},\ }\bibfield  {title} {\enquote {\bibinfo {title} {How
  to build hamiltonians that transport noncommuting charges in quantum
  thermodynamics},}\ }\href {\doibase 10.1038/s41534-022-00516-4} {\bibfield
  {journal} {\bibinfo  {journal} {npj Quantum Information}\ }\textbf {\bibinfo
  {volume} {8}},\ \bibinfo {pages} {10} (\bibinfo {year} {2022})}\BibitemShut
  {NoStop}%
\bibitem [{\citenamefont {Kranzl}\ \emph {et~al.}(2022)\citenamefont {Kranzl},
  \citenamefont {Lasek}, \citenamefont {Joshi}, \citenamefont {Kalev},
  \citenamefont {Blatt}, \citenamefont {Roos},\ and\ \citenamefont
  {Halpern}}]{Kranzl2022}%
  \BibitemOpen
  \bibfield  {author} {\bibinfo {author} {\bibfnamefont {Florian}\ \bibnamefont
  {Kranzl}}, \bibinfo {author} {\bibfnamefont {Aleksander}\ \bibnamefont
  {Lasek}}, \bibinfo {author} {\bibfnamefont {Manoj~K.}\ \bibnamefont {Joshi}},
  \bibinfo {author} {\bibfnamefont {Amir}\ \bibnamefont {Kalev}}, \bibinfo
  {author} {\bibfnamefont {Rainer}\ \bibnamefont {Blatt}}, \bibinfo {author}
  {\bibfnamefont {Christian~F.}\ \bibnamefont {Roos}}, \ and\ \bibinfo {author}
  {\bibfnamefont {Nicole~Yunger}\ \bibnamefont {Halpern}},\ }\href {\doibase
  10.48550/ARXIV.2202.04652} {\enquote {\bibinfo {title} {Experimental
  observation of thermalisation with noncommuting charges},}\ } (\bibinfo
  {year} {2022}),\ \Eprint {http://arxiv.org/abs/2202.04652} {arXiv:2202.04652
  [quant-ph]} \BibitemShut {NoStop}%
\bibitem [{\citenamefont {Lu}\ \emph {et~al.}(2021)\citenamefont {Lu},
  \citenamefont {Ba\~nuls},\ and\ \citenamefont {Cirac}}]{Sirui2021}%
  \BibitemOpen
  \bibfield  {author} {\bibinfo {author} {\bibfnamefont {Sirui}\ \bibnamefont
  {Lu}}, \bibinfo {author} {\bibfnamefont {Mari~Carmen}\ \bibnamefont
  {Ba\~nuls}}, \ and\ \bibinfo {author} {\bibfnamefont {J.~Ignacio}\
  \bibnamefont {Cirac}},\ }\bibfield  {title} {\enquote {\bibinfo {title}
  {{Algorithms for Quantum Simulation at Finite Energies}},}\ }\href {\doibase
  10.1103/PRXQuantum.2.020321} {\bibfield  {journal} {\bibinfo  {journal} {PRX
  Quantum}\ }\textbf {\bibinfo {volume} {2}},\ \bibinfo {pages} {020321}
  (\bibinfo {year} {2021})}\BibitemShut {NoStop}%
\bibitem [{\citenamefont {Yang}\ \emph {et~al.}(2022)\citenamefont {Yang},
  \citenamefont {Cirac},\ and\ \citenamefont {Ba\~nuls}}]{Yang2022}%
  \BibitemOpen
  \bibfield  {author} {\bibinfo {author} {\bibfnamefont {Yilun}\ \bibnamefont
  {Yang}}, \bibinfo {author} {\bibfnamefont {J.~Ignacio}\ \bibnamefont
  {Cirac}}, \ and\ \bibinfo {author} {\bibfnamefont {Mari~Carmen}\ \bibnamefont
  {Ba\~nuls}},\ }\bibfield  {title} {\enquote {\bibinfo {title} {Classical
  algorithms for many-body quantum systems at finite energies},}\ }\href
  {\doibase 10.1103/PhysRevB.106.024307} {\bibfield  {journal} {\bibinfo
  {journal} {Phys. Rev. B}\ }\textbf {\bibinfo {volume} {106}},\ \bibinfo
  {pages} {024307} (\bibinfo {year} {2022})}\BibitemShut {NoStop}%
\bibitem [{\citenamefont {Schuckert}\ \emph {et~al.}(2022)\citenamefont
  {Schuckert}, \citenamefont {Bohrdt}, \citenamefont {Crane},\ and\
  \citenamefont {Knap}}]{Schuckert2022}%
  \BibitemOpen
  \bibfield  {author} {\bibinfo {author} {\bibfnamefont {Alexander}\
  \bibnamefont {Schuckert}}, \bibinfo {author} {\bibfnamefont {Annabelle}\
  \bibnamefont {Bohrdt}}, \bibinfo {author} {\bibfnamefont {Eleanor}\
  \bibnamefont {Crane}}, \ and\ \bibinfo {author} {\bibfnamefont {Michael}\
  \bibnamefont {Knap}},\ }\href {\doibase 10.48550/ARXIV.2206.01756} {\enquote
  {\bibinfo {title} {{Probing finite-temperature observables in quantum
  simulators with short-time dynamics}},}\ } (\bibinfo {year} {2022}),\ \Eprint
  {http://arxiv.org/abs/2206.01756} {arXiv:2206.01756 [quant-ph]} \BibitemShut
  {NoStop}%
\end{thebibliography}%
\end{document}